\documentclass[%
 reprint,nofootinbib,
 amsmath,amssymb,
 aps,
]{revtex4-1}

\usepackage{graphicx}
\usepackage{dcolumn}
\usepackage{bm}
\usepackage{hyperref}
\usepackage{ulem}
 \usepackage{color}
     \usepackage{xcolor}

\begin{document}

\preprint{APS/123-QED}

\title{\boldmath Comprehensive study of Lorentz invariance violation in atmospheric and long-baseline experiments}%

\author{Deepak Raikwal}
\email{deepakraikwal@hri.res.in}
 \affiliation{Harish-Chandra Research Institute,  A CI of Homi Bhabha National Institute, Chhatnag Road, Jhunsi, Prayagraj - 211019}
\affiliation{Homi Bhabha National Institute, Anushakti Nagar, Mumbai 400094, India}

\author{Sandhya Choubey}%
 \email{choubey@kth.se}
\affiliation{%
 Department of Physics, School of Engineering Sciences, KTH Royal Institute of Technology,\\ AlbaNova University Center, Roslagstullsbacken 21, SE--106 91 Stockholm, Sweden}%
\affiliation{The Oskar Klein Centre, AlbaNova University Center, Roslagstullsbacken 21,\\ SE--106 91 Stockholm, Sweden}

\author{Monojit Ghosh}
\email{mghosh@irb.hr}
 \affiliation{Center of Excellence for Advanced Materials and Sensing Devices, Ruder Bo\v{s}kovi\'c Institute, 10000 Zagreb, Croatia}

\date{\today}

\begin{abstract}
In this paper, we have presented a comprehensive study of Lorentz Invariance Violation (LIV) in the context of the atmospheric neutrino experiment ICAL and the long-baseline experiments T2HK and DUNE. Our study consists of the full parameter space of the LIV parameters, i.e., six CPT-violating LIV parameters ($a_{\alpha \beta}$) and six CPT-conserving LIV parameters ($c_{\alpha \beta}$). In this study, our objective is to calculate the upper bound on all the LIV parameters with respect to the individual experiments as well as their combination. Our results show that DUNE gives the best sensitivity for the parameters $a_{ee}$, $a_{e\mu}$, $a_{e\tau}$, and $a_{\mu\tau}$ in its 7 years of running, whereas ICAL gives the best sensitivity on $a_{\mu\mu}$, $a_{\mu\tau}$, $c_{ee}$, $c_{\mu\mu}$, $c_{\tau\tau}$, and $c_{\mu\tau}$ in its 10 years of running. For $a_{\tau\tau}$, the sensitivities of DUNE and ICAL are almost same. The combination of T2HK, DUNE, and ICAL gives the best sensitivity for $a_{e\mu}$ and $a_{ee}$ with respect to all the existing bounds in the literature. For the CPT-even diagonal parameters $c_{ee}$ and $c_{\mu\mu}$, our work provides the first-ever bounds. 

\end{abstract}

\maketitle

\section{Introduction}
\label{sec:intro}

The Standard Model (SM) has been highly successful in explaining physics at low energies. However, at the Planck scale ($M_P \sim 10^{19}$ GeV), physics behaves differently, deviating from the rules set by the SM. It is widely accepted that the SM is a low-energy approximation of a more comprehensive theory that encompasses physics at both low energies and the Planck scale. At the Planck scale, various unusual physics scenarios can arise, such as Lorentz invariance violation (LIV) \cite{Kostelecky:1988zi} and CPT violation, indicating a departure from the exact symmetries preserved by the SM. Consequently, the laws governing physics at the Planck scale and low energies can be described by an effective quantum field theory that incorporates the SM. One example of such a theory is the Standard Model Extension (SME), which permits arbitrary coordinate-independent Lorentz invariance violation \cite{Colladay:1998fq}.

The SME extends the SM by including additional terms in its Lagrangian that can violate Lorentz symmetry in different ways. These violations can be realized through spontaneous Lorentz violation proposed by string and field theories of gravity, non-commutative field theories, quantum gravity, and other approaches. The SME encompasses various particle sectors, including quark, gluon, electron, muon, and neutrino sectors, and experimental bounds on Lorentz violation in these sectors have been studied and reported in the literature. Although no experimental evidence of Lorentz violation has been found so far, sensitivity limits have been established by experiments. These limits can be further improved through combined analyses of different experiments. While certain parameters related to Lorentz violation in the neutrino sector have not been extensively studied, ongoing research aims to explore and refine our understanding in this area.

The interference phenomenon of neutrino oscillation, in which active neutrinos oscillate among their flavors, provides a unique opportunity to probe LIV effects. In the presence of LIV, the neutrino oscillation Hamiltonian gets modified, and therefore terms describing LIV appear in the neutrino oscillation probabilities. The LIV parameters that appear in the neutrino oscillation probabilities can be either CPT-violating or CPT-conserving. In this paper, we present a comprehensive analysis of LIV in the context of upcoming long-baseline neutrino experiments T2HK~\cite{Hyper-Kamiokande:2016srs} in Japan and DUNE~\cite{dune} in the USA, and the upcoming atmospheric neutrino experiment ICAL~\cite{ICAL:2015stm} at the INO facility in India. In particular, we study the capability of these experiments to put an upper bound on the parameters of LIV. Note that the study of LIV in the context of DUNE and ICAL has been performed earlier. In what follows next, we mention the studies that have been carried out in the past and the novel features in our work, which we will show for the first time. In the context of DUNE, Ref.~\cite{Barenboim:2018ctx} obtained the upper bounds on the CPT-violating LIV parameters, and Ref.~\cite{KumarAgarwalla:2019gdj} studied the effect of LIV on the determination of the octant of the atmospheric mixing angle $\theta_{23}$ and on the determination of the leptonic phase $\delta_{\rm CP}$. A combined bound on the CPT-violating LIV parameters for DUNE and the long-baseline experiment option at KM3NeT~\cite{KM3Net:2016zxf}, namely P2O~\cite{Akindinov:2019flp}, has been obtained in Ref.~\cite{Fiza:2022xfw}. For ICAL, Ref.~\cite{Sahoo:2021dit} calculated bounds on the three off-diagonal CPT-violating LIV parameters, using events calculated in the energy window 1 GeV to 25 GeV, with the assumption that the phases associated with the LIV parameters have less than 1$\%$ effect on the results. To the best of our knowledge, the study of LIV in the context of T2HK has not been performed in the past\footnote{When our paper was in the final stage of preparation, two papers appeared in the arXiv which studied LIV in long-baseline experiments. Ref.~\cite{Sarker:2023mlz} studied LIV in the context of DUNE, and Ref.~\cite{Agarwalla:2023wft} studied LIV in the context of DUNE and T2HK. In our paper, we will not discuss the results of these two papers.}. Apart from T2HK, DUNE, and ICAL, LIV has been studied in the context of Super-Kamiokande~\cite{Super-Kamiokande:2014exs}, IceCube~\cite{IceCube:2017qyp, Crivellin:2020oov}, T2K, and NO$\nu$A~\cite{Majhi:2019tfi, Lin:2021cst, Rahaman:2021leu}. In the present work, we aim to perform a detailed study of LIV in the context of T2HK, DUNE, and ICAL and compare our results with the existing bounds on the LIV parameters. In our study, we will consider the complete LIV parameter space, namely, six CPT conserving LIV parameters and six CPT violating LIV parameters. In addition, for ICAL we calculate sensitivities for an extended energy range of up to 100 GeV and consider the full effect of the phases associated with the parameters of LIV. We will show how our results for ICAL improve in comparison to the old results. This improvement comes mainly from the higher energy events included in our analysis. Furthermore, our results will also show that the phases associated with the LIV parameters have a non-trivial effect on the final results in ICAL. Apart from the individual sensitivities of each experiment, we will also show the combined sensitivity of these three experiments. Therefore, for the very first time, our work demonstrates: (i) a study of LIV in the context of T2HK, (ii) a study of CPT conserving LIV parameters in the context of DUNE and ICAL, (iii) an analysis of the diagonal CPT violating LIV parameters in the context of ICAL, (iv) a study of LIV with an extended energy window up to 100 GeV with the full effect of the phases associated with the parameters of LIV in ICAL, (v) combined sensitivity for long-baseline and atmospheric experiments, and (vi) a detailed comparison of all the available bounds.

The paper is organized as follows. In the next section, we will briefly discuss the theory of LIV and show how it changes the Hamiltonian of the neutrino oscillations. In section~\ref{exp}, we will discuss the experimental configurations that we use in our analysis. In section~\ref{prob}, we study the effect of LIV at the probability level, and in section~\ref{sens}, we present the sensitivity of the experiments in terms of their capability to put an upper bound on the LIV parameters. After that, in section~\ref{comp}, we compare our results with the results obtained in the previous works. Finally, in section~\ref{conc}, we will summarize our results and then conclude.

\section{LIV and its effect on neutrino oscillations}

Lorentz Invariance violating neutrinos and antineutrinos are effectively described by the Lagrangian density,
\begin{equation}
    \mathcal{L}=\frac{1}{2}\bar{\psi}(i\partial-M-\hat{\mathcal{Q}})\psi + h.c.
\label{lag}
\end{equation}
where, $\hat{\mathcal{Q}}$ is generic Lorentz invariance violating operator and fields related to neutrino and antineutrino are introduced by fermionic spinors $\psi$ and $\bar{\psi}$. The first term of the Eq.~(\ref{lag}) is the kinetic term, the second term is the mass term and the third term is the Lorentz invariance violating term. The Lorentz invariance violating part of the Lagrangian can be written for renormalizable Dirac coupling as \cite{Kosteleck__2012}

  \begin{equation}
    \begin{aligned}[b]
        & \mathcal{L}_{\rm LIV} =\frac{-1}{2}\left[a^{\mu}_{\alpha\beta} \bar{\psi_{\alpha}}\gamma_{\mu}\psi_{\beta}+ b^{\mu}_{\alpha\beta} \bar{\psi_{\alpha}}\gamma_{5}\gamma_{\mu}\psi_{\beta}\right]\\
        & \frac{-1}{2}\left[-ic^{\mu\nu}_{\alpha\beta} \bar{\psi_{\alpha}}\gamma_{\mu}\partial_{\nu}\psi_{\beta} - id^{\mu\nu}_{\alpha\beta} \bar{\psi_{\alpha}}\gamma_{5}\gamma_{\mu}\partial_{\nu}\psi_{\beta} \right] + h.c
    \end{aligned}
  \end{equation}

The observable effects on the left handed neutrinos are controlled by 
\begin{equation}
(a_{L})^{\mu}_{\alpha\beta}=(a + b )^{\mu}_{\alpha\beta}, \hfill     (c_{L})^{\mu\nu}_{\alpha\beta}=(c  + d )^{\mu\nu}_{\alpha\beta},
\end{equation}
which are constant Hermitian matrices in the flavor space that can modify the standard vacuum Hamiltonian. The first term is CPT-odd LIV term and second term is CPT-even LIV term. In this work we will focus on the isotropic component of the Lorentz invariance violating terms and we will fix the $(\mu,\nu)$ indices to zero. To simplify our notation, from now on, we will denote the parameters
$(a_{L})^{0}_{\alpha\beta}$ and $(c_{L})^{00}_{\alpha\beta}$ as $a_{\alpha\beta}$ and $c_{\alpha\beta}$.
Explicitly, one can write the Lorentz–violating contribution to the full oscillation Hamiltonian as,
\begin{equation}
    H=UMU^{\dagger}+V_{e}+H_{\rm LIV},
\end{equation}
where U is the PMNS(Pontecorvo–Maki–Nakagawa–Sakata) \cite{pmns} mixing matrix having three mixing angles $\theta_{12}$, $\theta_{13}$ ,$\theta_{23}$ and one CP phase $\delta_{\rm CP}$, $M$ is the neutrino mass matrix given by
\begin{equation}
    M=\frac{1}{2E}\begin{pmatrix}
0 & 0 & 0\\
0 & \Delta m^{2}_{21} & 0\\
0 & 0 & \Delta m^{2}_{31}
\end{pmatrix},
\end{equation}
with $\Delta m^2_{21} = m_2^2 - m_1^2$ and $\Delta m^2_{31} = m_3^2 - m_1^2$ where $m_1$, $m_2$ and $m_3$ are the masses of the active neutrinos, $V_{e}$ is the matter potential 
\begin{equation}
    V_{e}=\pm\sqrt{2}G_{F}\begin{pmatrix}
N_{e} & 0 & 0\\
0 & 0 & 0\\
0 & 0 & 0
\end{pmatrix},
\end{equation}
where $G_{F}$ is the Fermi constant and $N_{e}$ is the electron density. The $+$ sign in $V_{e}$ is for neutrinos and $-$ sign is for antineutrinos. The term $H_{\rm LIV}$ is given by
\begin{equation}
    H_{\rm LIV}=\pm\begin{pmatrix}
a_{ee} & a_{e\mu} & a_{e\tau}\\
a_{e\mu}^{\star} & a_{\mu\mu} & a_{\mu\tau}\\
a_{e\tau}^{\star} & a_{\mu\tau}^{\star} & a_{\tau\tau}
\end{pmatrix}-\frac{4}{3}E\begin{pmatrix}
c_{ee} & c_{e\mu} & c_{e\tau}\\
c_{e\mu}^{\star} & c_{\mu\mu} & c_{\mu\tau}\\
c_{e\tau}^{\star} & c_{\mu\tau}^{\star} & c_{\tau\tau}
\end{pmatrix}.
\label{equ:hliv}
\end{equation}
Note that the diagonal parameters in Eq.~\ref{equ:hliv} are real while the off-diagonal parameters are complex with a phase $\phi^{a/c}_{\alpha \beta}$ ($\alpha \neq \beta$) associated with them. In the CPT violating first term, the $+$ sign in $V_{e}$ is for neutrinos and $-$ sign is for antineutrinos. The second term being CPT conserving is the same for neutrinos and antineutrinos.

\section{Experimental configuration}
\label{exp}

In this section, we discuss the experimental configurations of T2HK, DUNE, and ICAL that we use in our analysis. As these experiments have different combinations of energy, baselines, and matter effects, the sensitivity of one experiment can be complementary to another. We simulate T2HK and DUNE using GLoBES~\cite{Huber:2004ka,Huber:2007ji}, and for ICAL, we use the software from the INO collaboration. In order to implement LIV in GLoBES, we have written an independent probability engine and then included it in GLoBES. We also use the same probability engine for ICAL.

\subsection{ICAL detector at the INO facility}
\subsubsection{Detector}

 The India-based neutrino observatory (INO) will house a 50 kton magnetized iron calorimeter (ICAL) detector to study oscillations of the atmospheric neutrinos. The resistive plate chambers (RPCs)~\cite{ICAL:2015stm} will be the active detector elements in ICAL, while iron will be the target for atmospheric neutrinos. ICAL will be optimized to be sensitive primarily to atmospheric $\nu_{\mu}$ and $\bar{\nu}_{\mu}$. The detector's structure, with horizontal layers of iron interspersed with RPCs, allows it to have nearly complete coverage to the direction of incoming neutrinos, except for those that produce nearly horizontally traveling $\mu^{\pm}$. As a result, it is sensitive to a wide range of neutrino path lengths. ICAL will be sensitive to muon energy, muon  angle, and total hadron energy \cite{hadron-en} deposited. RPCs with fast response times can distinguish between upward and downward moving $\mu^{\pm}$.  ICAL will be magnetized to about 1.5 T and hence will be able to distinguish between $\nu_{\mu}$ and $\bar{\nu}_{\mu}$. Our analysis uses a muon energy range of $E_{\mu}$ = 1 GeV to 100 GeV. The ICAL detector simulations use the CERN GEANT4~\cite{geant4}-based package developed by the ICAL collaboration. In our analysis, we will consider a run-time of 10 years.

\subsubsection{Flux}

When primary cosmic ray protons and nuclei interact in the atmosphere, they produce secondary "atmospheric" cosmic rays made of hadrons and their decay products. The spectrum of these secondaries peaks in the GeV range but extends to high energy with a power-law spectrum. Neutrinos are the final component of secondary cosmic radiation because they interact weakly, and they include $\nu_{e}$, $\bar{\nu}_{e}$, $\nu_{\mu}$, and $\bar{\nu}_{\mu}$. The neutrino flux varies with location and also has seasonal variations, as described in Ref~\cite{honda}. We used the atmospheric flux generated for the Theni site by Honda et al.~\cite{honda}. The flux was generated for both solar maximum and solar minimum with rock on top of the detector. We use the average of both fluxes for our simulations.

\subsubsection{Event Generator}

The GENIE event generator uses Monte Carlo simulations to generate neutrino nucleon interactions. It supports various neutrino flux formats, including HAKKM, BGLRS, and FLUKA for atmospheric neutrino flux, as well as any flux distribution in the form of a polynomial equation. It can produce cross-section files for any material or neutrino with energies ranging from a few MeV to several hundred GeV, and it also includes the cross-section files generated and distributed by the Fermi collaboration with GENIE MC. In GENIE, it's possible to specify the detector geometry using GEANT4. To adapt GENIE for ICAL, the ICAL collaboration has made the necessary modifications to the software \cite{AJMI2017}.

\subsubsection{Binning Scheme}

We use data binned in three variables: the energy of the observed muon $E^{obs}_{\mu^{\pm}}$, the direction of the observed muon $\cos(\theta^{obs}_{ \mu^{\pm}})$, and the energy of the observed hadron $E^{obs}_{had}$. We calculate the hadron energy as $E^{true}_{had}$ = $E_{\nu}$ - $E_{\mu}$. At the generator level, we obtain the true muon energy and angle, which may differ from the observed quantities due to the detector response. To incorporate neutrino oscillations, we apply the re-weighting algorithm~\cite{barger-code} and then implement the detector response to the events. To account for detector efficiency, we use muon and hadron lookup~\cite{muon-res,Devi:2013wxa} tables provided by the INO collaboration. We assume a Gaussian distribution for the muon energy and angle to account for detector resolution. We use the Vavilov distribution function for the hadron energy~\cite{Devi:2013wxa}. We bin the data in the scheme presented in Table-\ref{table:bin}.

\begin{table}[h!]
\begin{center}
\scalebox{1.1}{
\begin{tabular}{ |c|c|c|c| } 
 \hline
 Observable & Range & Bin width & No. of bins\\ 
\hline
 $E^{obs}_{\mu}$(GeV)(15 bins) & [1,11] & 1 & 10\\
						& [11,21] & 5 & 2\\
						&[21,25] & 4&1\\
						&[25,50] & 25&1\\
						&[50,100] & 50&1\\
$\cos\theta^{obs}_{\mu}$ (15 bins)       & [-1.0,0.0]&0.1&10\\
                        &[0.0,1]&0.2&5\\
$E^{obs}_{had}$ (GeV)  (4bins)                  &[0,2]&1&2\\
						&[2,4]&2&1\\
						&[4,15]&11&1\\
\hline
\end{tabular}}
\caption{The number of bins and bin widths of the three observables $E_{\mu}$, $\cos\theta^{obs}_{\mu}$ and $E_{had}^{obs}$ are given in the table which we use in our analysis for ICAL.}
\label{table:bin}
\end{center}

\end{table}

\subsubsection{Systematic uncertainties}

For the atmospheric neutrino studies, the dominant uncertainties come from uncertainties in the atmospheric neutrino flux calculations. For the fluxes we take the following systematic uncertainties into account: an overall flux normalization error of 20$\%$, correlated tilt error of 5$\%$ and  correlated zenith angle error of 5$\%$. In addition, we take a cross-section error of 10\% to account for the uncertainty in the neutrino-nucleon cross-sections. Finally, we should have systematic uncertainties coming from the ICAL detector configuration.  The ICAL collaboration takes this to be a consolidated 5$\%$  systematic uncertainty, uncorrelated amongst the bins. These errors are put separately for neutrinos and antineutrinos. The magnitude of the systematic errors are taken to be the same for neutrino and antineutrino events.

\subsection{DUNE}

DUNE (Deep Underground Neutrino Experiment) is a versatile neutrino experiment with multiple scientific objectives. It is suitable for long baseline studies, atmospheric neutrino studies, and indirect dark matter detection studies. To simulate DUNE for the long-baseline case, we use the most recent DUNE configuration files provided by the collaboration~\cite{DUNE:2021cuw}. The neutrino source is located at Fermilab and has a beam power of 1.2 MW with a total exposure of (1.1 - 1.9) x 10$^{21}$ POT/year. The beam power will be upgraded to 2.4 MW after 7 years of running. We assume a total run-time of 7 years, with 3.5 years in neutrino mode and the remaining 3.5 years in antineutrino mode. The on-axis 40 kt liquid argon far detector (FD) is housed at the Homestake Mine in South Dakota over a 1300 km baseline. For both the appearance and disappearance channels, we consider neutrino and antineutrino energies ranging from 0 to 20 GeV. The reconstructed energy is divided into a total of 71 bins, with 64 bins having widths of 0.125 GeV in the energy range of 0 to 8 GeV, and 7 bins with variable widths beyond 8 GeV. We include the "wrong-sign" components in the beam when calculating events, whether signal or background, for both $\nu_e/\bar{\nu}_e$ and $\nu_{\mu}/\bar{\nu}_{\mu}$ candidate events. The flux uncertainty for DUNE is approximately 8$\%$ and is dominated by hadron production uncertainties. The flux uncertainty for the near detector/far detector (ND/FD) is approximately 0.5$\%$ at the peak but rises to approximately 2$\%$ in the falling edge and is dominated by focusing effects. 

\subsection{T2HK}

For T2HK, we have utilized the configuration described in Ref.~\cite{Abe:2016ero}. The experiment consists of two water-Cerenkov detector tanks, each with a fiducial volume of 187 kt, located at Kamioka, which is 295 km away from the neutrino source at J-PARC. The beam power is 1.3 MW, and the total exposure is $27\times 10^{21}$ protons on target, which corresponds to ten years of operation. The total run-time is divided into five years in neutrino mode and five years in antineutrino mode. For systematic errors, we considered an overall normalization error of 4.71$\%$ (4.13$\%$) for the appearance (disappearance) channel in neutrino mode and 4.47$\%$ (4.15$\%$) for the appearance (disappearance) channel in antineutrino mode. Both the signal and the background have the same systematic error. In T2HK, we have used 98 equispaced bins in the energy range of 0.2 - 10 GeV.

\section{Discussion at the probability level}
\label{prob}

In this section, we will analyze the impact of LIV on the probability levels for all three experiments. For our analysis, we have considered one parameter at a time and studied the sensitivity of each individual parameter. Therefore, in our study, we have six independent CPT-conserving parameters and six independent CPT-violating parameters. Note that if we consider all the parameters at the same time instead of one at a time, the number of independent parameters in each CPT-conserving case and CPT-violating case would be five. This is because, when we consider all the parameters at the same time, we can subtract a matrix proportional to identity without changing the oscillation probabilities. Therefore, only two of the diagonal parameters are left to be independent. The values of the LIV parameters used in the probability plots are mentioned in the panel titles/legends. The values of the oscillation parameters used in these figures are given in Table~\ref{table:para}. For $\delta_{\rm CP}$, we consider the value $0^\circ$ for ICAL and $0^\circ$ and $-90^\circ$ for T2HK and DUNE. Our choices of $\delta_{\rm CP}$ are motivated by the current best-fit value of this parameter as obtained by the currently running experiment T2K~\cite{con_talk_t2k} and NO$\nu$A~\cite{con_talk}. All the figures are given for normal ordering of the neutrino masses, i.e., $\Delta m^2_{31} > 0$.

\subsection{ICAL}

\begin{figure*}

\begin{minipage}[t]{0.3\textwidth}
  \includegraphics[width=\linewidth]{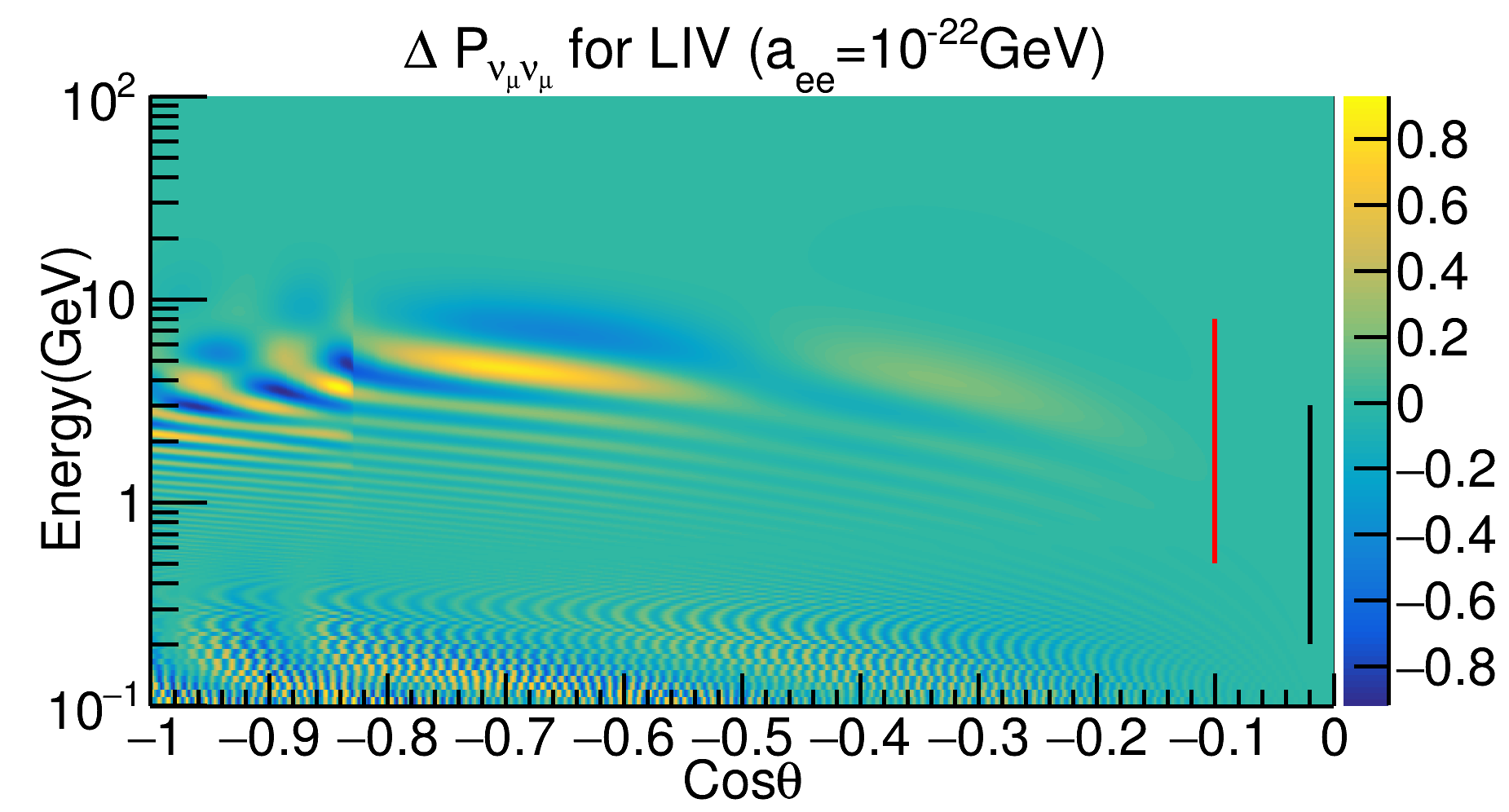}
 \end{minipage}
\begin{minipage}[t]{0.3\textwidth}
  \includegraphics[width=\linewidth]{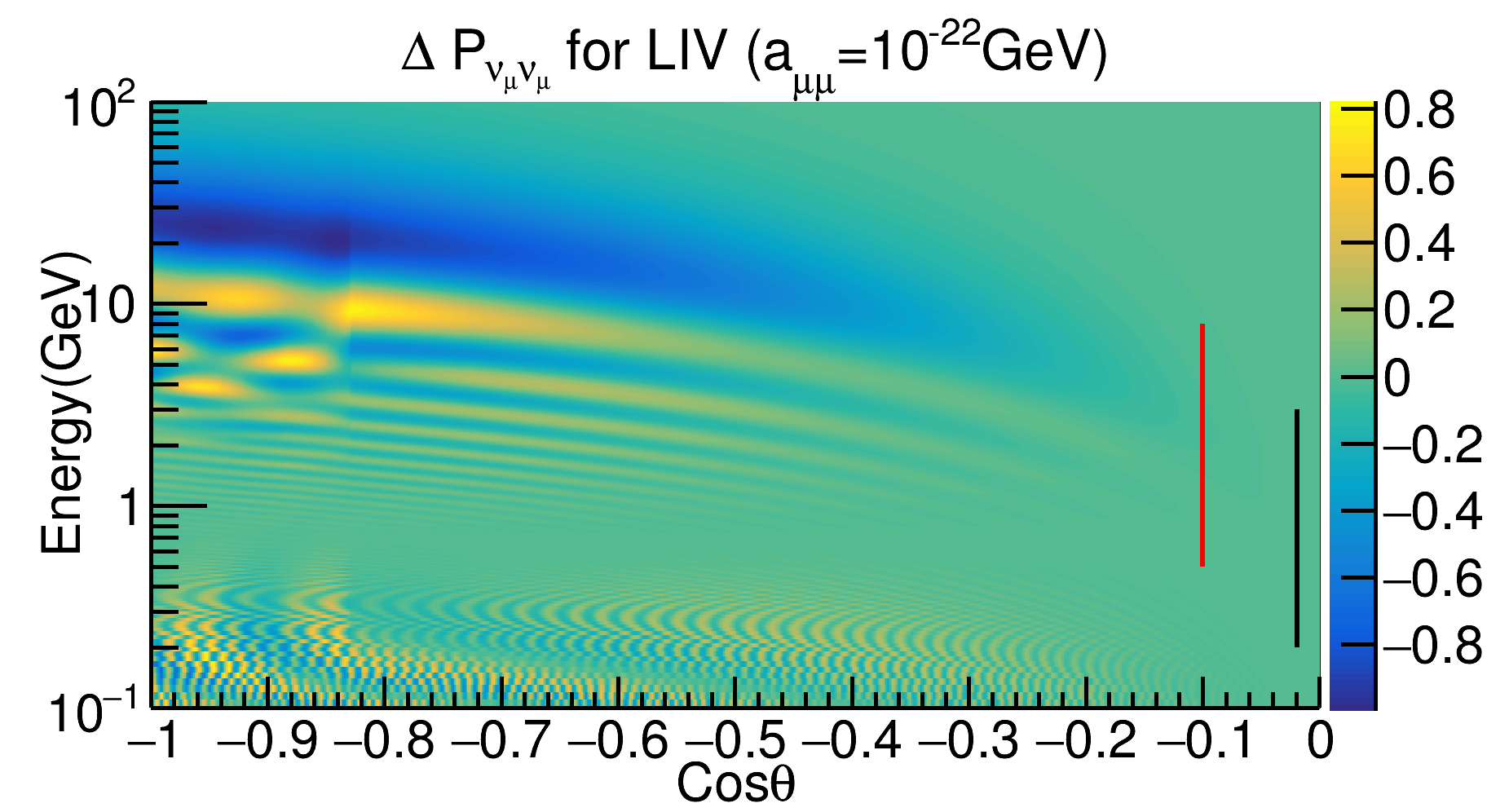}
 \end{minipage}
\begin{minipage}[t]{0.3\textwidth}
  \includegraphics[width=\linewidth]{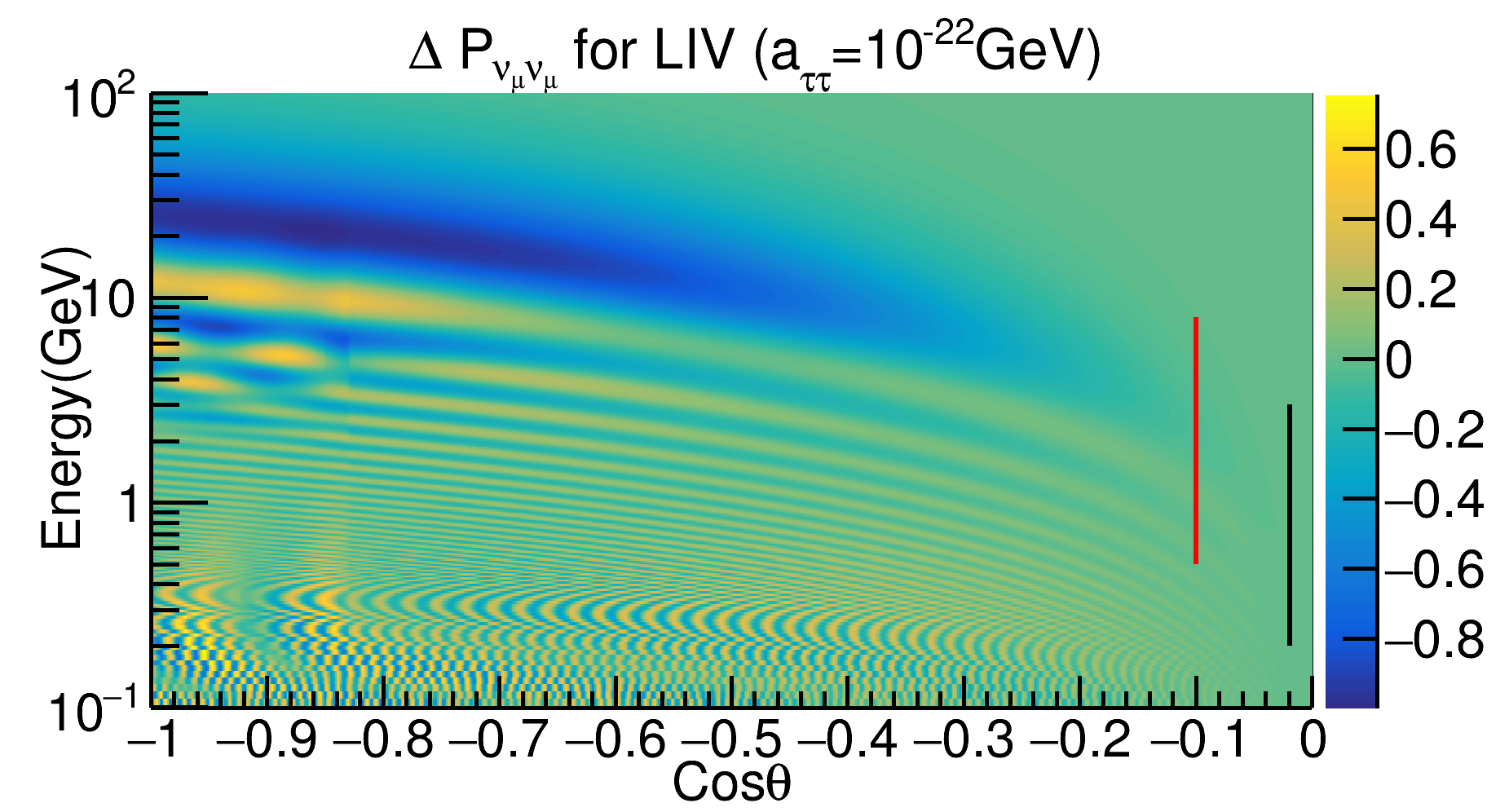}
\end{minipage}%

\begin{minipage}[t]{0.3\textwidth}
  \includegraphics[width=\linewidth]{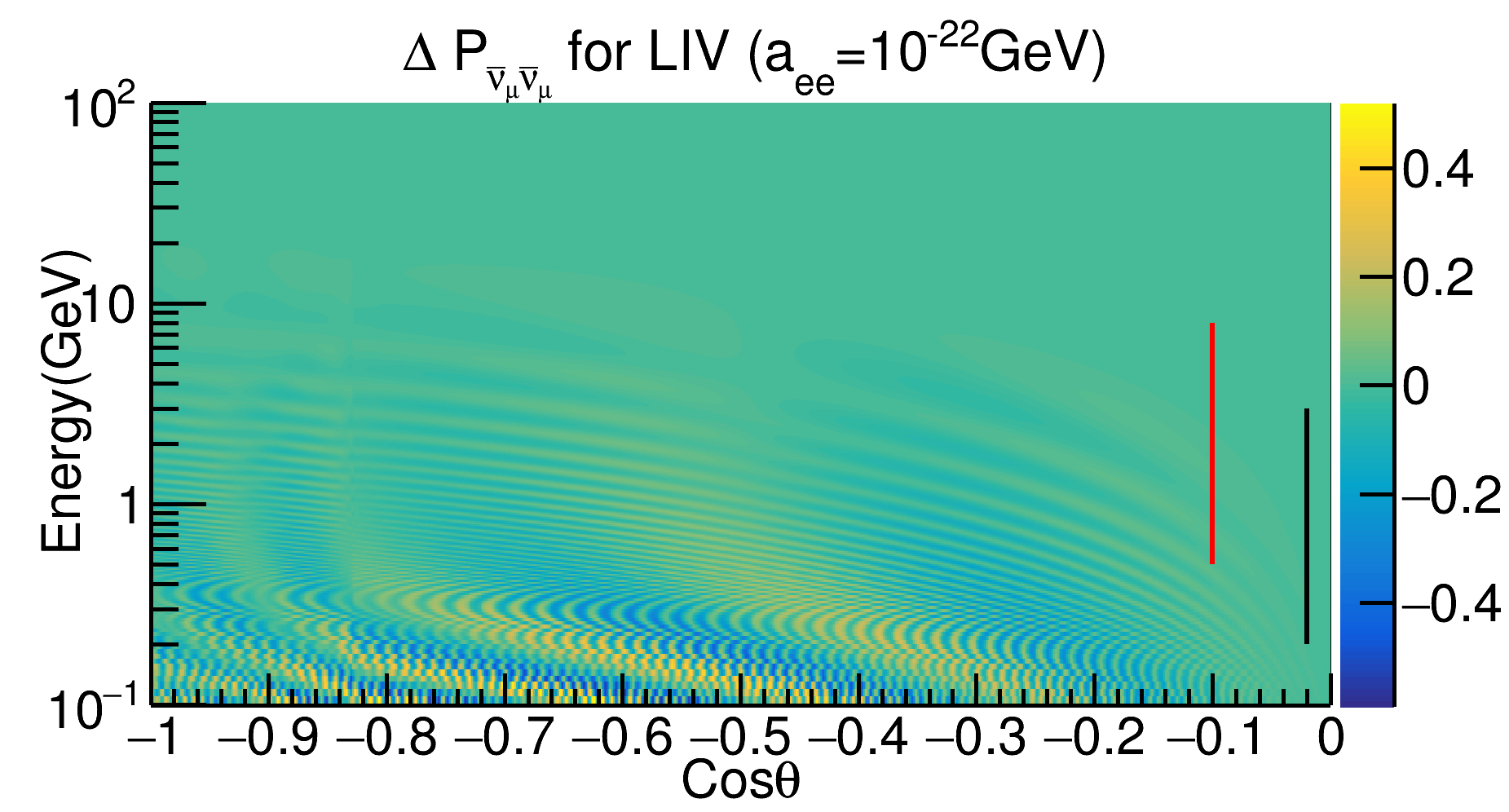}
 \end{minipage}
\begin{minipage}[t]{0.3\textwidth}
  \includegraphics[width=\linewidth]{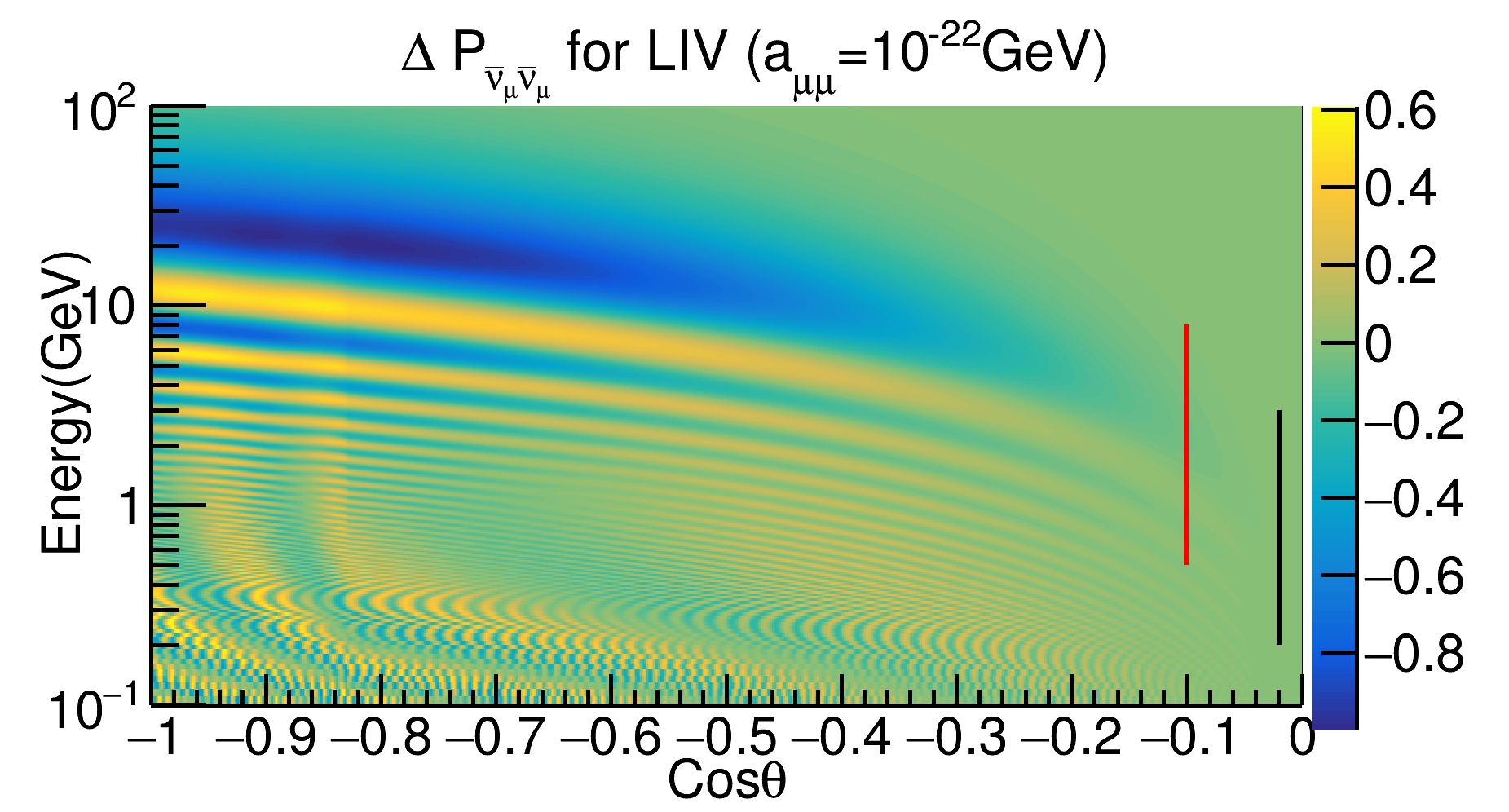}
 \end{minipage}
\begin{minipage}[t]{0.3\textwidth}
  \includegraphics[width=\linewidth]{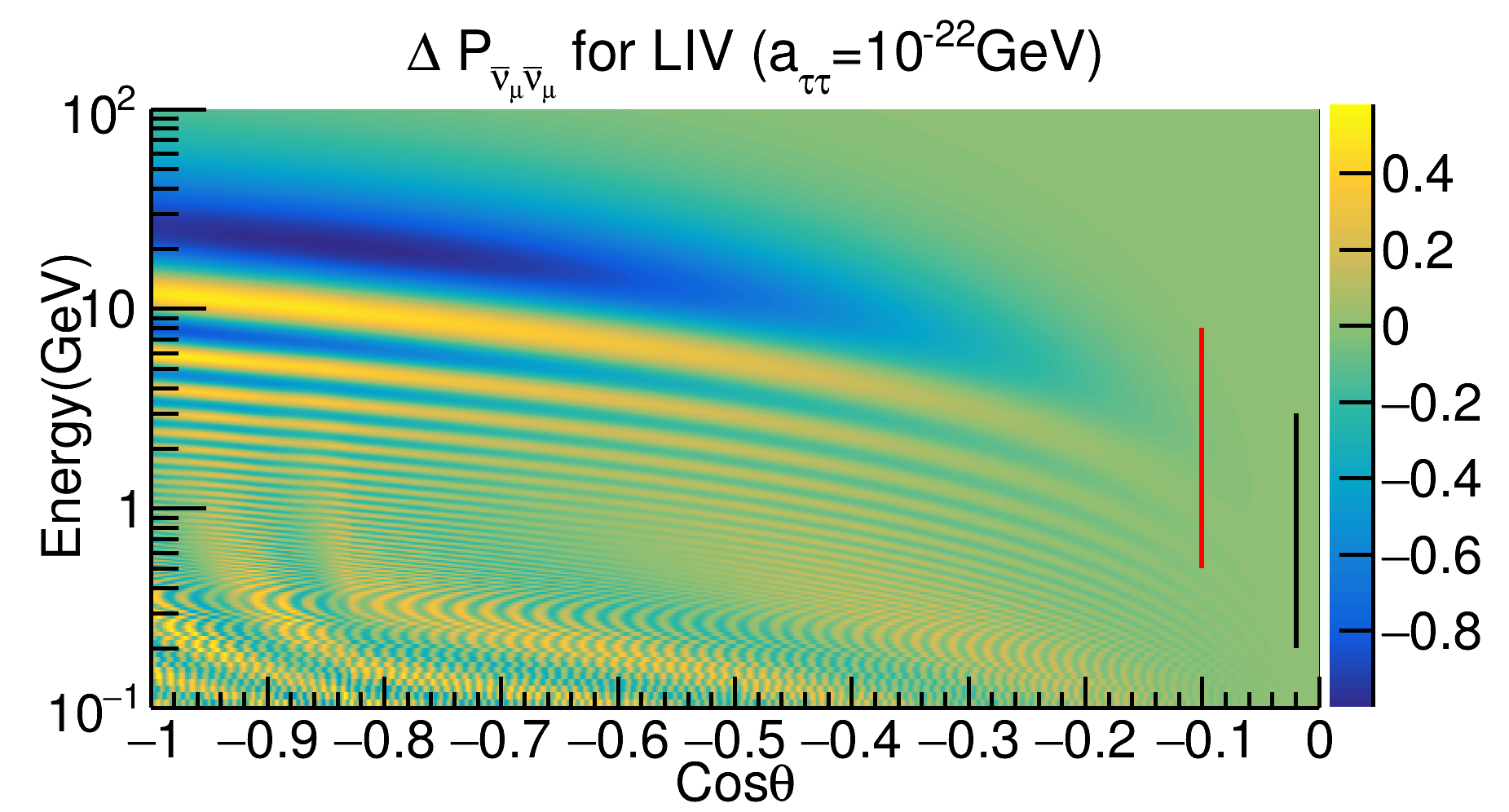}
\end{minipage}%

\begin{minipage}[t]{0.3\textwidth}
  \includegraphics[width=\linewidth]{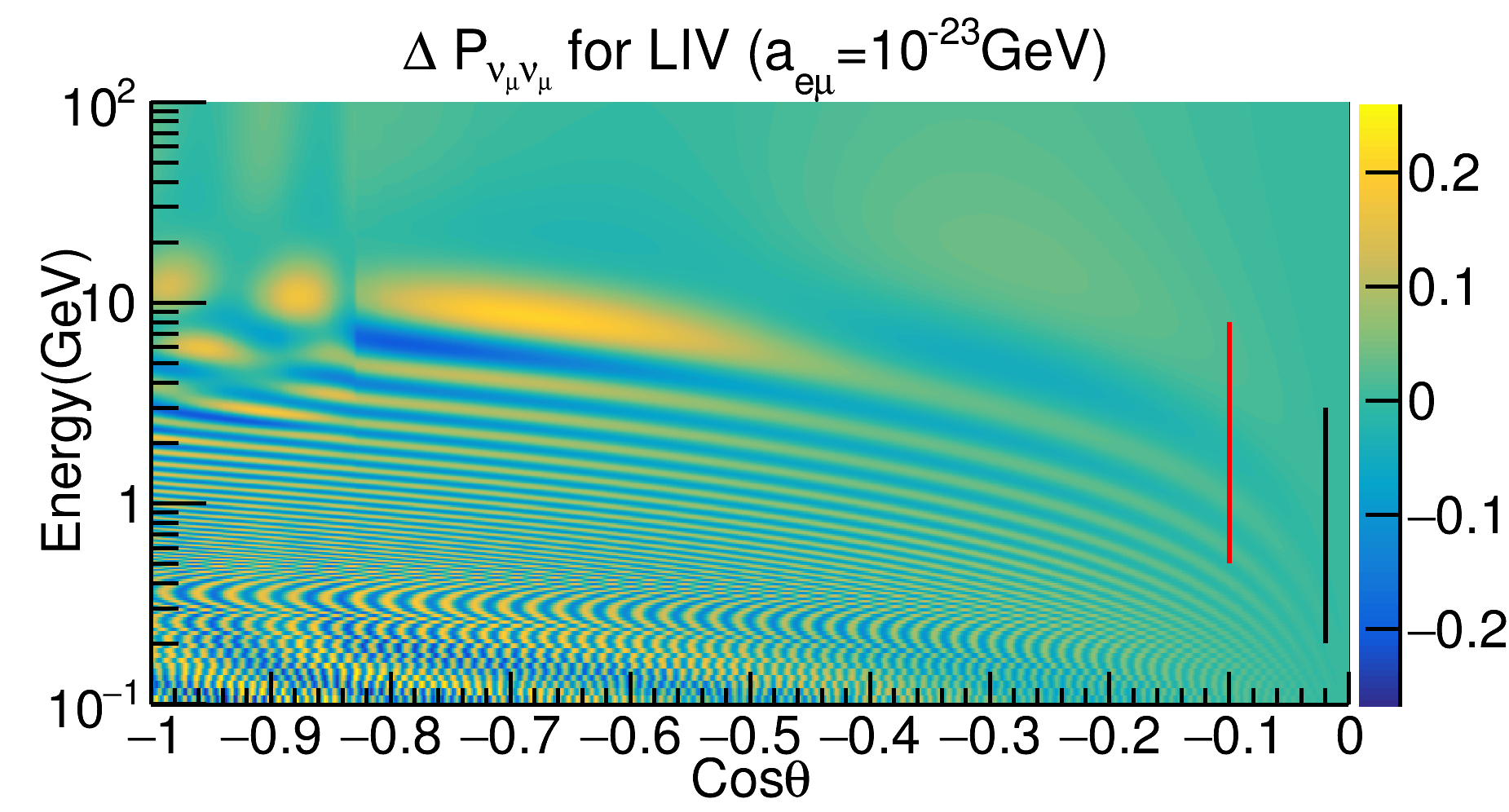}
 \end{minipage}
\begin{minipage}[t]{0.3\textwidth}
  \includegraphics[width=\linewidth]{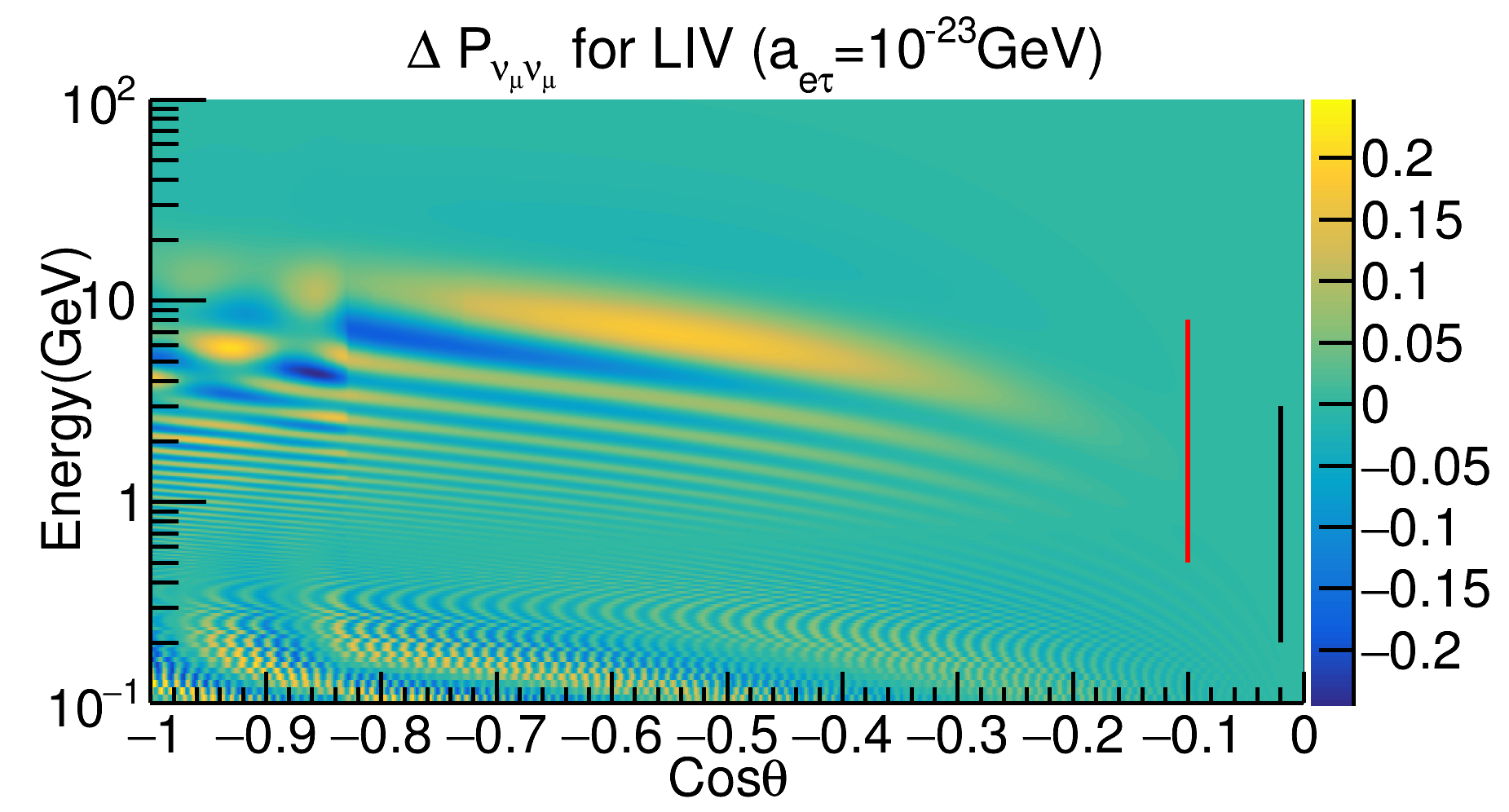}
 \end{minipage}
\begin{minipage}[t]{0.3\textwidth}
  \includegraphics[width=\linewidth]{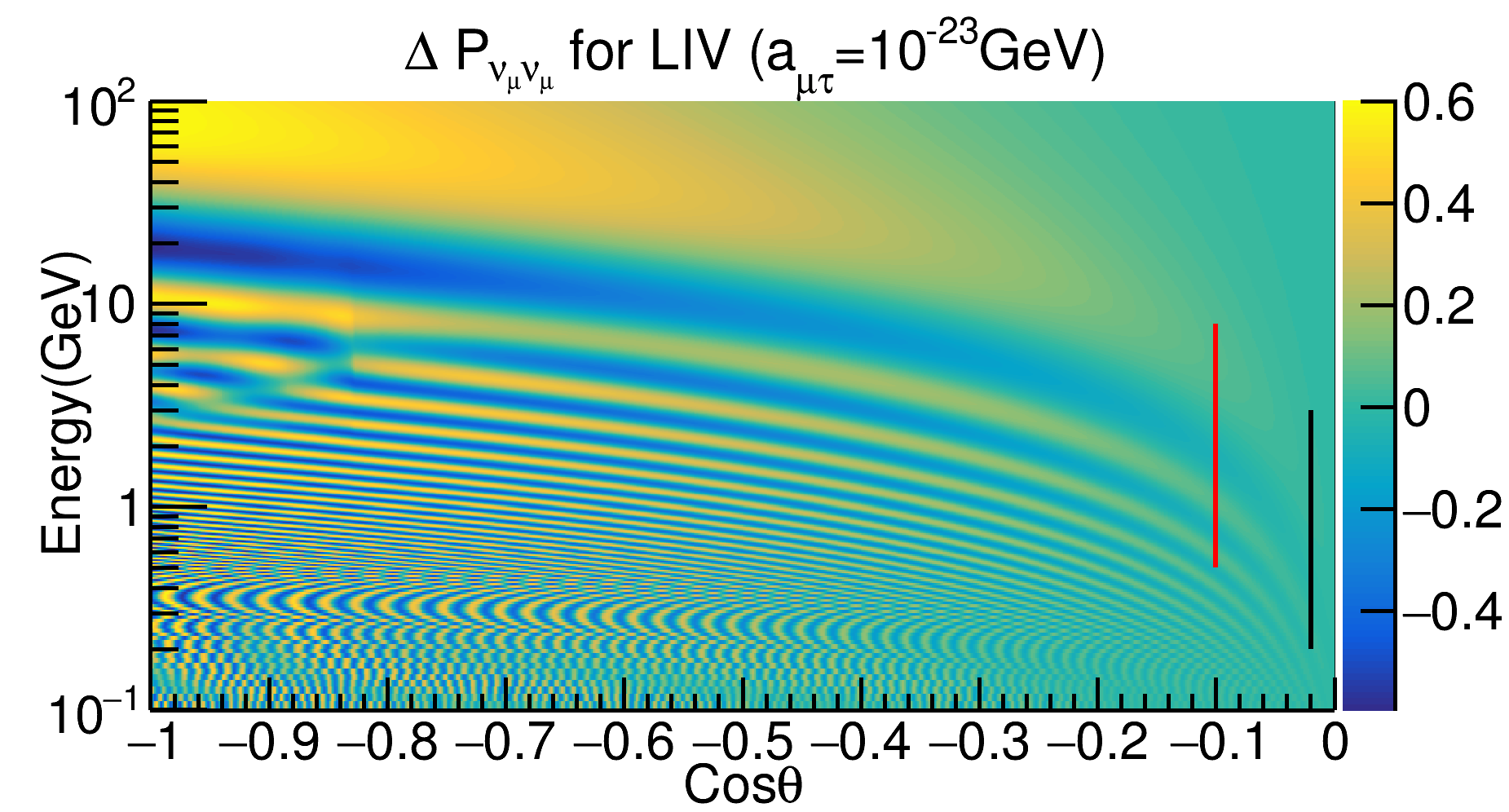}
\end{minipage}%

\begin{minipage}[t]{0.3\textwidth}
  \includegraphics[width=\linewidth]{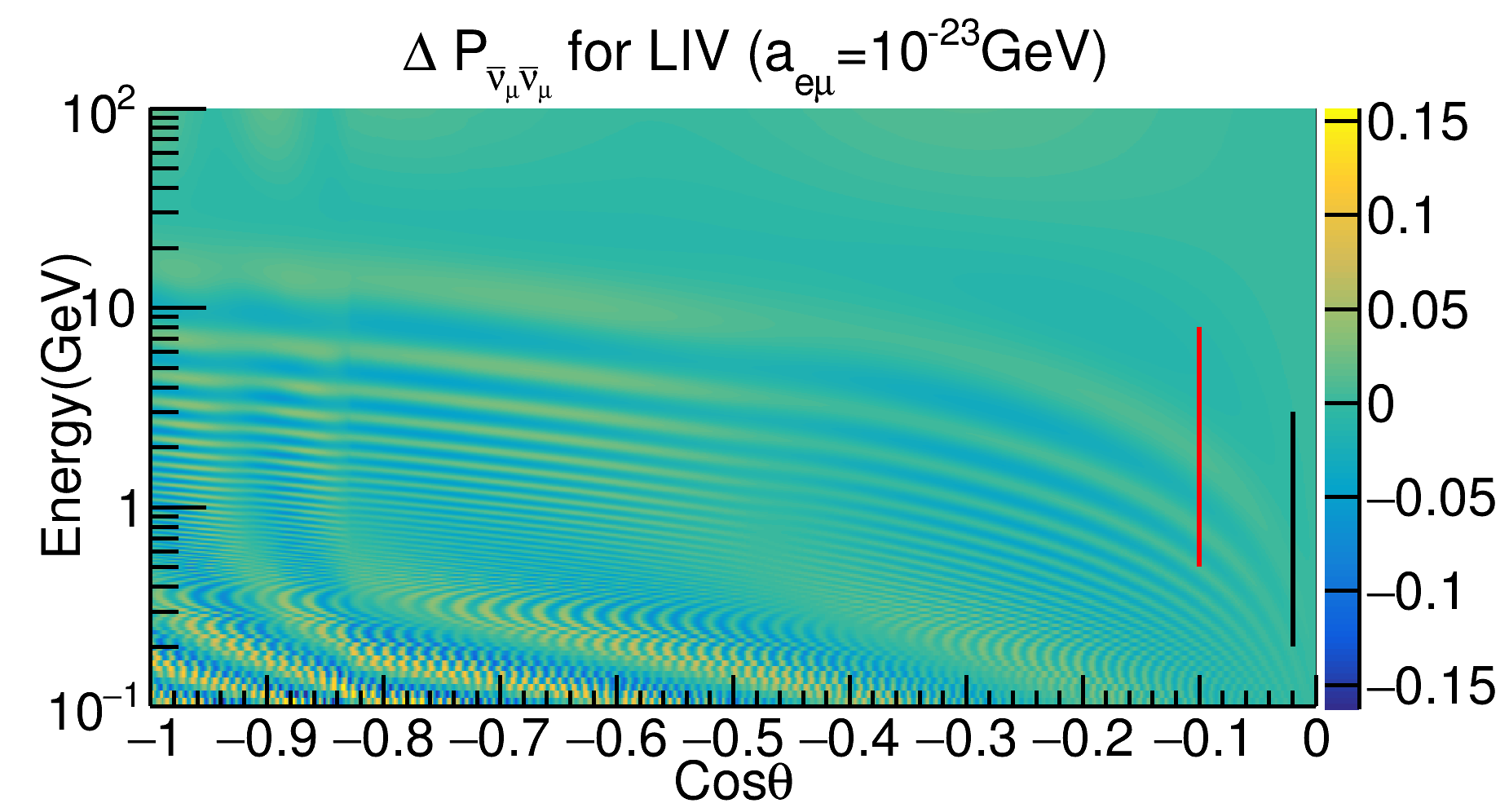}
 \end{minipage}
\begin{minipage}[t]{0.3\textwidth}
  \includegraphics[width=\linewidth]{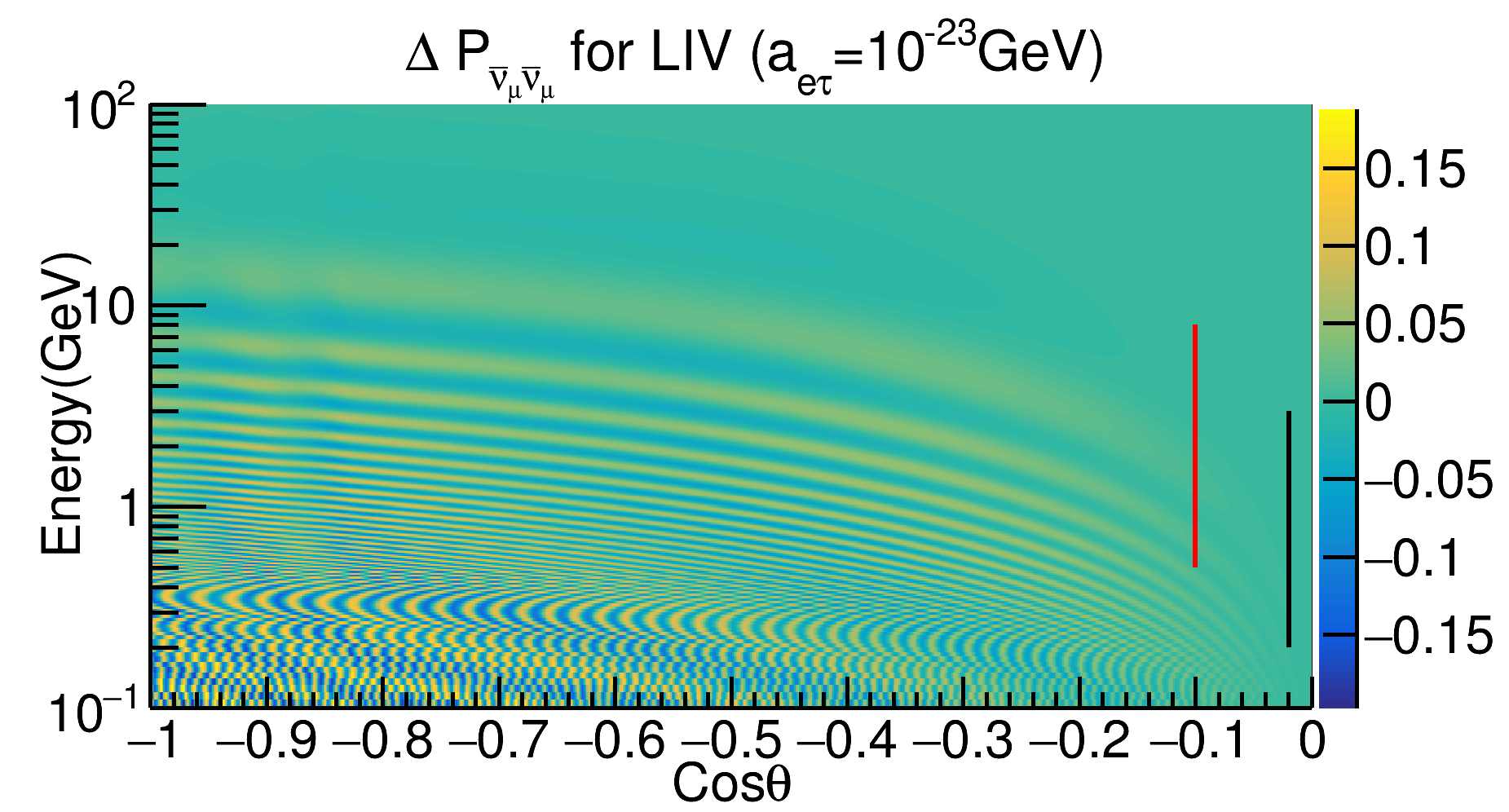}
 \end{minipage}
\begin{minipage}[t]{0.3\textwidth}
  \includegraphics[width=\linewidth]{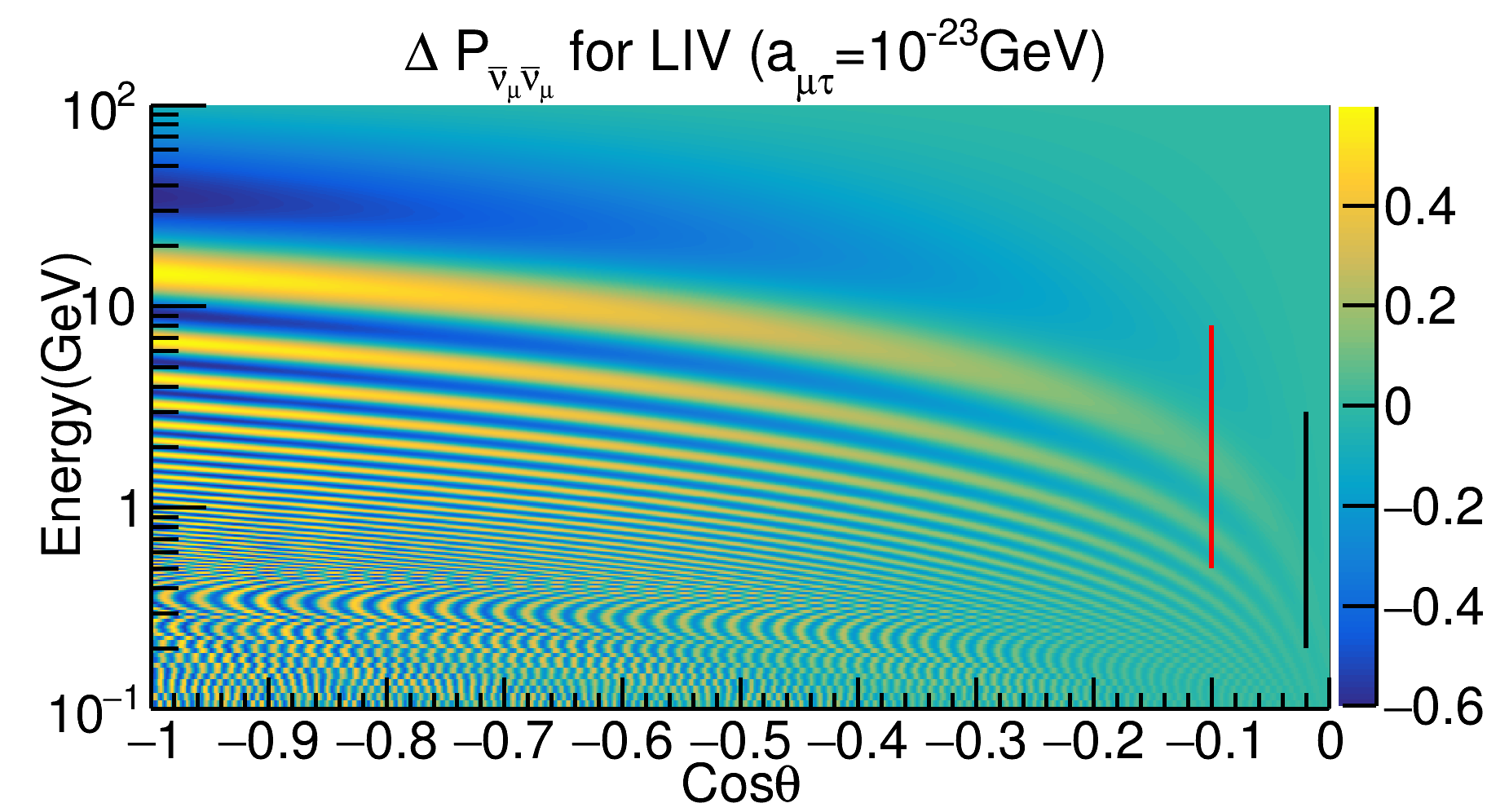}
\end{minipage}%
\caption{Probability oscillogram for $\Delta P_{\nu_{\mu}\nu_{\mu}}$ and $\Delta P_{\bar{\nu}_{\mu}\bar{\nu}_{\mu}}$ channels for $a_{ee}$, $a_{\mu\mu}$, $a_{\tau\tau}$, $a_{e\mu}$, $a_{e\tau}$ and $a_{\mu\tau}$. For ICAL, the relevant energy region of the oscillograms is 1-100 GeV. The red line shows the region relevant for DUNE and the black line shows the region relevant for T2HK.} 
\label{fig:inooscillogrm_a}
\end{figure*}

\begin{figure*}

\begin{minipage}[t]{0.3\textwidth}
  \includegraphics[width=\linewidth]{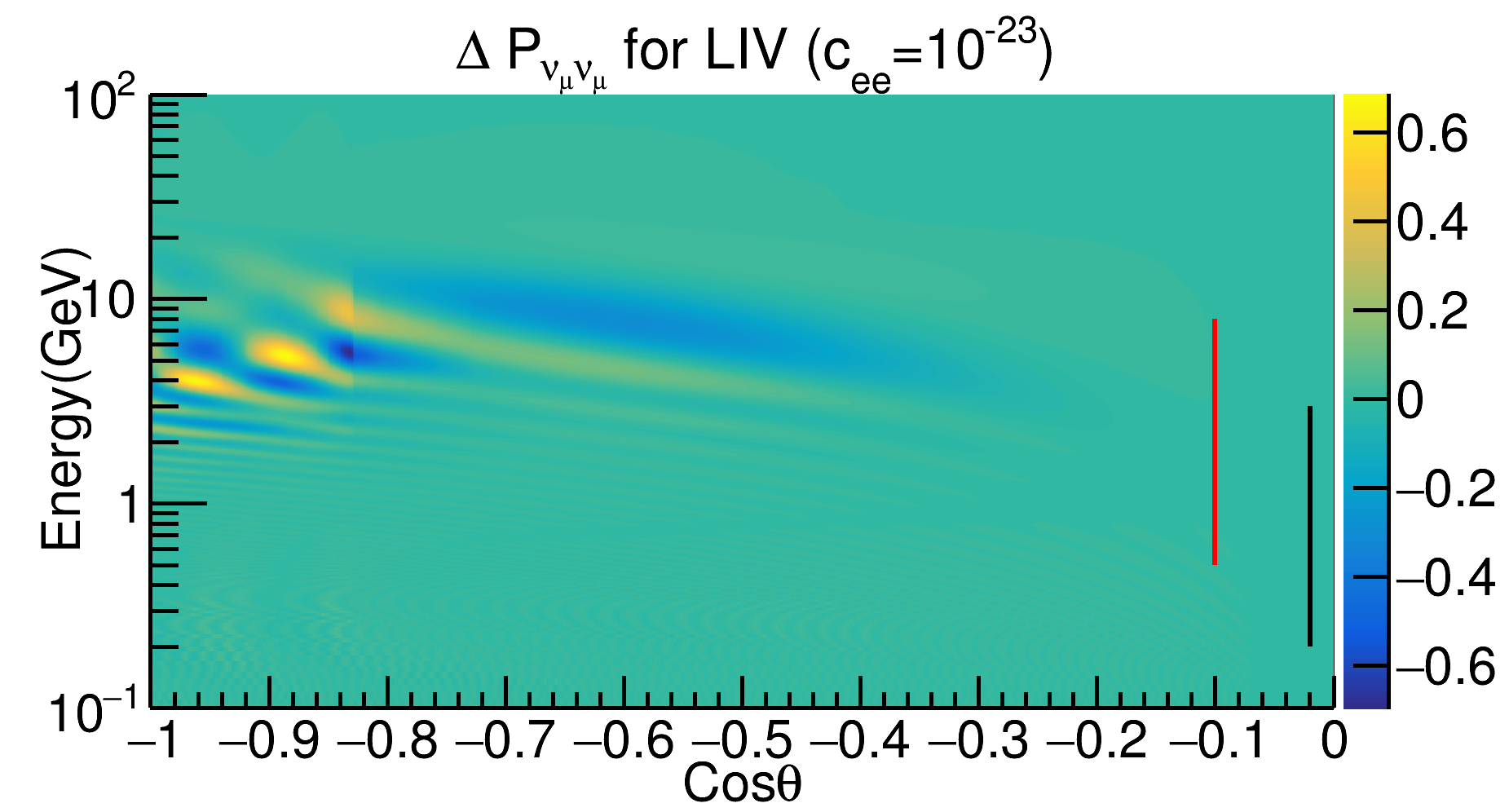}
 \end{minipage}
\begin{minipage}[t]{0.3\textwidth}
  \includegraphics[width=\linewidth]{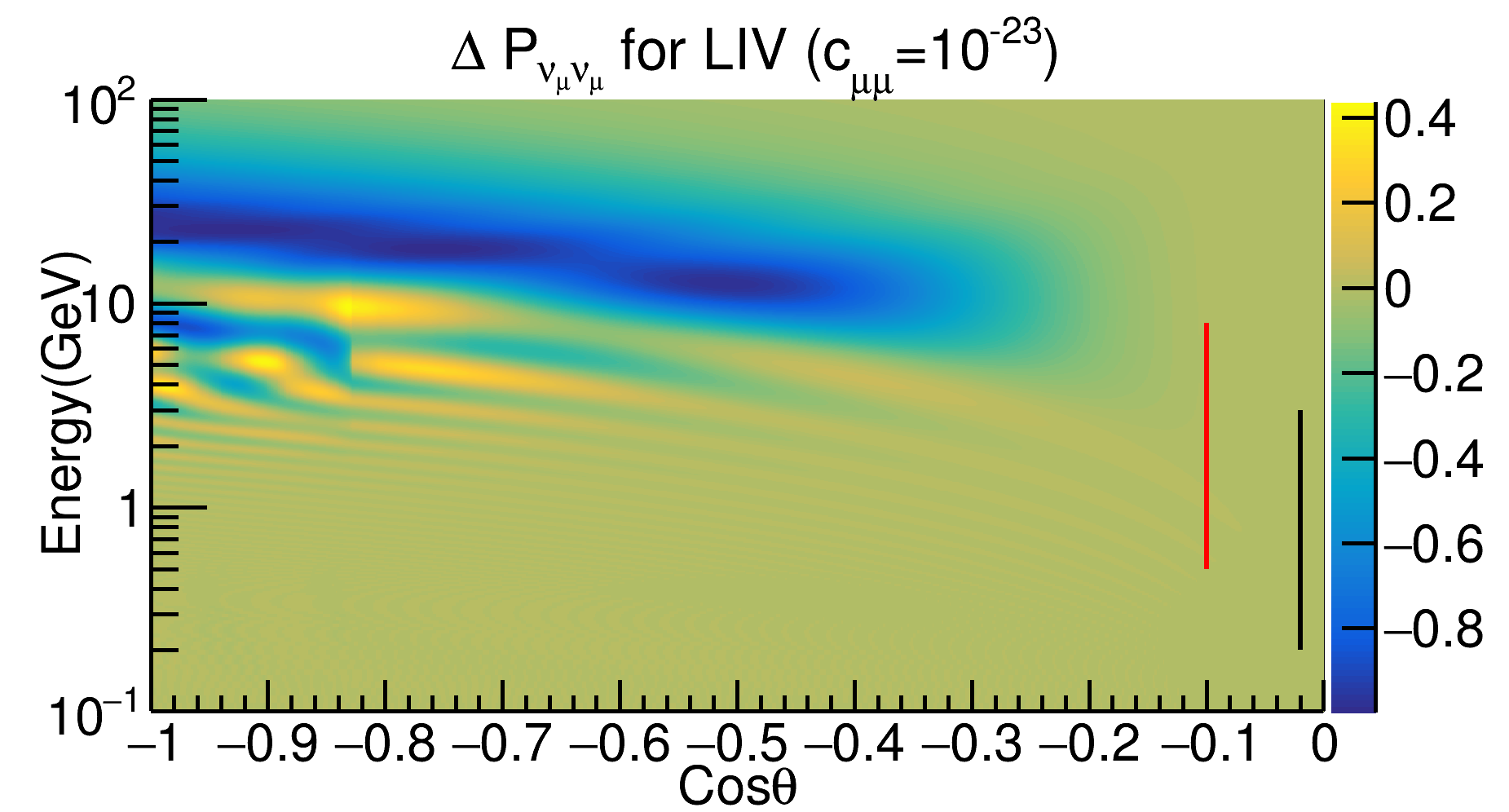}
 \end{minipage}
\begin{minipage}[t]{0.3\textwidth}
  \includegraphics[width=\linewidth]{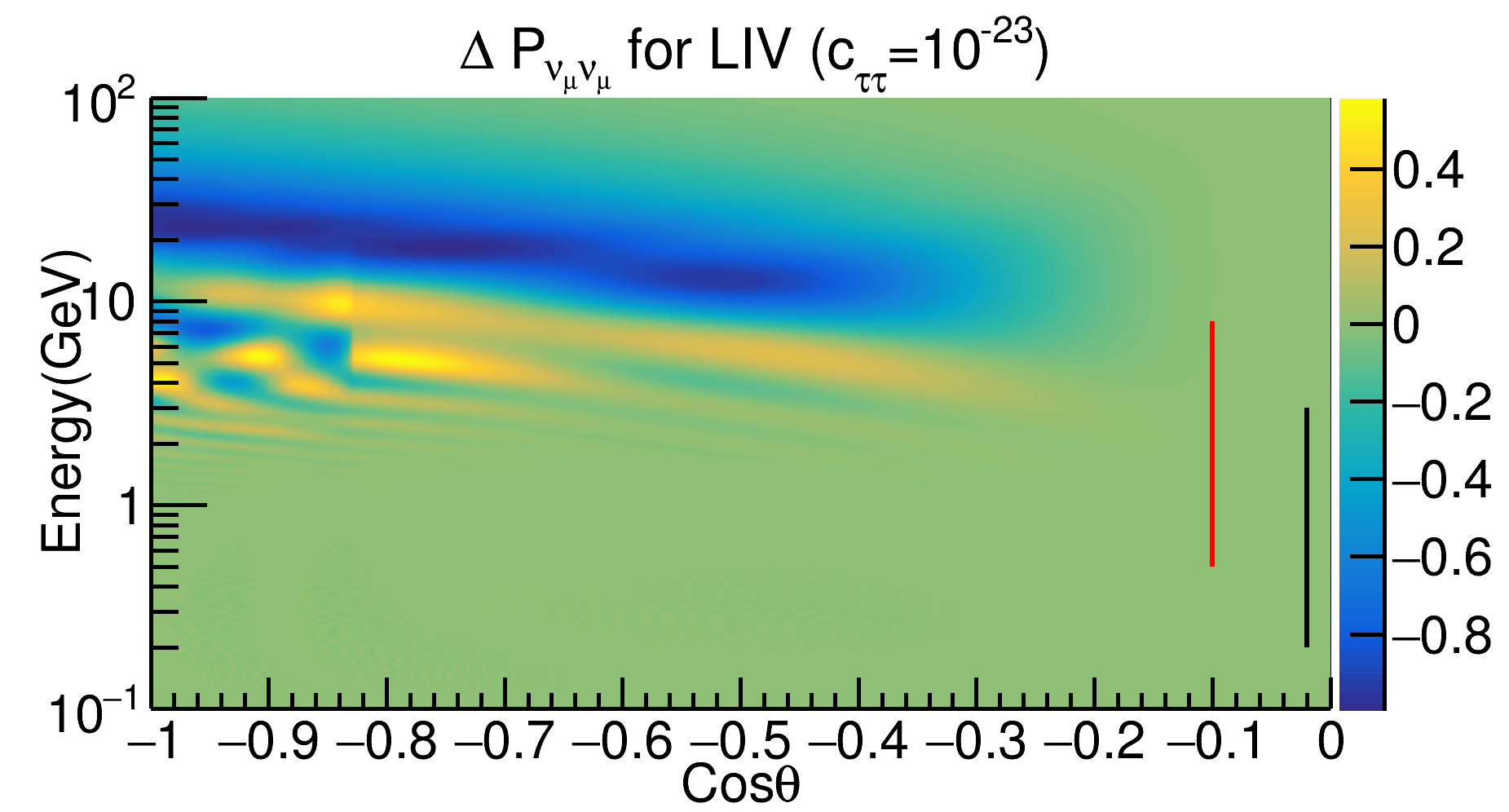}
\end{minipage}%

\begin{minipage}[t]{0.3\textwidth}
  \includegraphics[width=\linewidth]{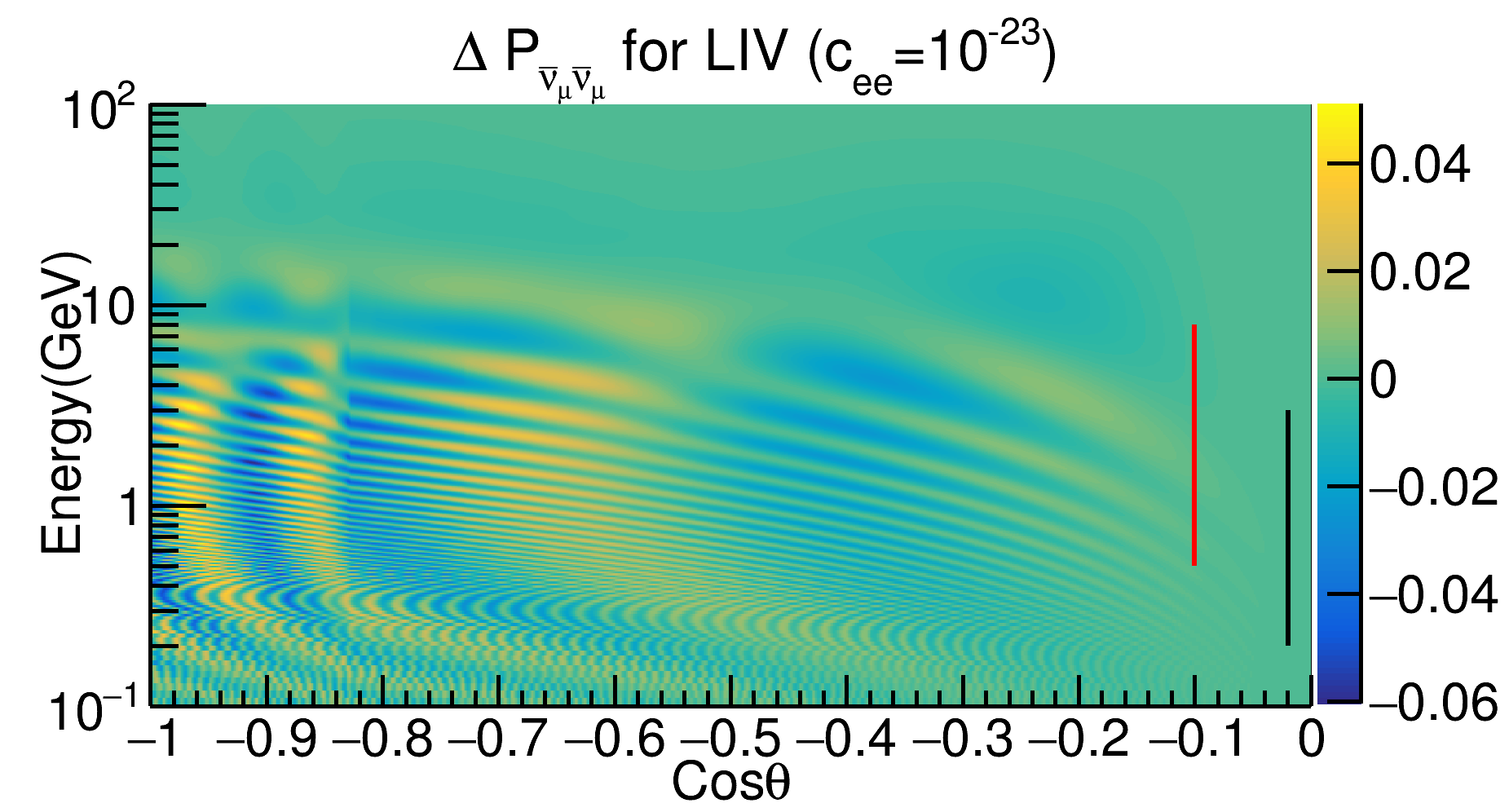}
 \end{minipage}
\begin{minipage}[t]{0.3\textwidth}
  \includegraphics[width=\linewidth]{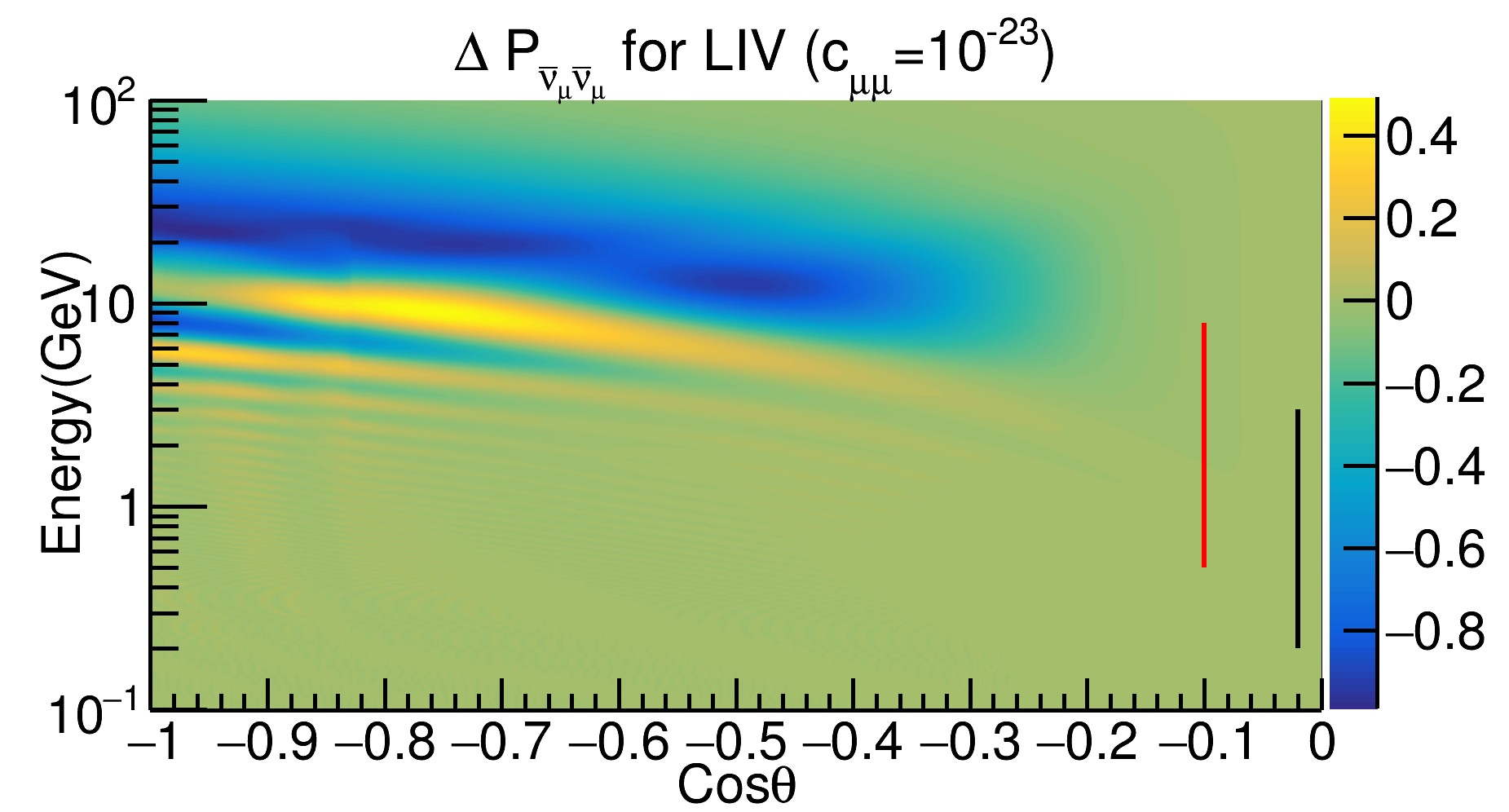}
 \end{minipage}
\begin{minipage}[t]{0.3\textwidth}
  \includegraphics[width=\linewidth]{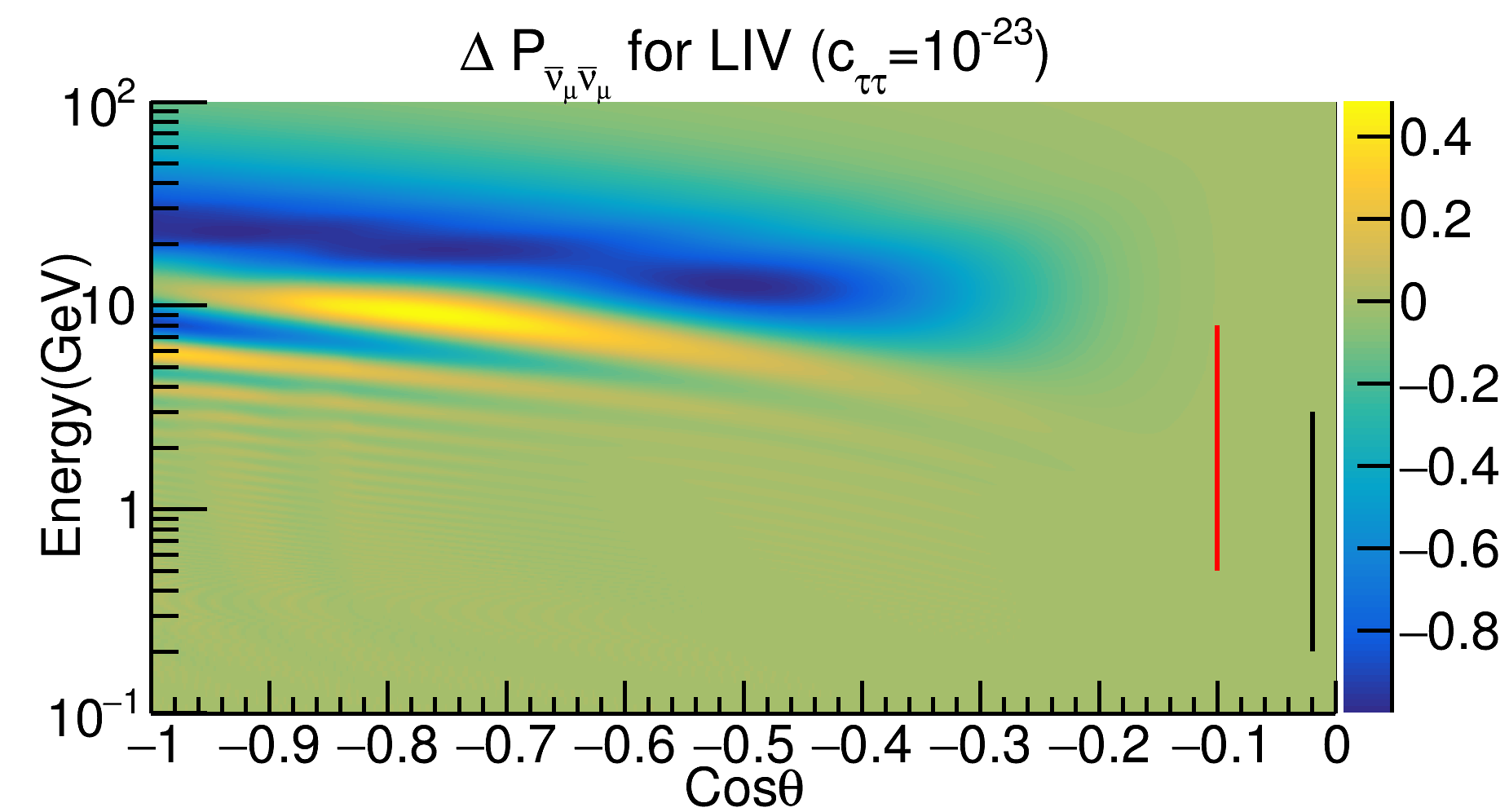}
\end{minipage}%

\begin{minipage}[t]{0.3\textwidth}
  \includegraphics[width=\linewidth]{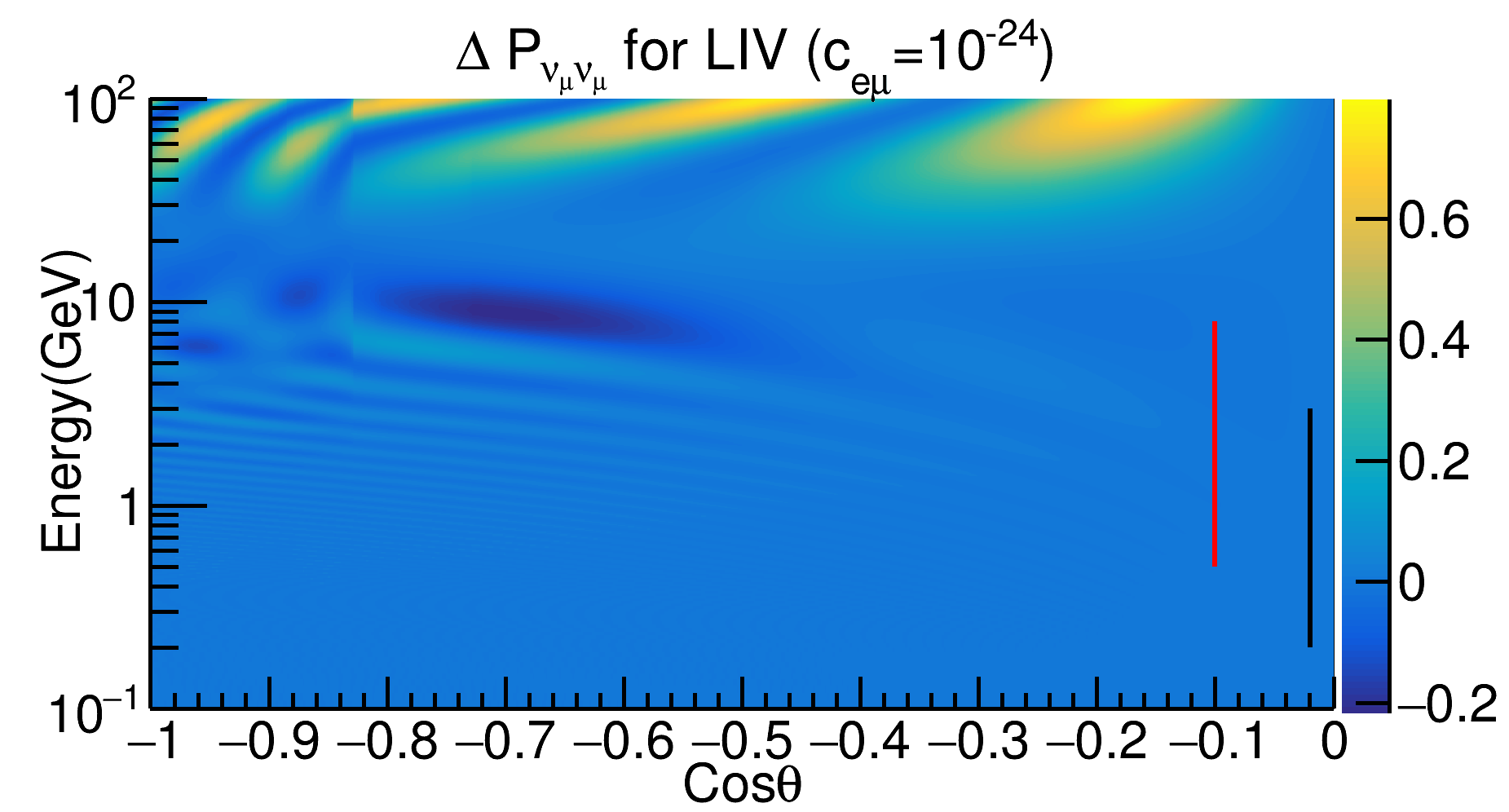}
 \end{minipage}
\begin{minipage}[t]{0.3\textwidth}
  \includegraphics[width=\linewidth]{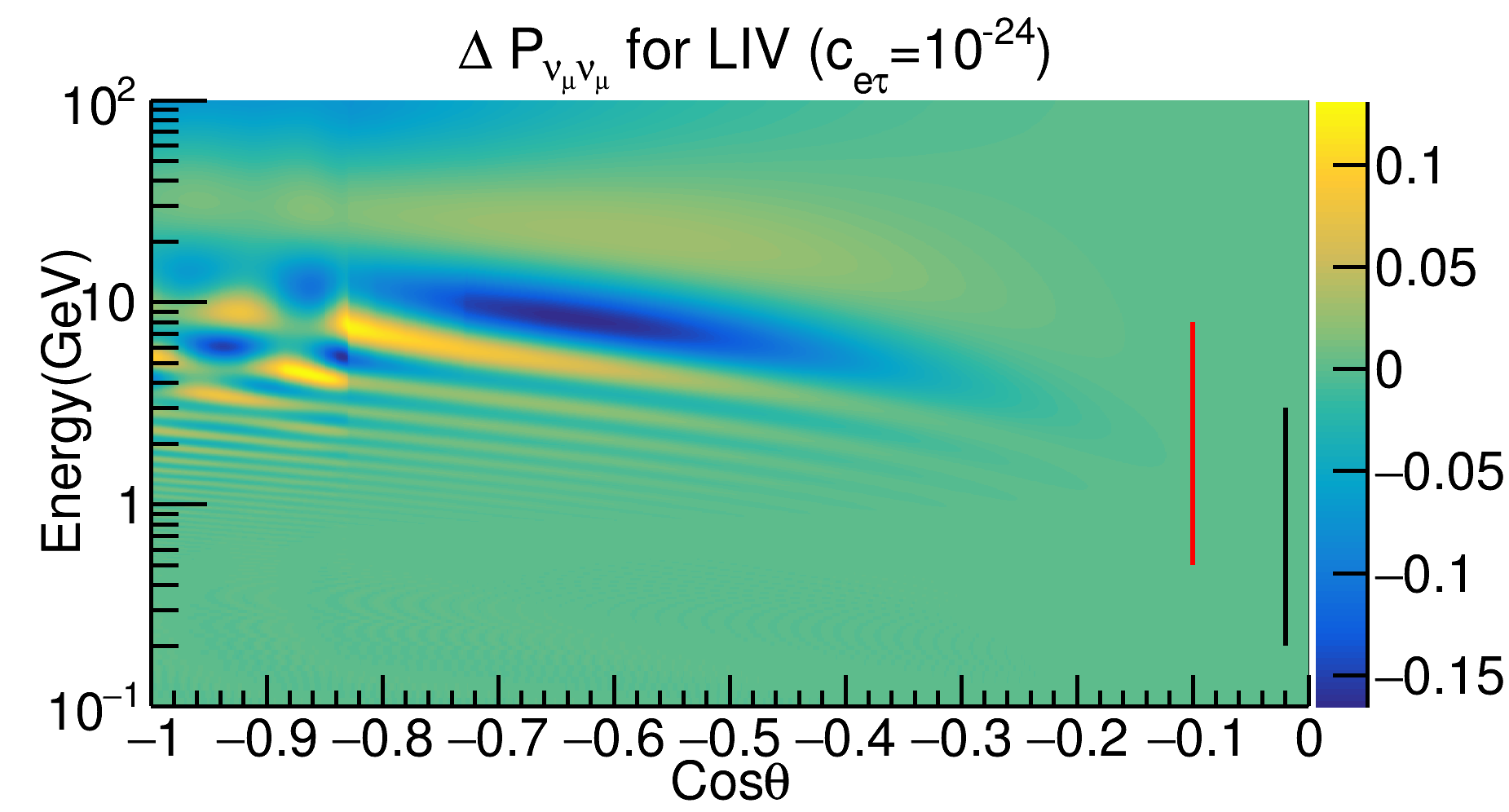}
 \end{minipage}
\begin{minipage}[t]{0.3\textwidth}
  \includegraphics[width=\linewidth]{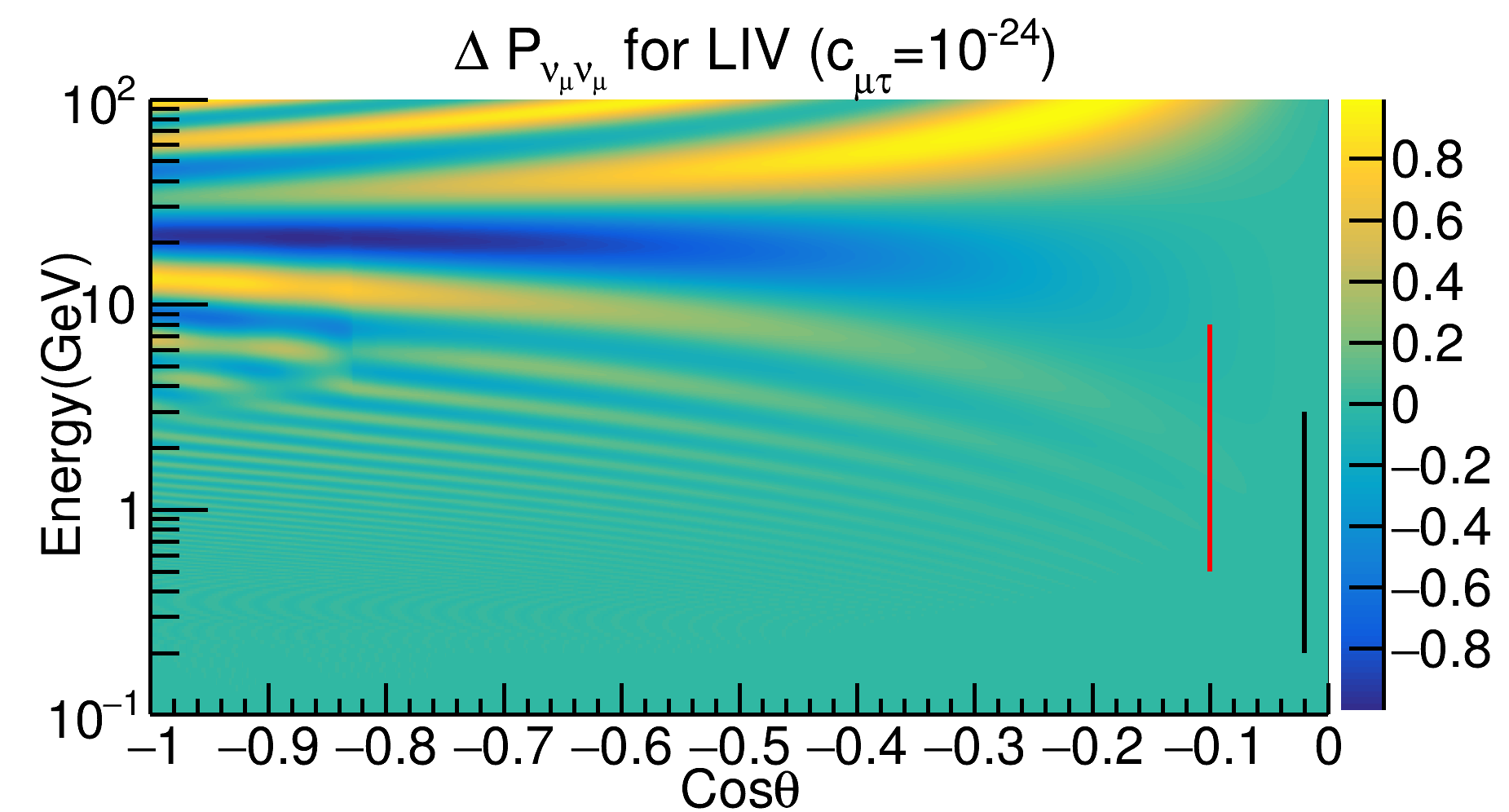}
\end{minipage}%

\begin{minipage}[t]{0.3\textwidth}
  \includegraphics[width=\linewidth]{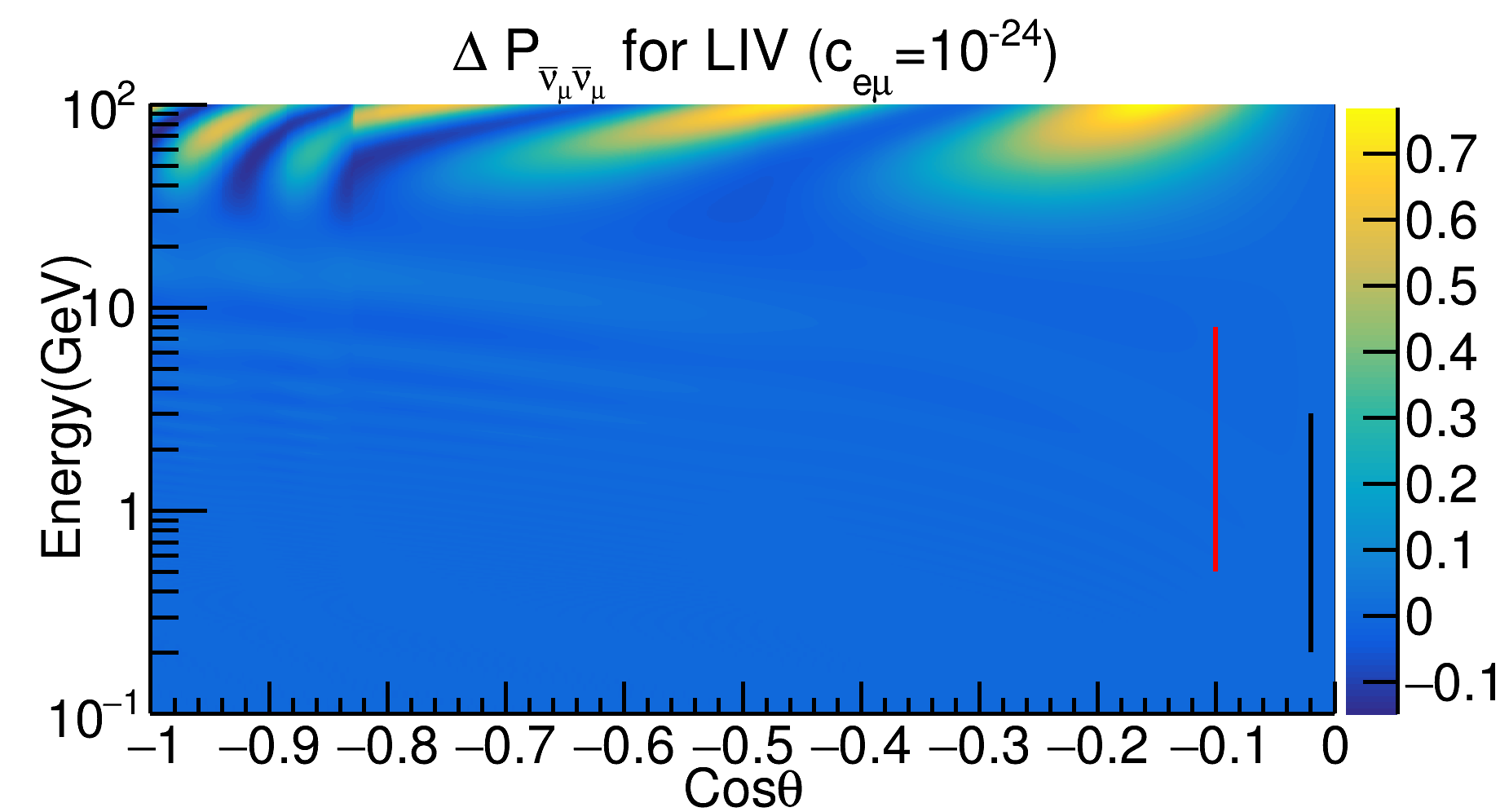}
 \end{minipage}
\begin{minipage}[t]{0.3\textwidth}
  \includegraphics[width=\linewidth]{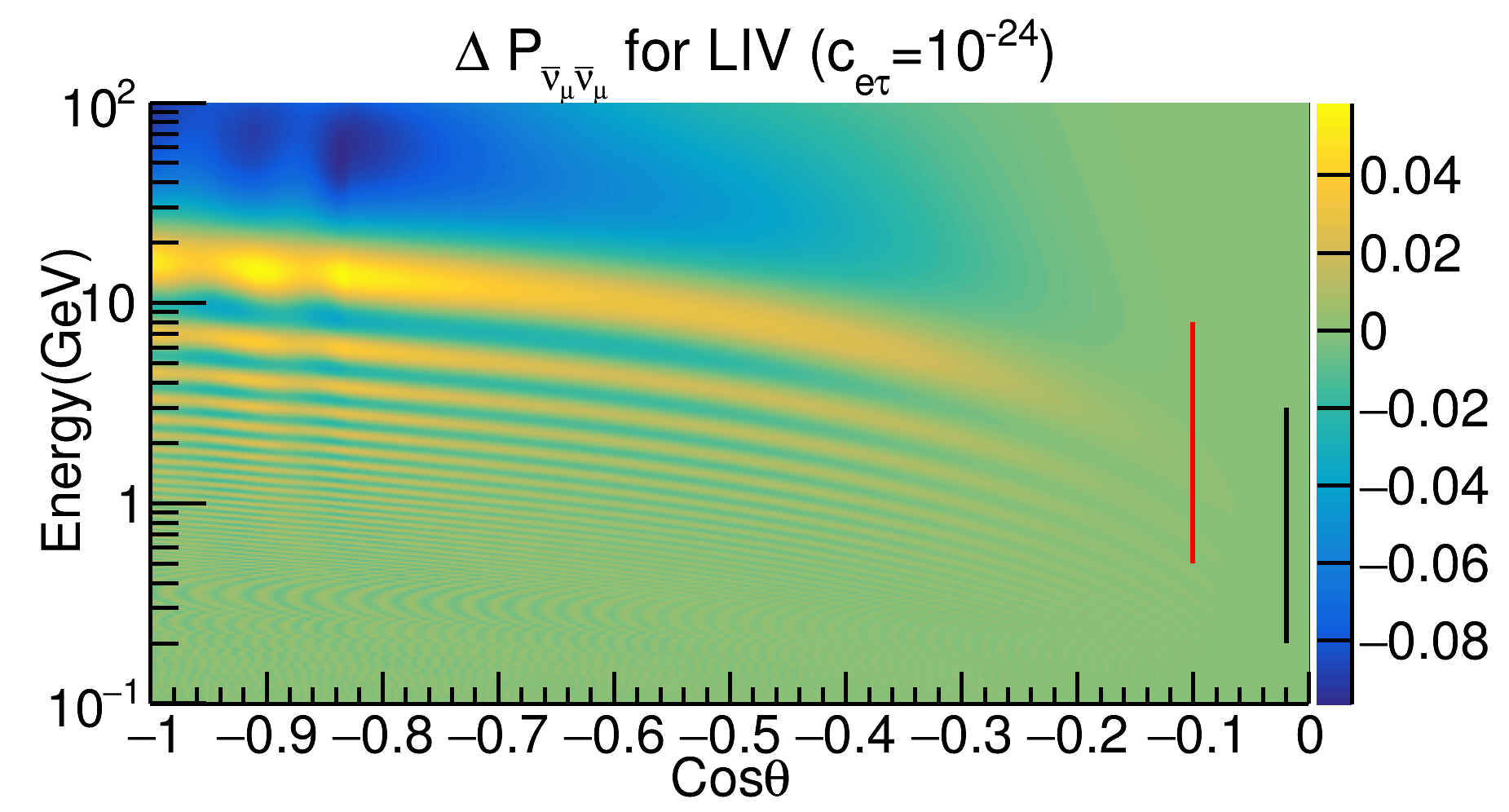}
 \end{minipage}
\begin{minipage}[t]{0.3\textwidth}
  \includegraphics[width=\linewidth]{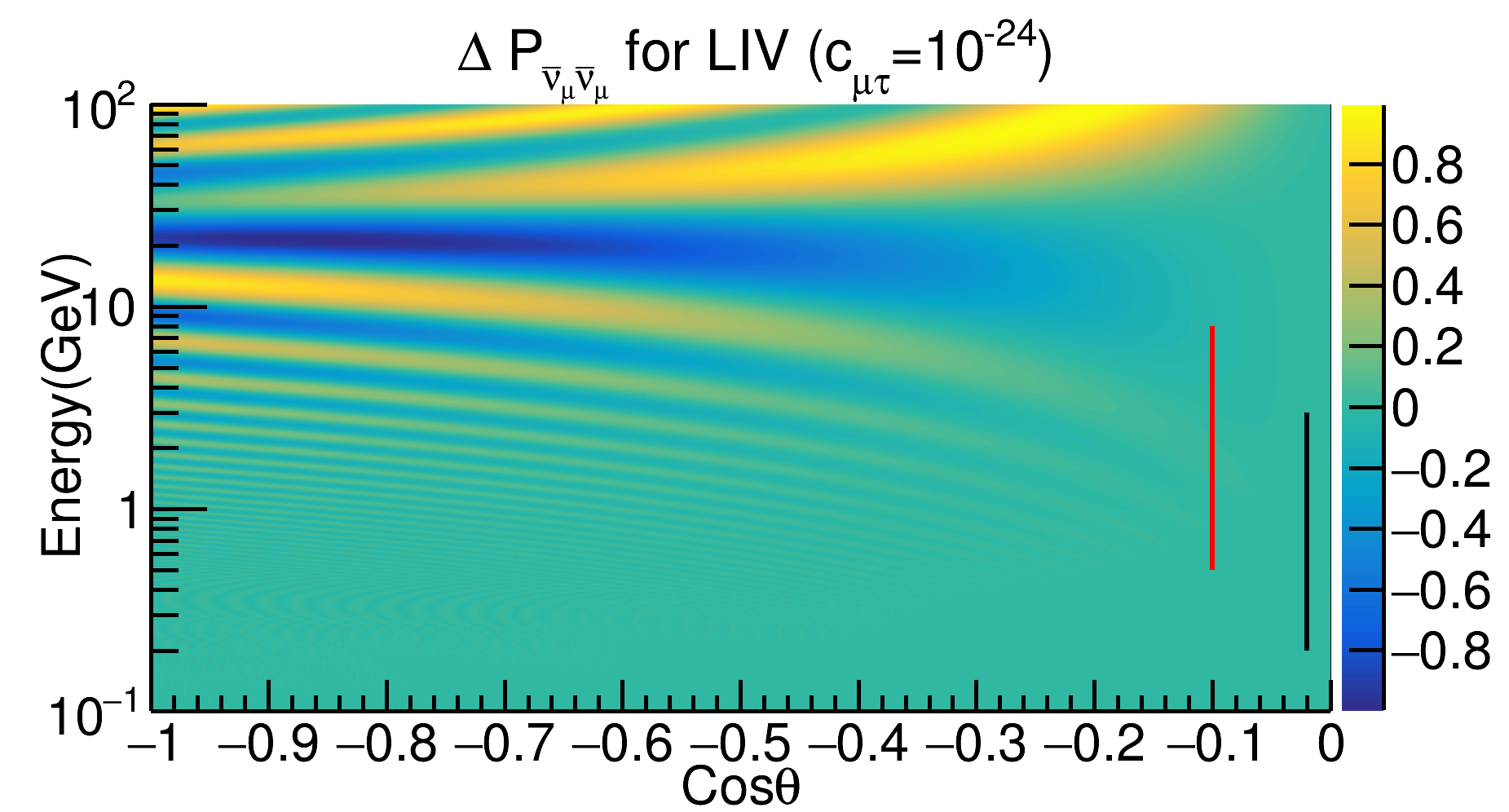}
\end{minipage}%
\caption{Probability oscillograms for $\Delta P_{\nu_{\mu}\nu_{\mu}}$ and $\Delta P_{\bar{\nu}_{\mu}\bar{\nu}_{\mu}}$ channels for $c_{ee}$, $c_{\mu\mu}$, $c_{\tau\tau}$, $c_{e\mu}$, $c_{e\tau}$ and $c_{\mu\tau}$. For ICAL, the relevant energy region of the oscillograms is 1-100 GeV. The red line shows the region relevant for DUNE and the black line shows the region relevant for T2HK. }
\label{fig:inooscillogrm_c}
\end{figure*}

To illustrate the effect of LIV on ICAL, we calculate the disappearance channel probability ($\nu_\mu \rightarrow \nu_\mu$) with and without LIV, and present the difference between them as oscillograms in the $\cos\theta$ - $E$ plane, where $E$ and $\theta$ here refer to the neutrino energy and neutrino zenith angle. For ICAL the relevant energy region of the oscillograms is 1 - 100 GeV. In Fig. \ref{fig:inooscillogrm_a} (\ref{fig:inooscillogrm_c}), we present the probability oscillogram for $a_{\alpha \beta}$ ($c_{\alpha \beta}$), respectively. The first and second rows are for the diagonal LIV parameters, while the third and fourth rows are for the off-diagonal LIV parameters. The first and third rows correspond to neutrinos, while the second and fourth rows correspond to antineutrinos. The value of $\phi^{a/c}_{\alpha \beta}$ for the off-diagonal LIV parameters is taken as zero in these figures.

Observations from these oscillograms are as follows. For the diagonal CPT-violating LIV parameters, we observe that the effect of LIV is greater in $a_{\mu\mu}$ and $a_{\tau\tau}$ than in $a_{ee}$. In $a_{ee}$, the effect is more pronounced in neutrinos than in antineutrinos. For neutrinos, $a_{ee}$ can be probed in the energy range of 1 GeV to 20 GeV, while for antineutrinos, it can be probed below 1 GeV. For $a_{\mu\mu}$ and $a_{\tau\tau}$, the effect of LIV is similar, with some differences along the $\cos\theta$ axis. The effect of LIV on these two parameters is more significant in antineutrinos than in neutrinos.

Regarding the off-diagonal CPT-violating LIV parameters, we see that the effect of LIV on $a_{e\mu}$ and $a_{e\tau}$ is similar, with some differences along the $\cos\theta$ axis. The effect of LIV on these two parameters is more prominent in neutrinos than in antineutrinos. Among the three off-diagonal LIV parameters, $a_{\mu\tau}$ is more sensitive to LIV. For this parameter, the effect of LIV is opposite in neutrinos and antineutrinos, with the blue and yellow colors interchanged in the panels. Thus, we expect to have strong bounds on $a_{\mu\mu}$, $a_{\tau\tau}$, and $a_{\mu\tau}$ in ICAL as far as CPT-violating LIV parameters are concerned, compared to the other three parameters. This is consistent with our analytical understanding of these parameters. It has been shown that in the disappearance channel probabilities, the parameters $a_{\mu\mu}$, $a_{\tau\tau}$, and $a_{\mu\tau}$ contribute at the leading order~\cite{Sahoo:2021dit}. 

Now, let us discuss the case for CPT-conserving parameters. From Eq.~\ref{equ:hliv}, we see that there is a factor $\frac{4}{3} E$ that appears with the CPT conserving LIV parameters. Therefore, we expect that ICAL will be sensitive to these parameters for higher values of $E$. For the diagonal parameters, the behavior of $c_{\alpha \alpha}$ is similar to that of $a_{\alpha \alpha}$. From the panels, we see that the effect of LIV is weak in $c_{ee}$ compared to the other two parameters. For $c_{ee}$, the effect of LIV is visible in the energy range of 2 GeV to 20 GeV in neutrinos and 0.1 GeV to 10 GeV for antineutrinos. For $c_{\mu\mu}$ and $c_{\tau\tau}$, the effect is similar for neutrinos and antineutrinos, with some small differences along the $\cos\theta$ axis. For the off-diagonal parameter $c_{e\mu}$, the effect of LIV is mostly the same in neutrinos and antineutrinos. For this parameter, the effect of LIV is visible in the energy range of 1-10 GeV, which has the most matter effect for neutrinos but is absent for antineutrinos. For $c_{e\tau}$, the effect of LIV is more prominent in neutrinos than in antineutrinos. The parameter $c_{\mu\tau}$ has a similar effect in neutrinos and antineutrinos. From the above discussion, we understand that ICAL will have weaker bounds on $c_{ee}$ as compared to the other CPT-conserving LIV parameters.

\subsection{DUNE}

\begin{figure*}

\begin{minipage}[t]{0.3\textwidth}
  \includegraphics[width=\linewidth]{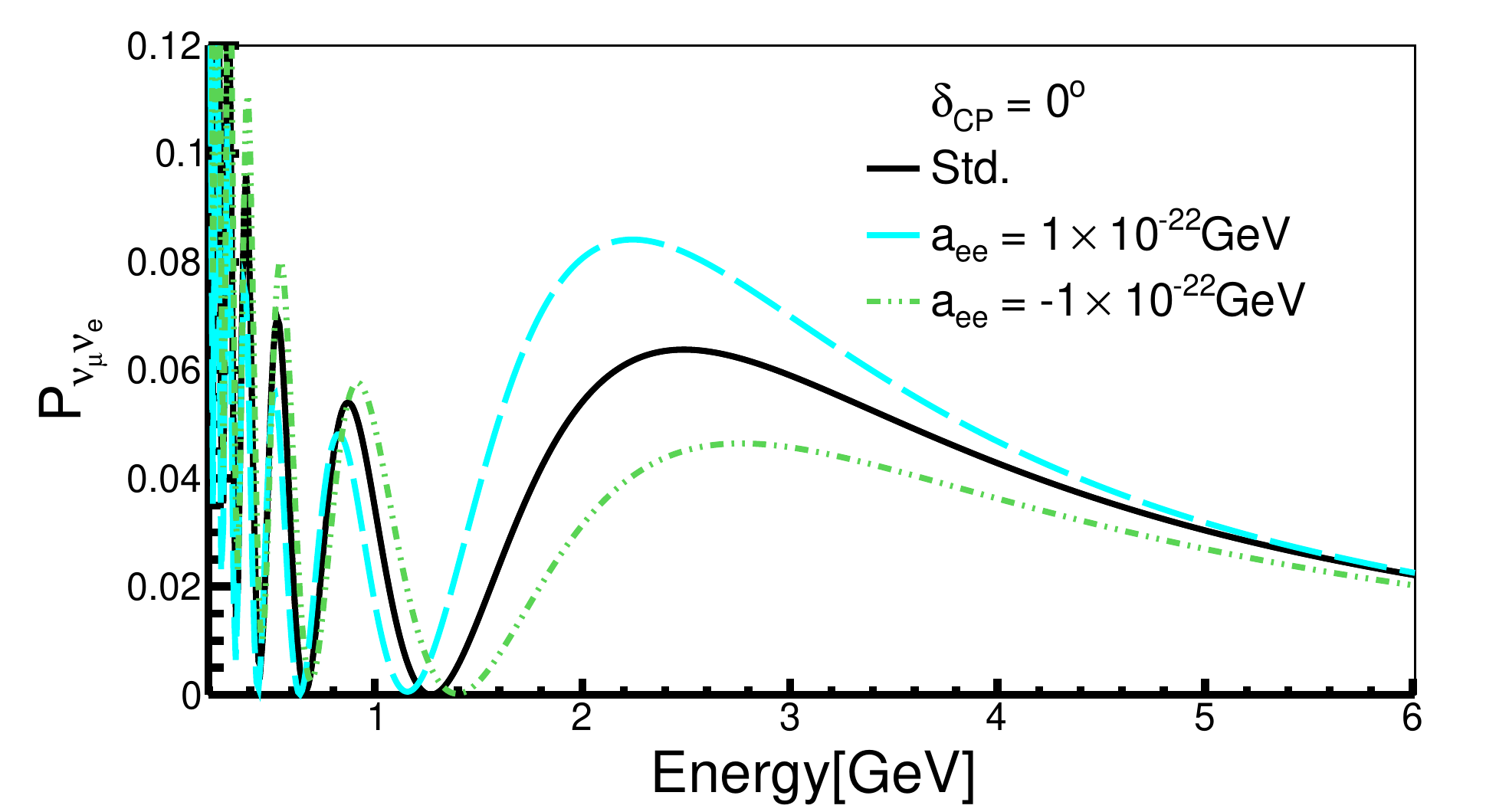}
 \end{minipage}
\begin{minipage}[t]{0.3\textwidth}
  \includegraphics[width=\linewidth]{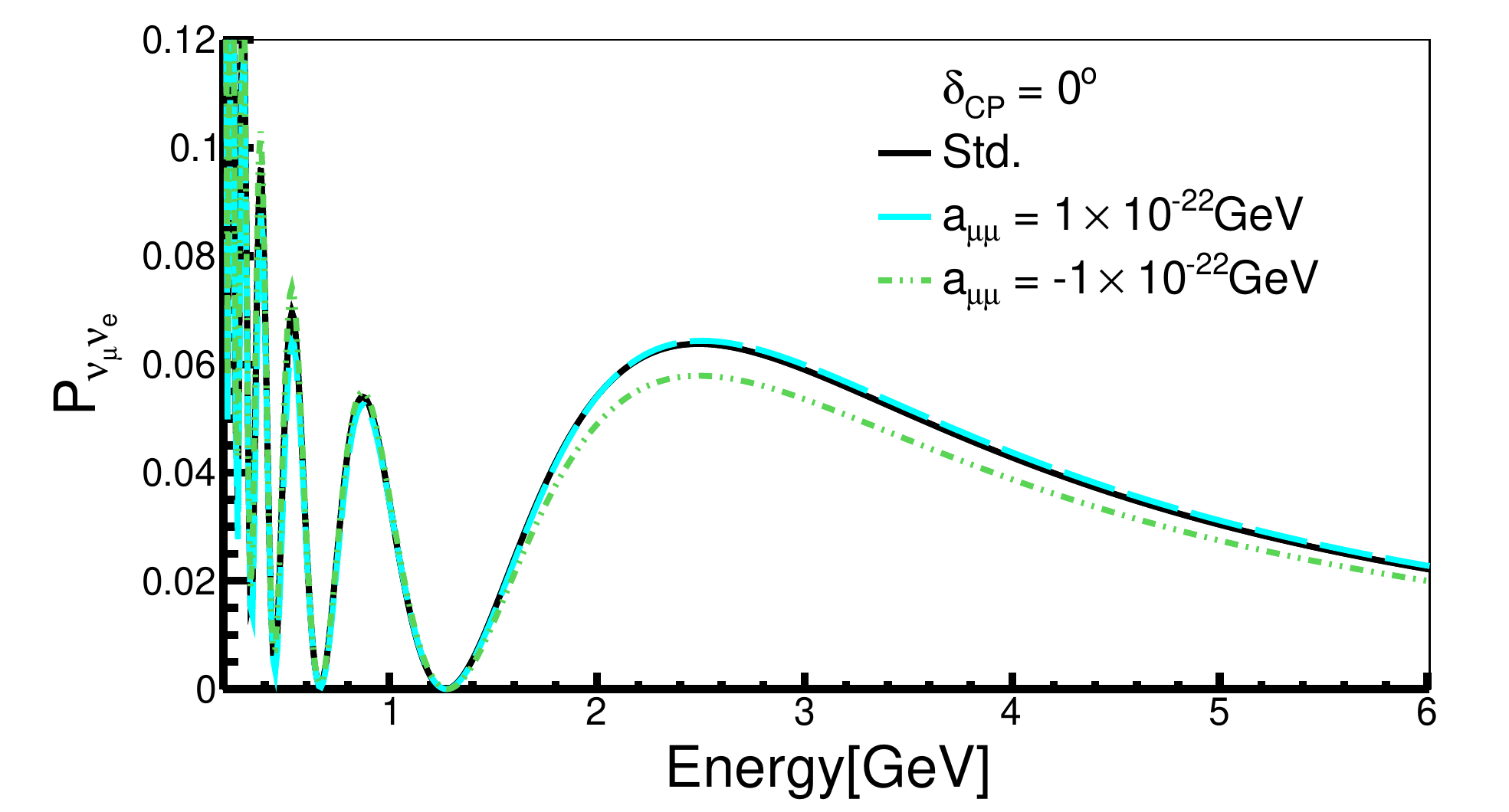}
 \end{minipage}
\begin{minipage}[t]{0.3\textwidth}
  \includegraphics[width=\linewidth]{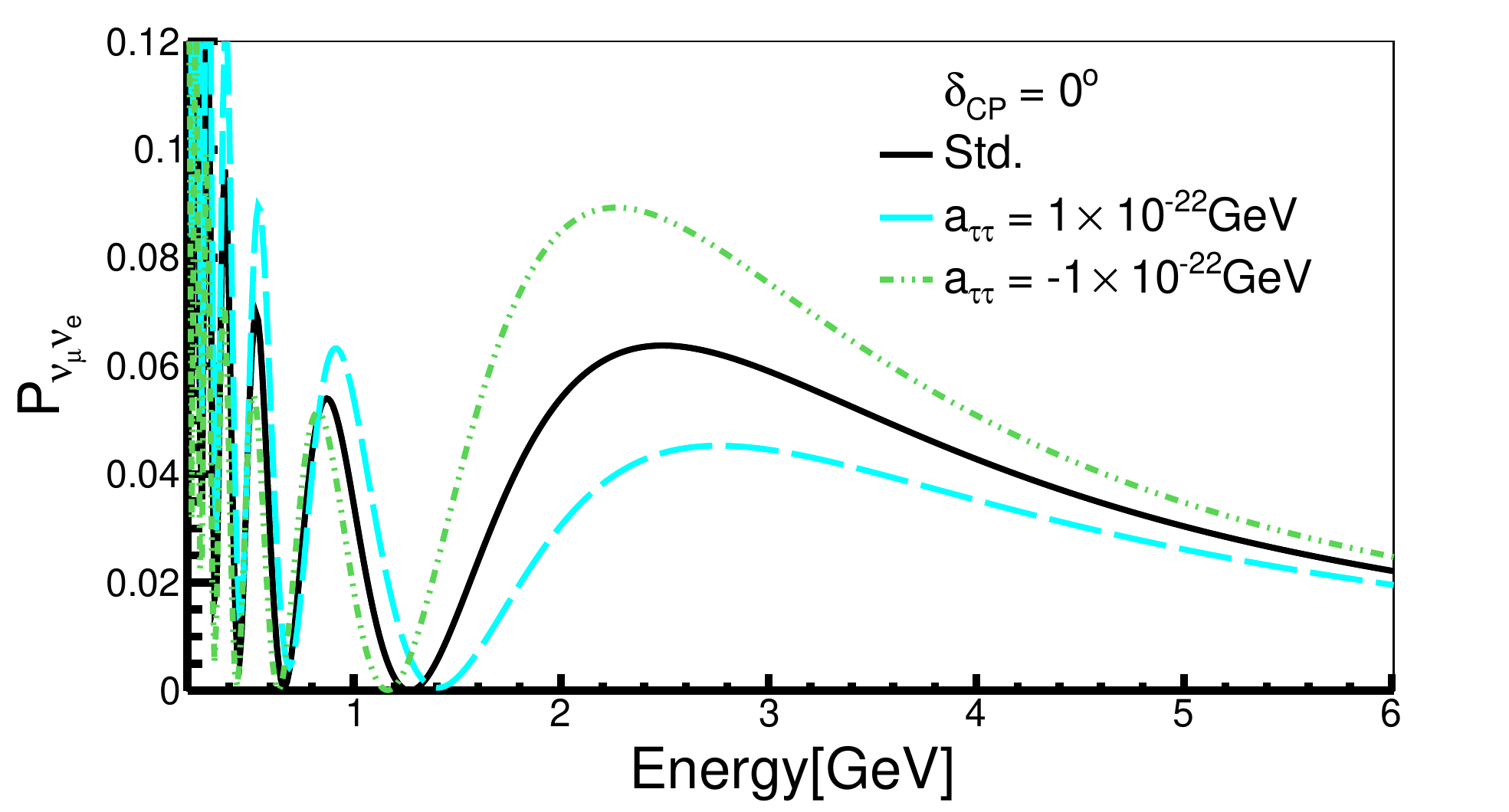}
\end{minipage}%

\begin{minipage}[t]{0.3\textwidth}
  \includegraphics[width=\linewidth]{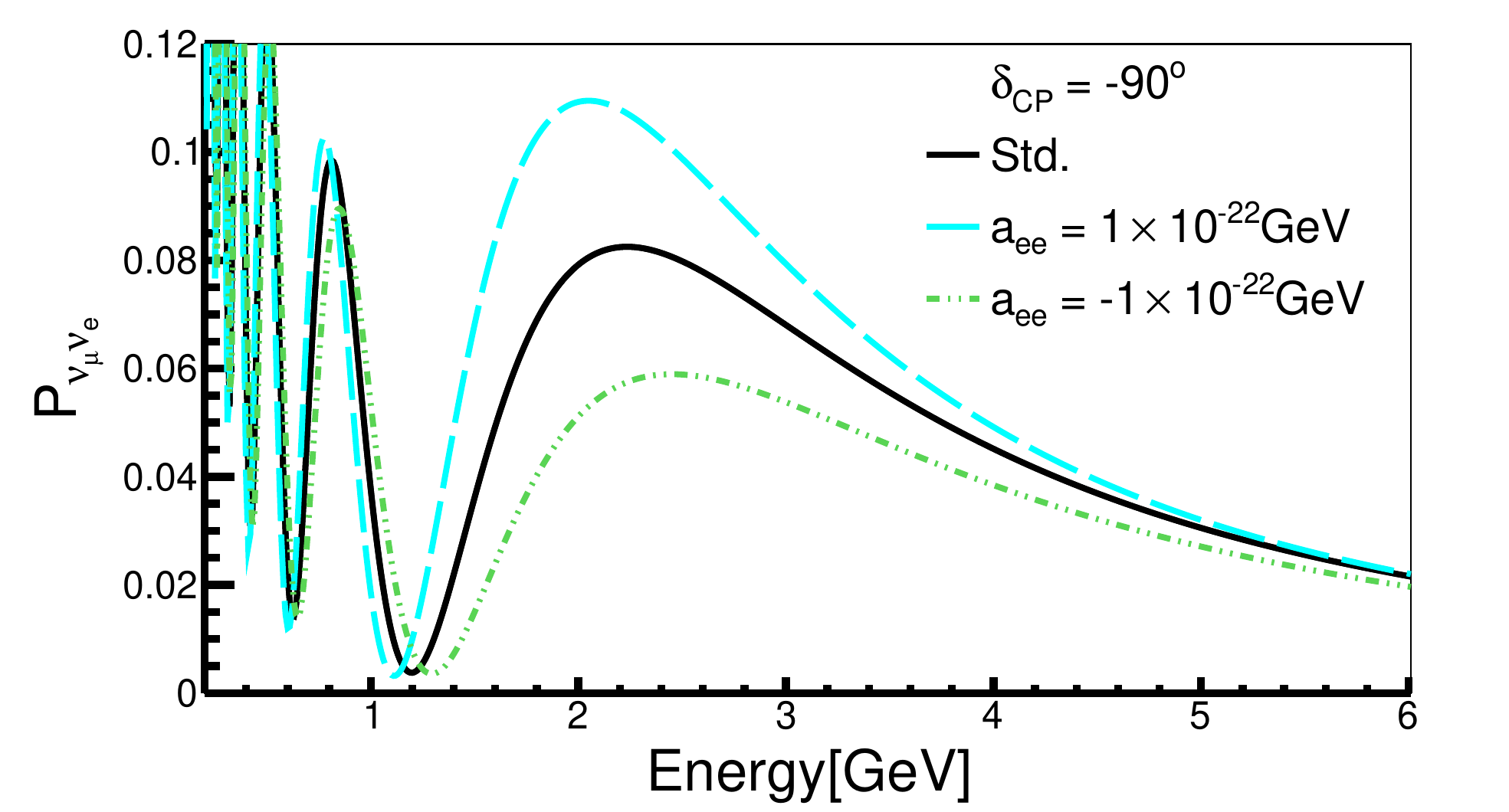}
 \end{minipage}
\begin{minipage}[t]{0.3\textwidth}
  \includegraphics[width=\linewidth]{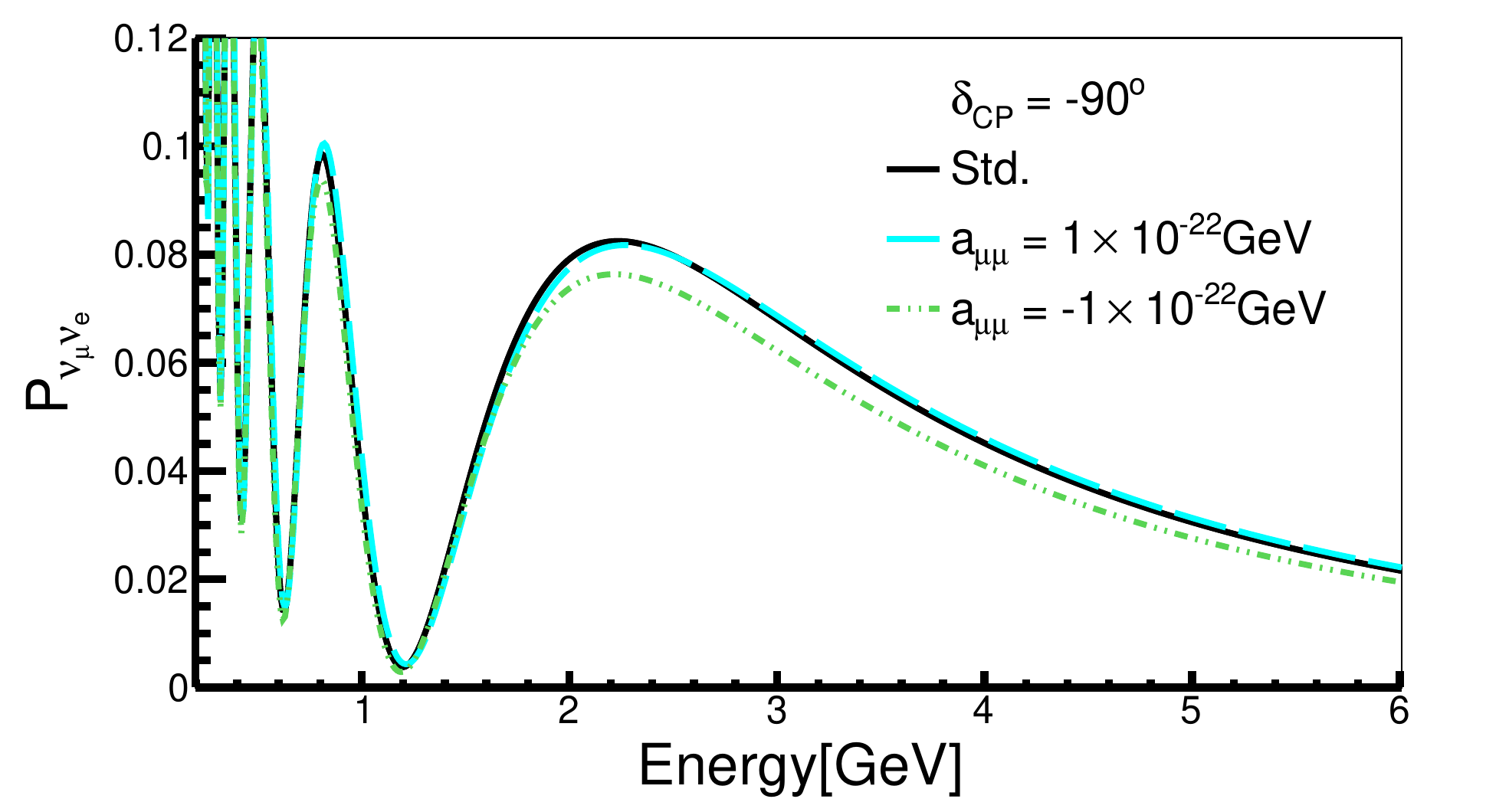}
 \end{minipage}
\begin{minipage}[t]{0.3\textwidth}
  \includegraphics[width=\linewidth]{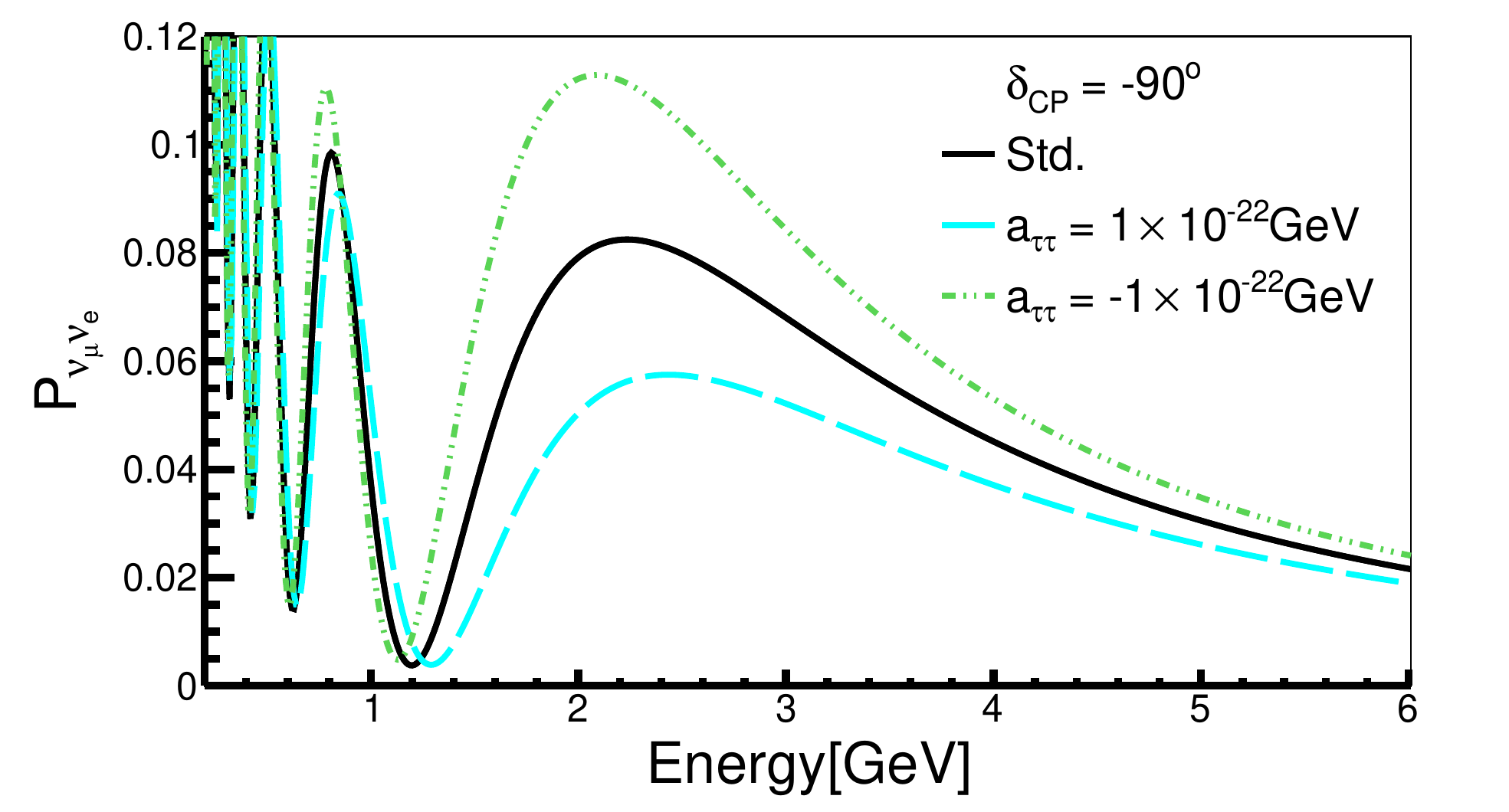}
\end{minipage}%

\begin{minipage}[t]{0.3\textwidth}
  \includegraphics[width=\linewidth]{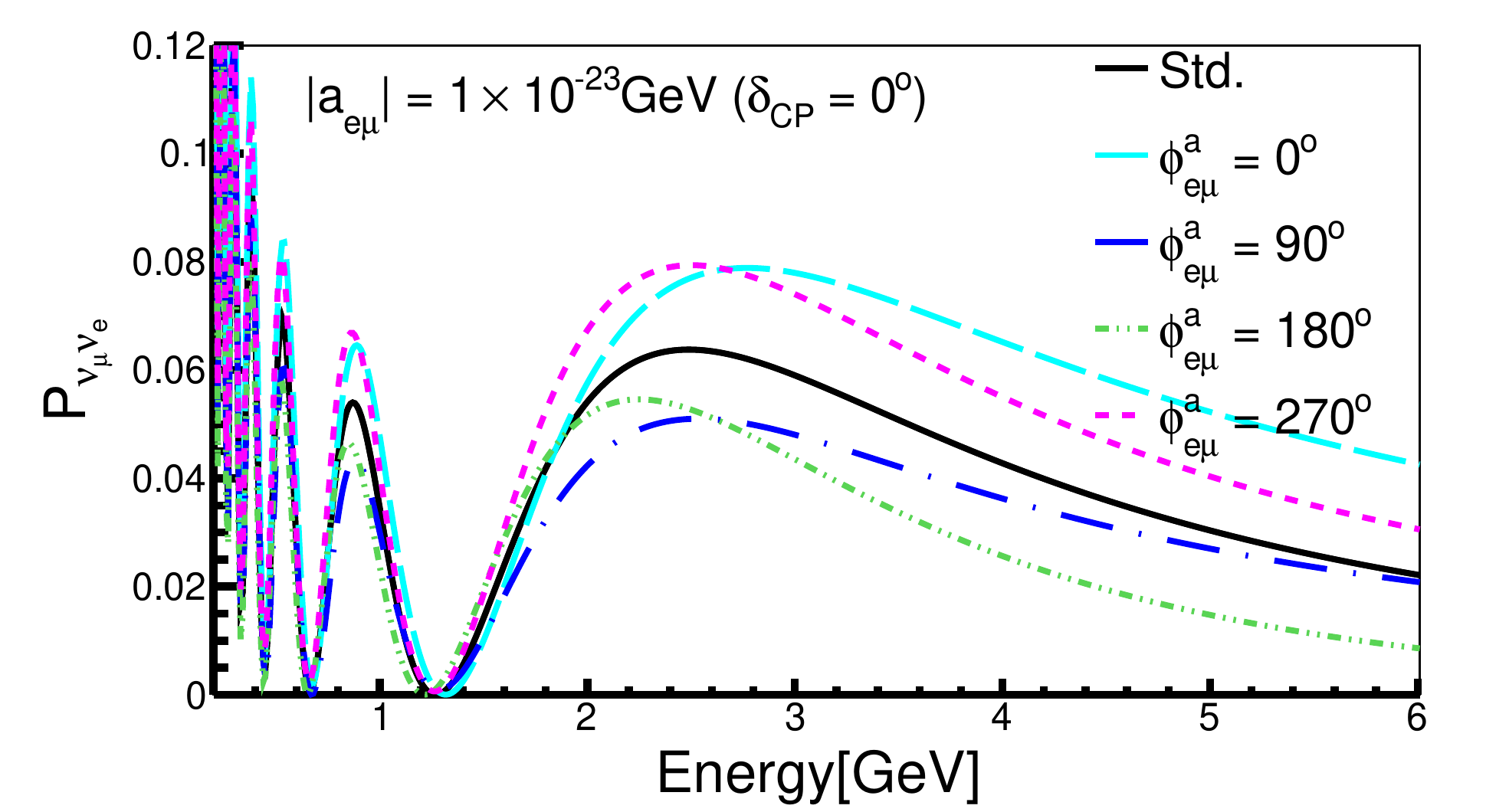}
 \end{minipage}
\begin{minipage}[t]{0.3\textwidth}
  \includegraphics[width=\linewidth]{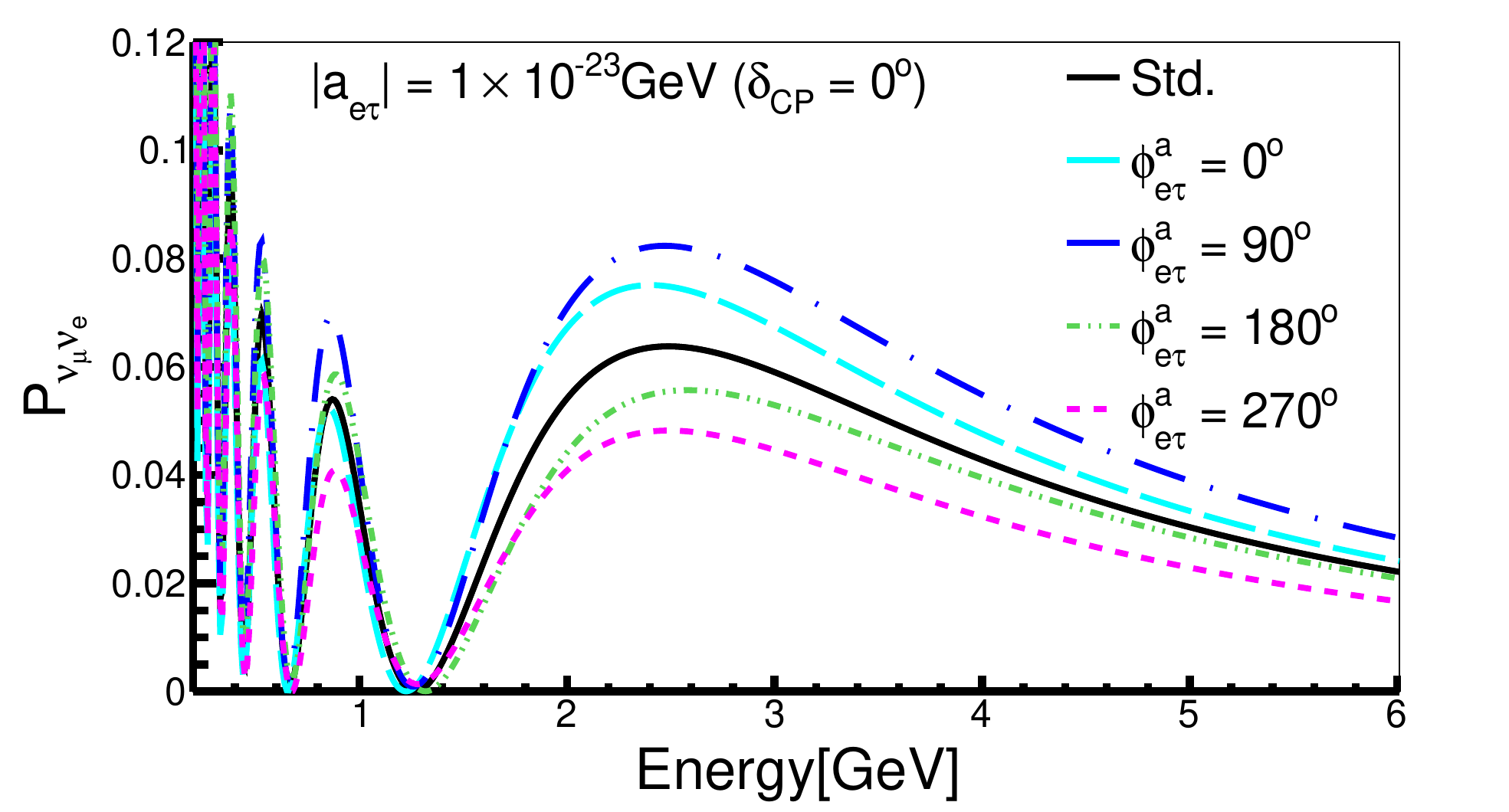}
 \end{minipage}
\begin{minipage}[t]{0.3\textwidth}
  \includegraphics[width=\linewidth]{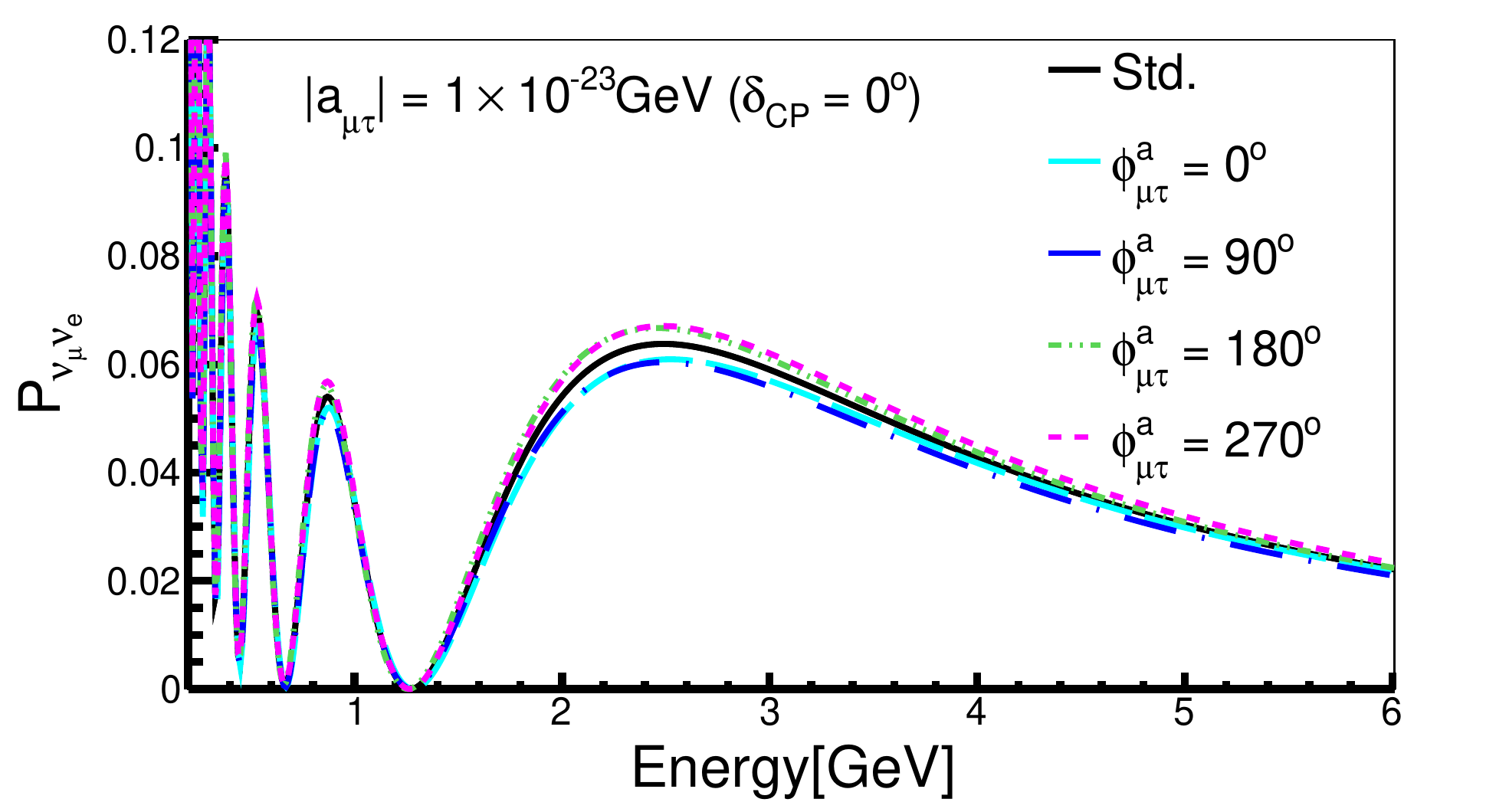}
\end{minipage}%

\begin{minipage}[t]{0.3\textwidth}
  \includegraphics[width=\linewidth]{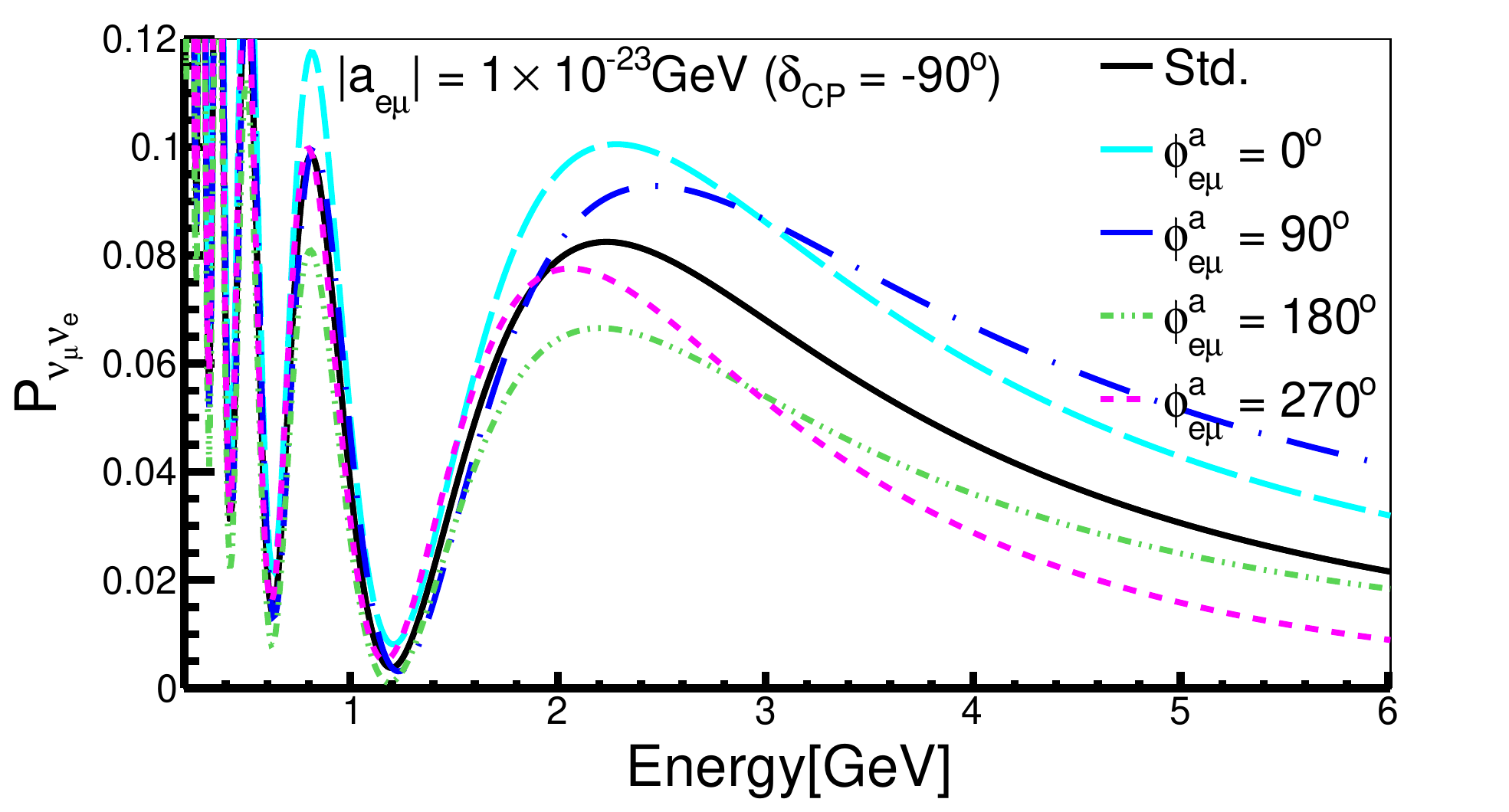}
 \end{minipage}
\begin{minipage}[t]{0.3\textwidth}
  \includegraphics[width=\linewidth]{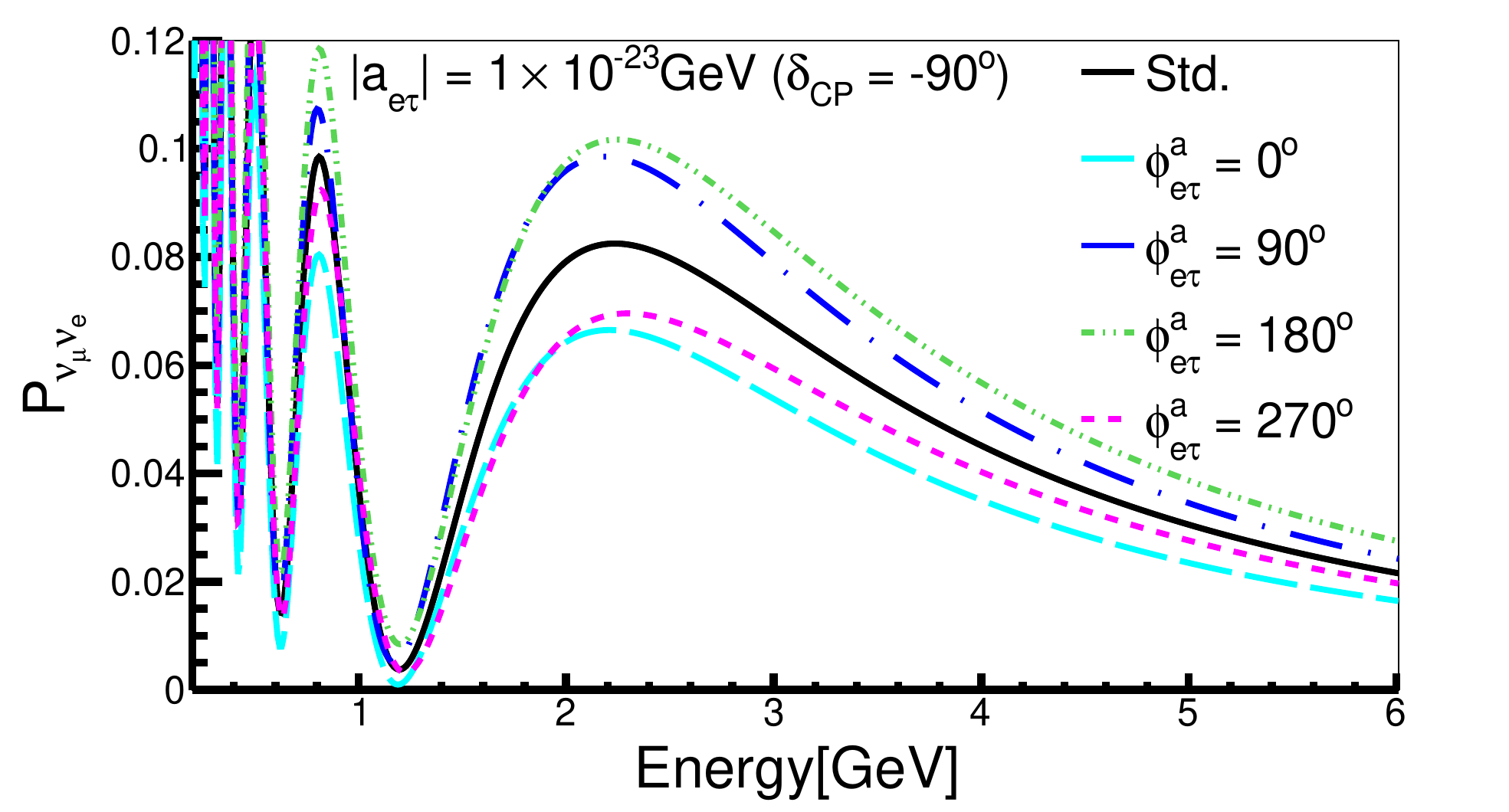}
 \end{minipage}
\begin{minipage}[t]{0.3\textwidth}
  \includegraphics[width=\linewidth]{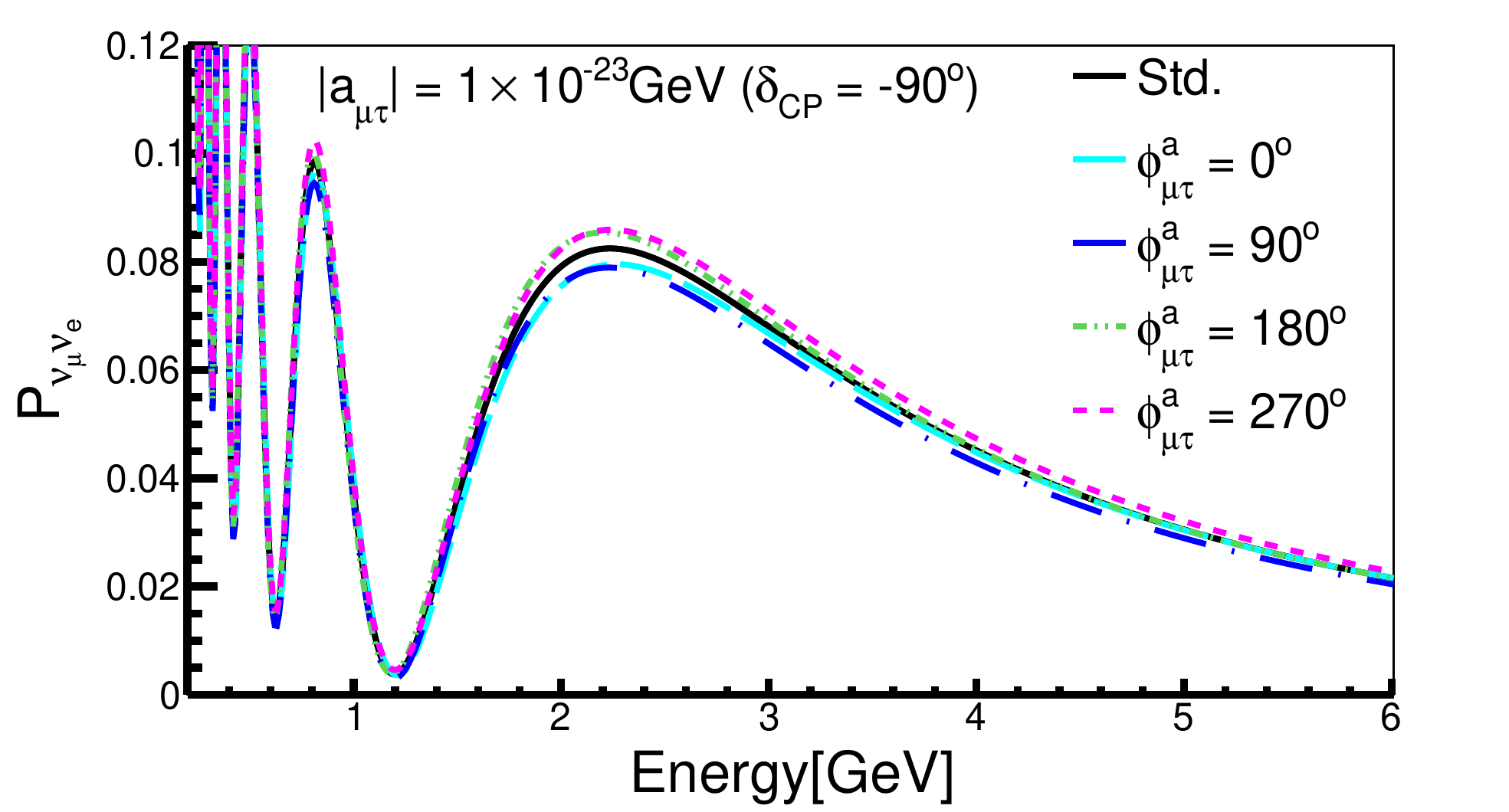}
\end{minipage}%
\caption{The $\nu_{e}$ appearance probability plots for $a_{ee}$, $a_{\mu\mu}$, $a_{\tau\tau}$ (first/second row is for $\delta_{\rm CP}=0^\circ/-90^\circ$) and  $a_{e\mu}$, $a_{e\tau}$, $a_{\mu\tau}$ (third/fourth row is for $\delta_{\rm CP}=0^\circ/-90^\circ$) for DUNE setup. In all panels the probabilities for standard three generation oscillations are shown by the solid black curves.}
\label{fig:dune-prob-a}
\end{figure*}

\begin{figure*}

\begin{minipage}[t]{0.3\textwidth}
  \includegraphics[width=\linewidth]{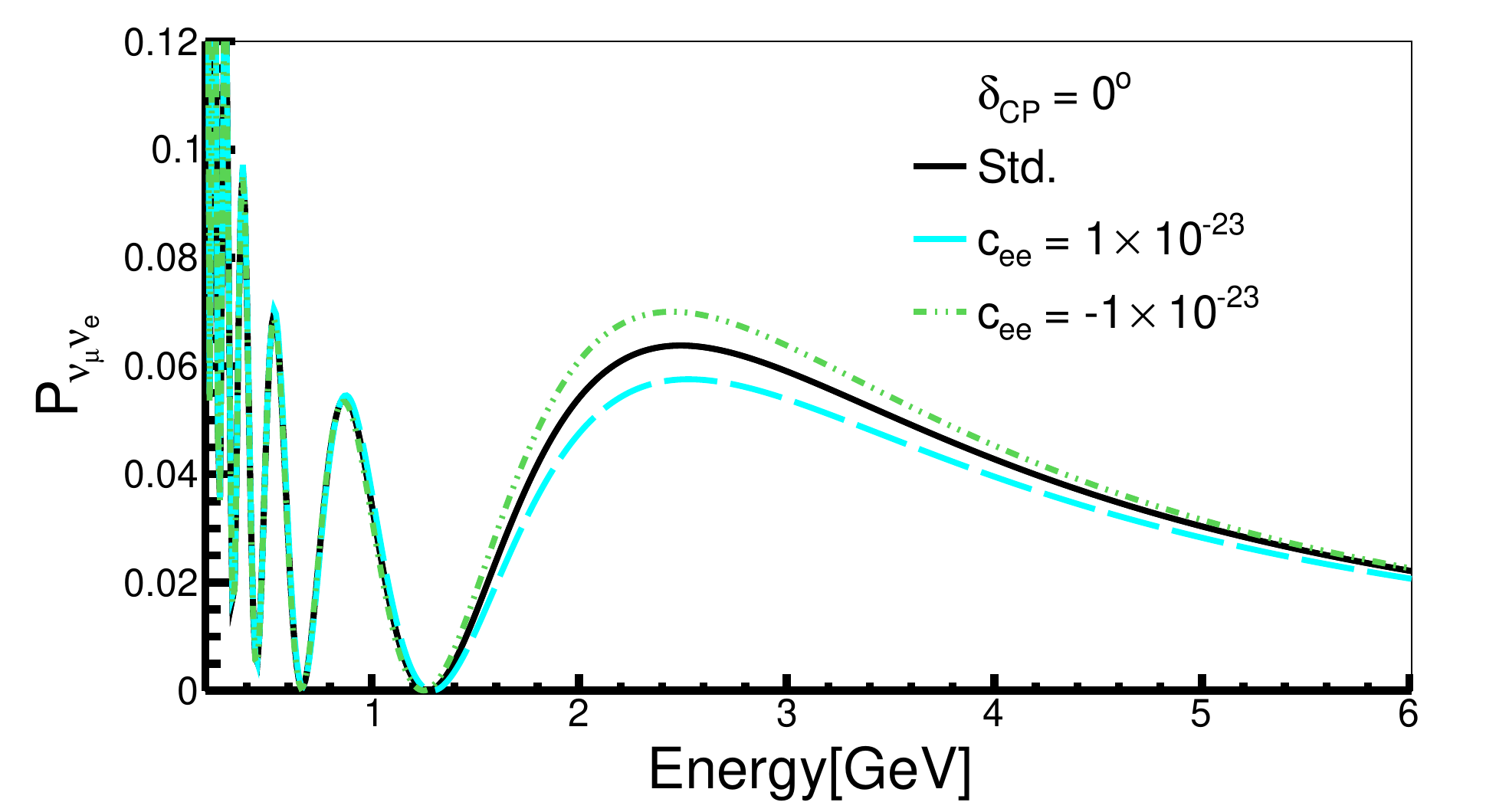}
 \end{minipage}
\begin{minipage}[t]{0.3\textwidth}
  \includegraphics[width=\linewidth]{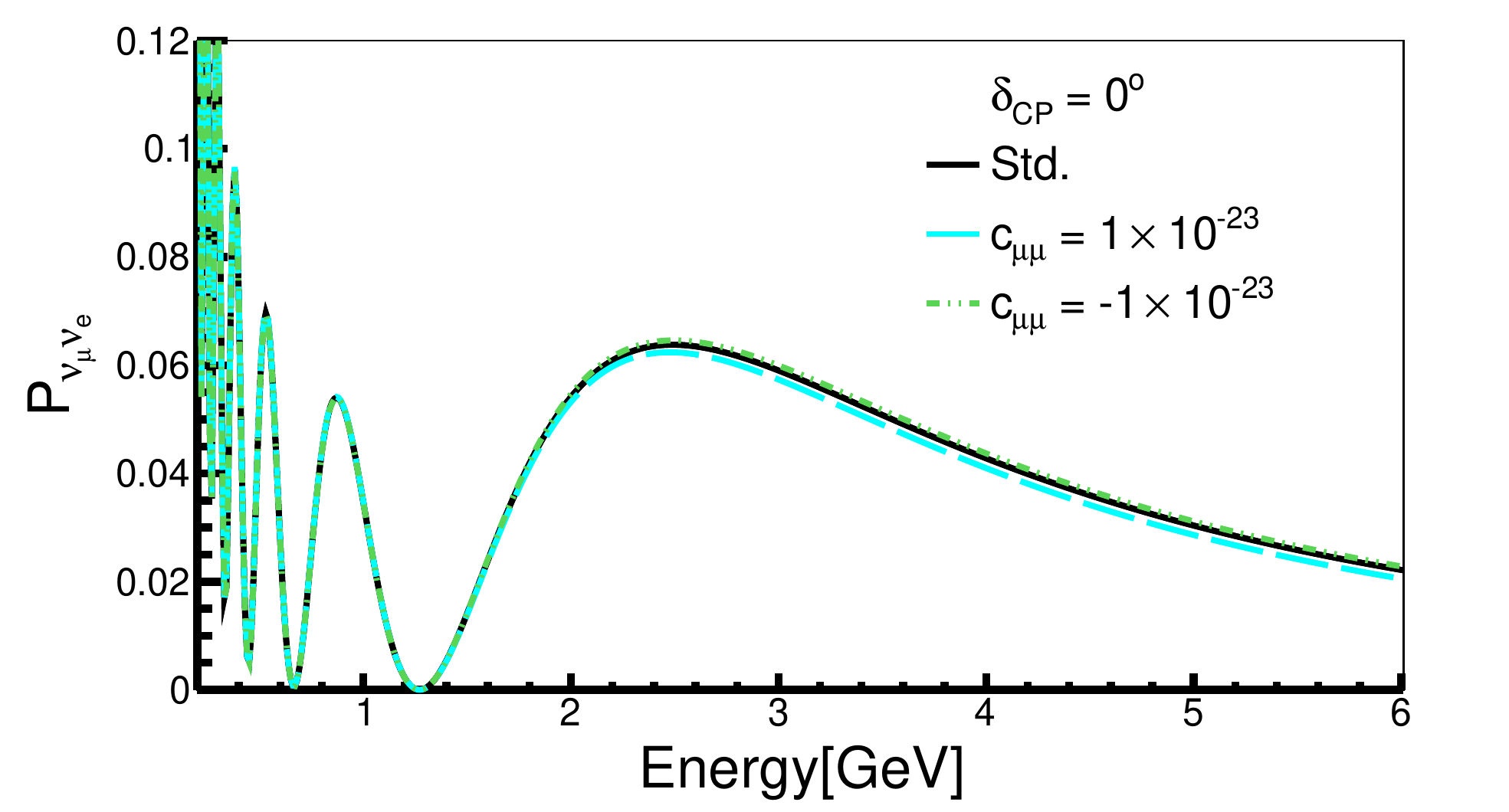}
 \end{minipage}
\begin{minipage}[t]{0.3\textwidth}
  \includegraphics[width=\linewidth]{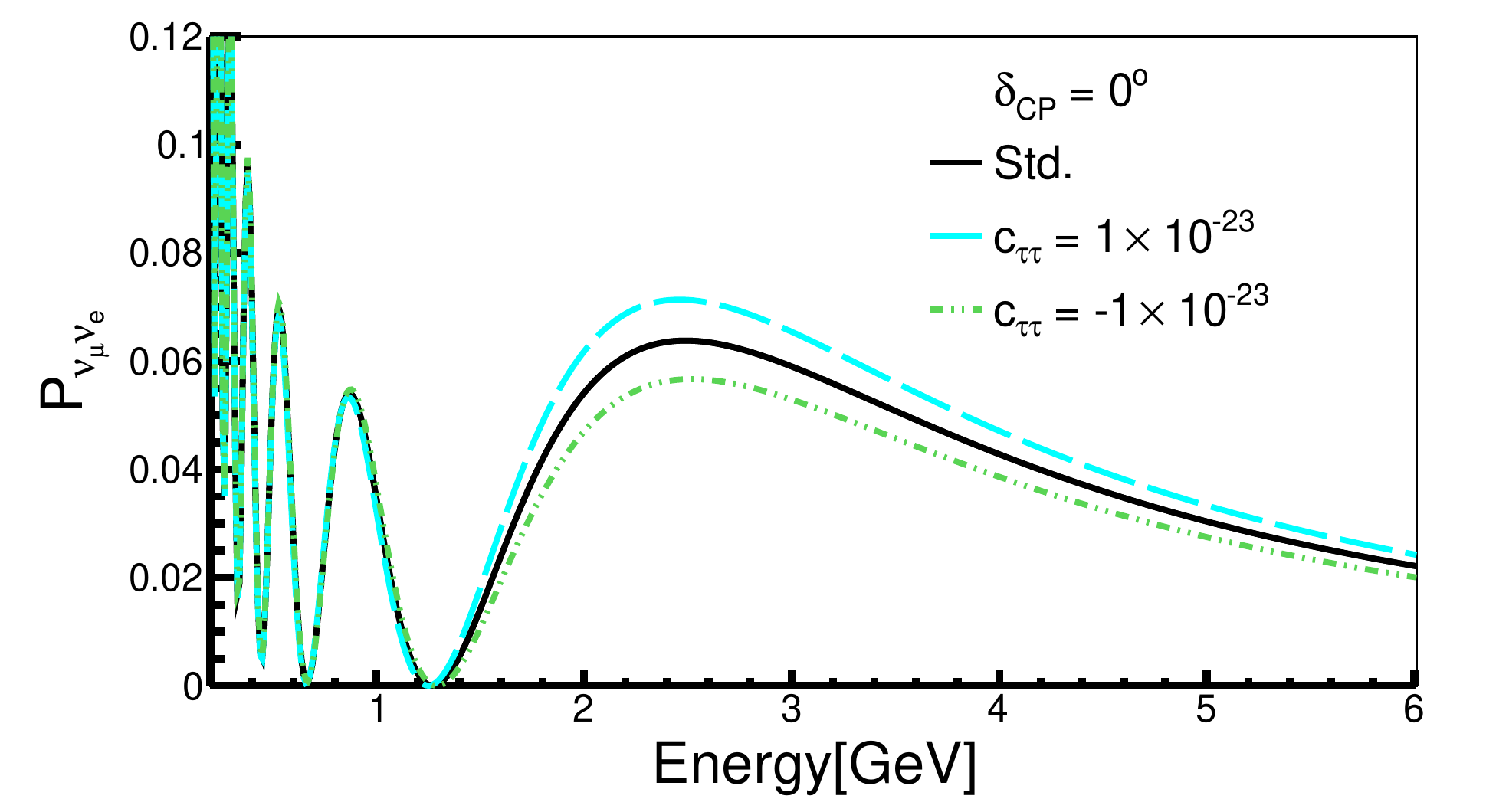}
\end{minipage}%

\begin{minipage}[t]{0.3\textwidth}
  \includegraphics[width=\linewidth]{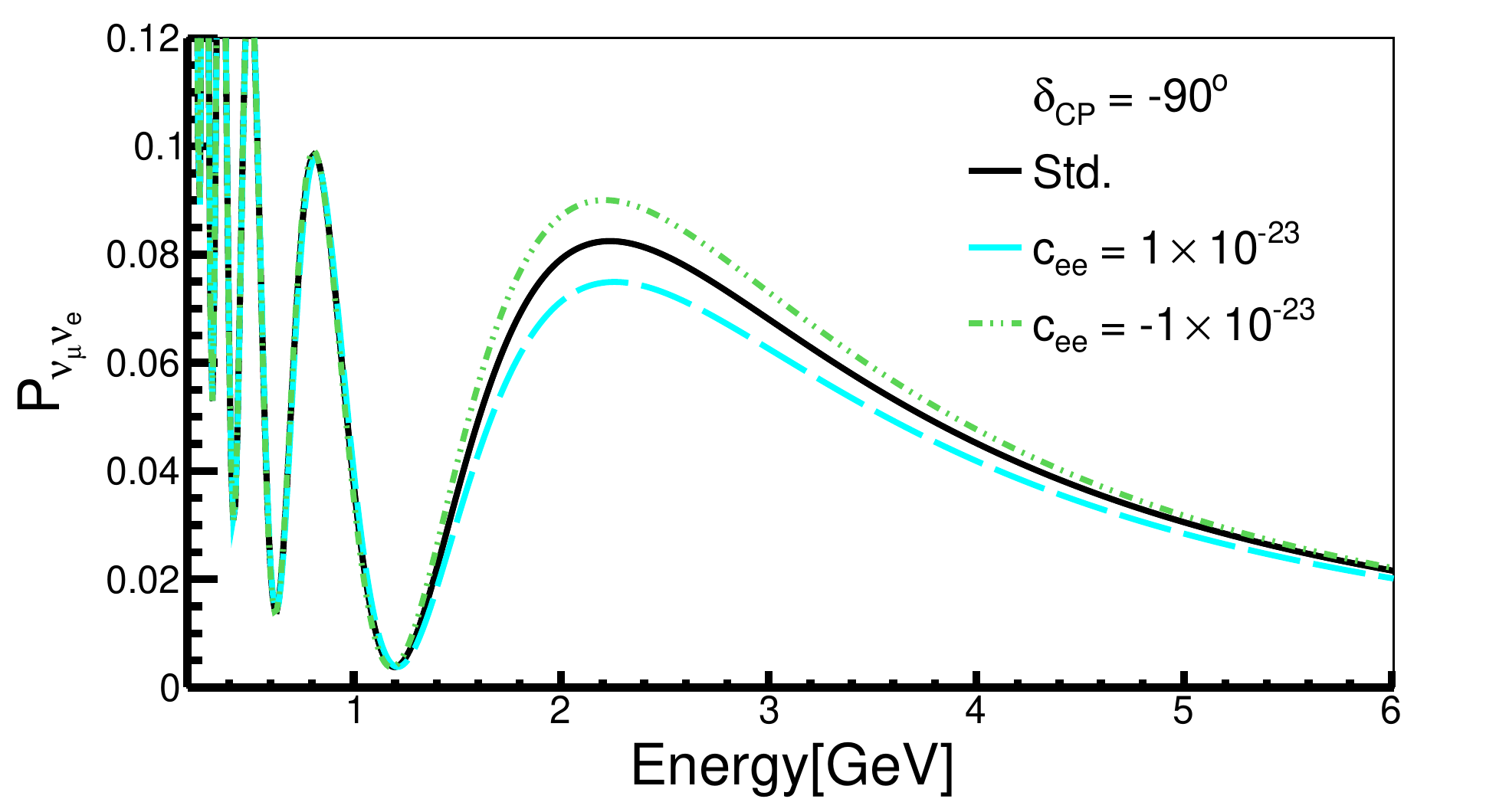}
 \end{minipage}
\begin{minipage}[t]{0.3\textwidth}
  \includegraphics[width=\linewidth]{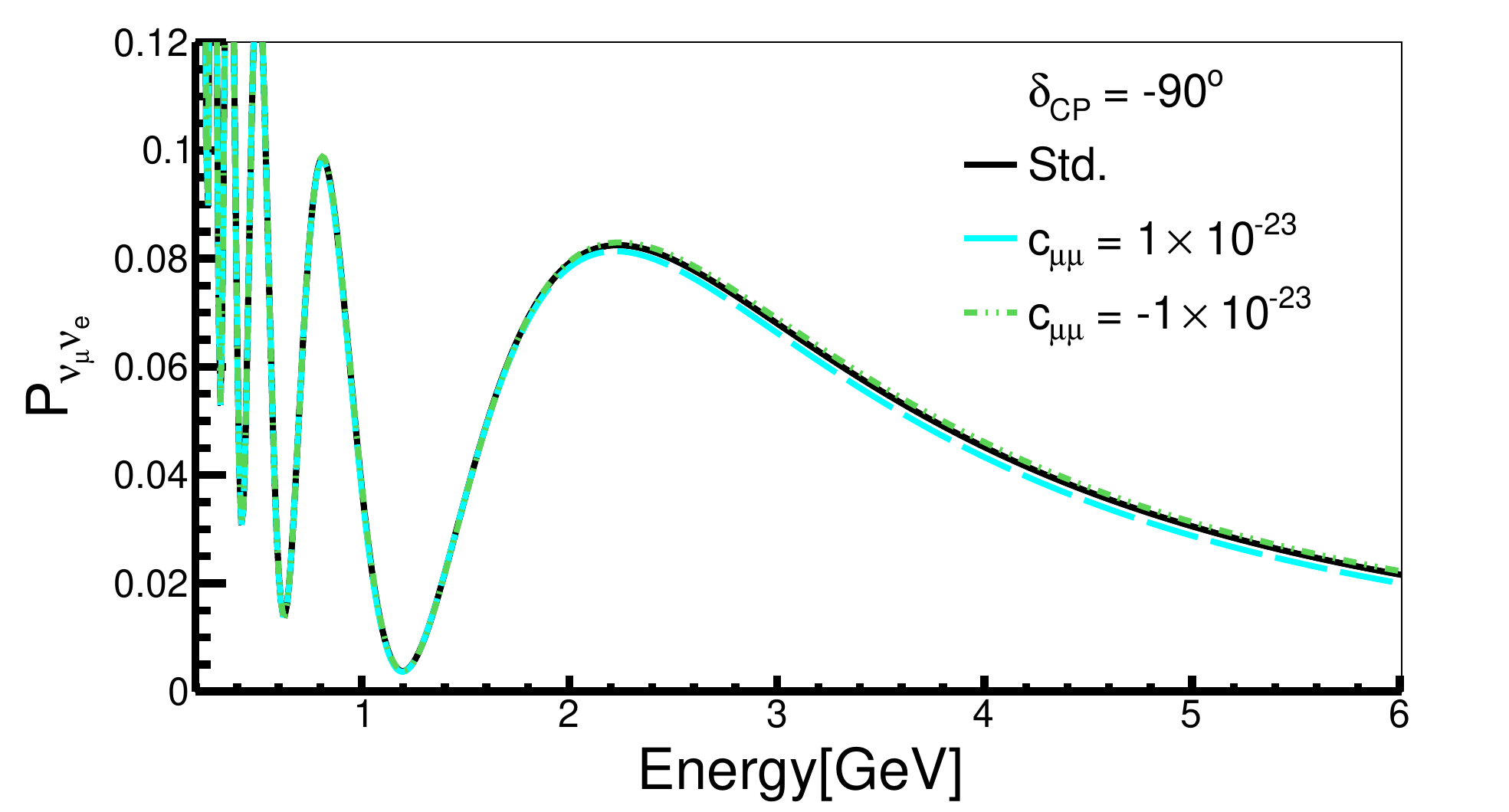}
 \end{minipage}
\begin{minipage}[t]{0.3\textwidth}
  \includegraphics[width=\linewidth]{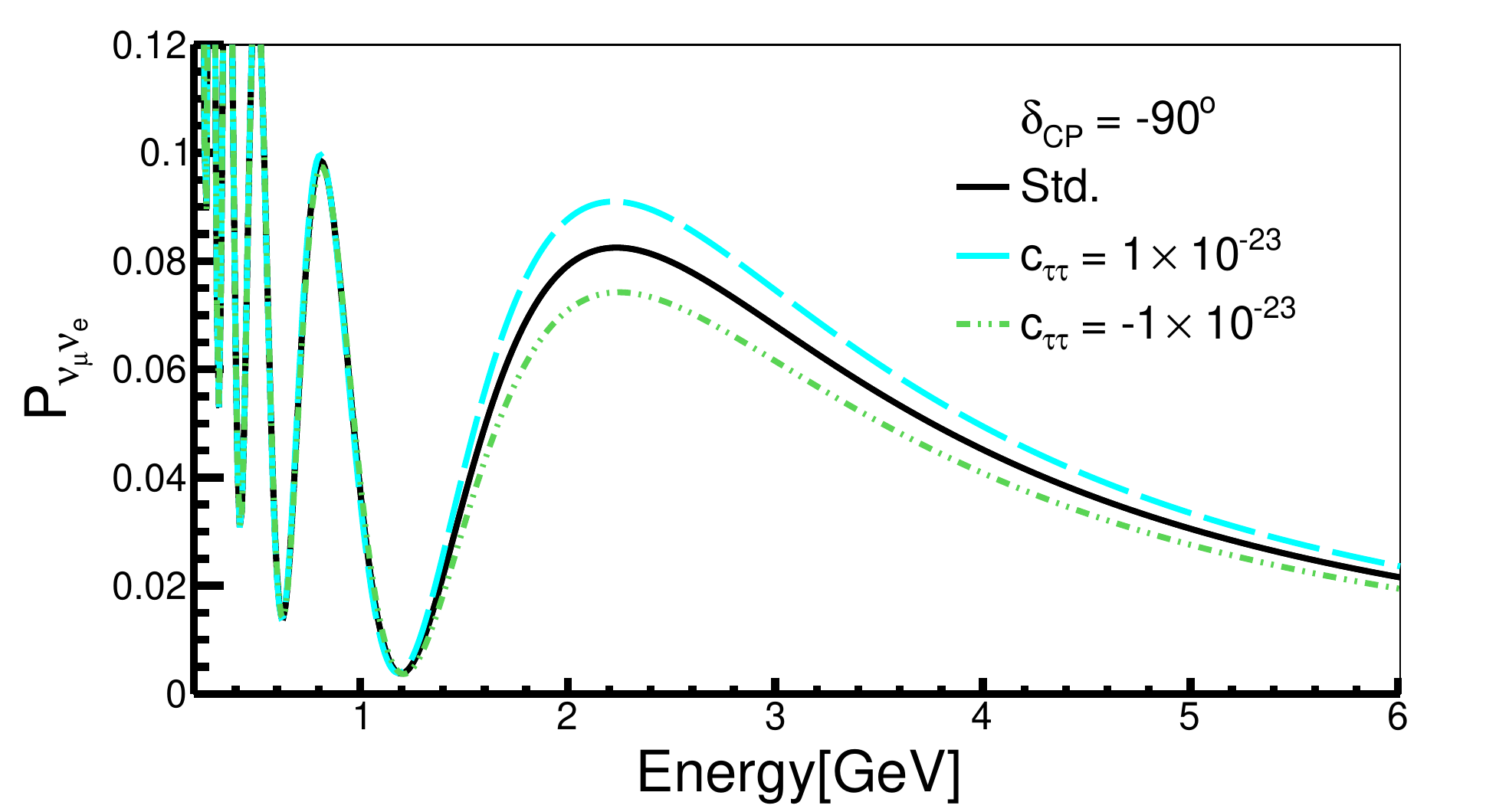}
\end{minipage}%

\begin{minipage}[t]{0.3\textwidth}
  \includegraphics[width=\linewidth]{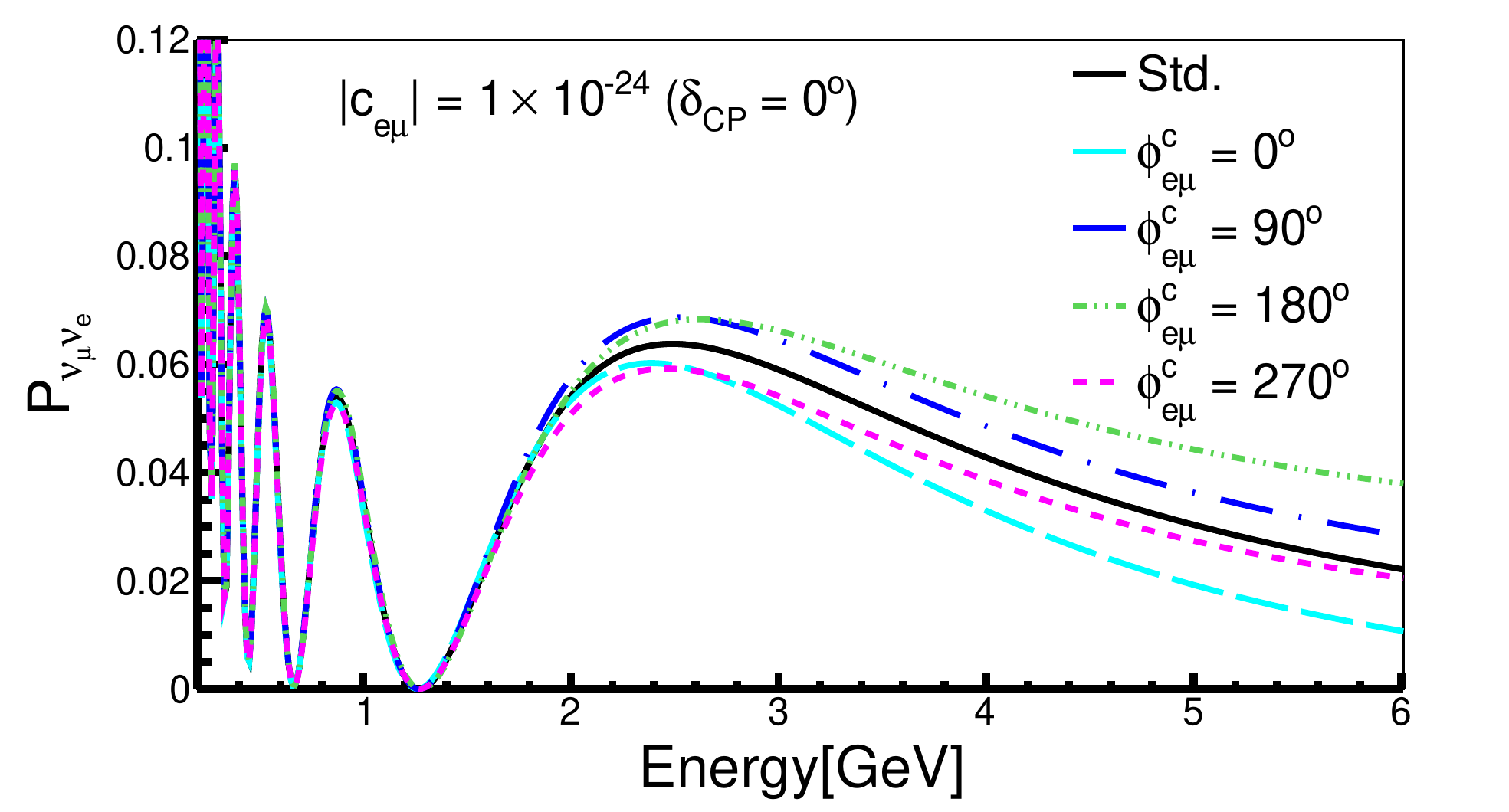}
 \end{minipage}
\begin{minipage}[t]{0.3\textwidth}
  \includegraphics[width=\linewidth]{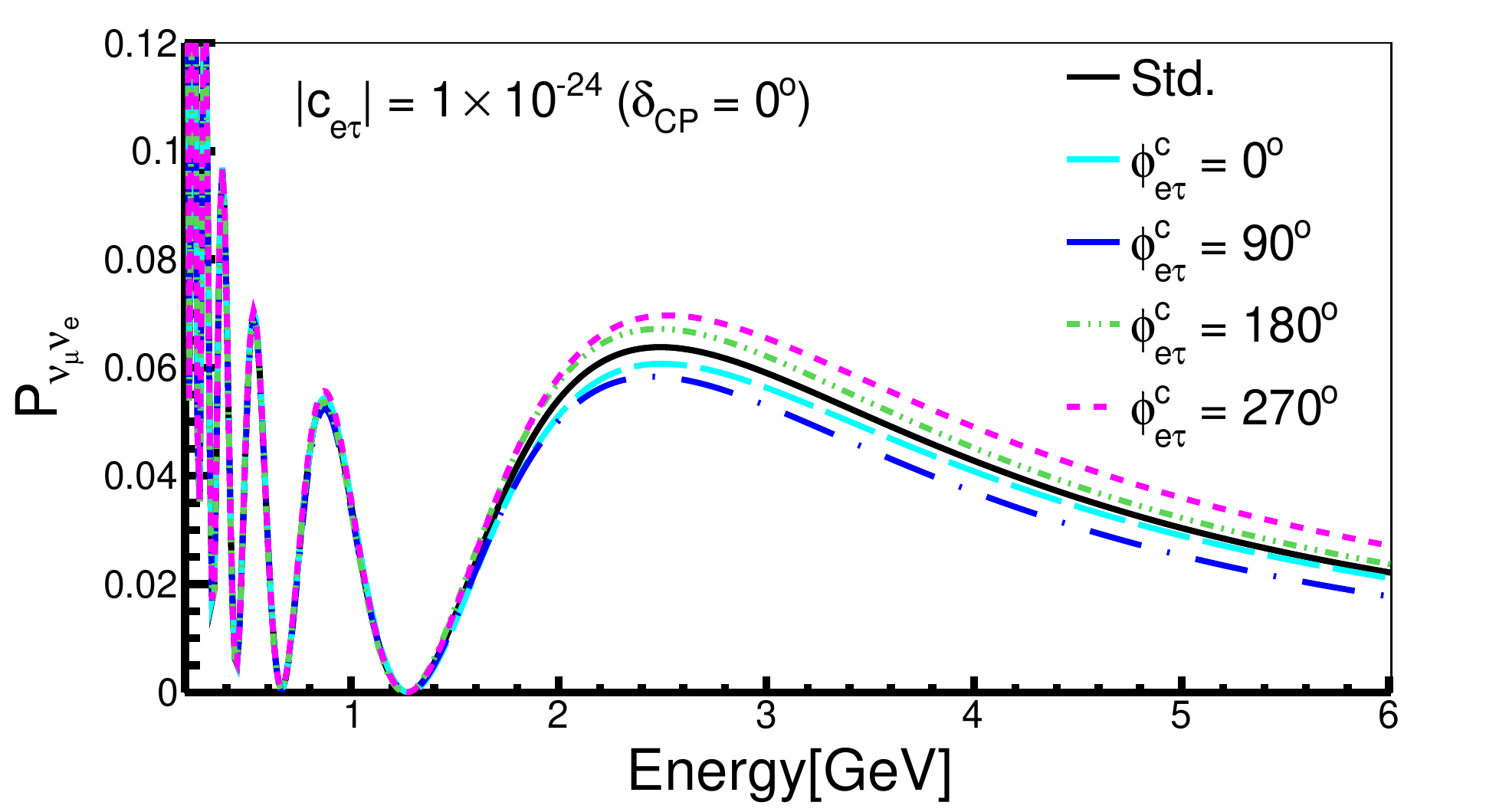}
 \end{minipage}
\begin{minipage}[t]{0.3\textwidth}
  \includegraphics[width=\linewidth]{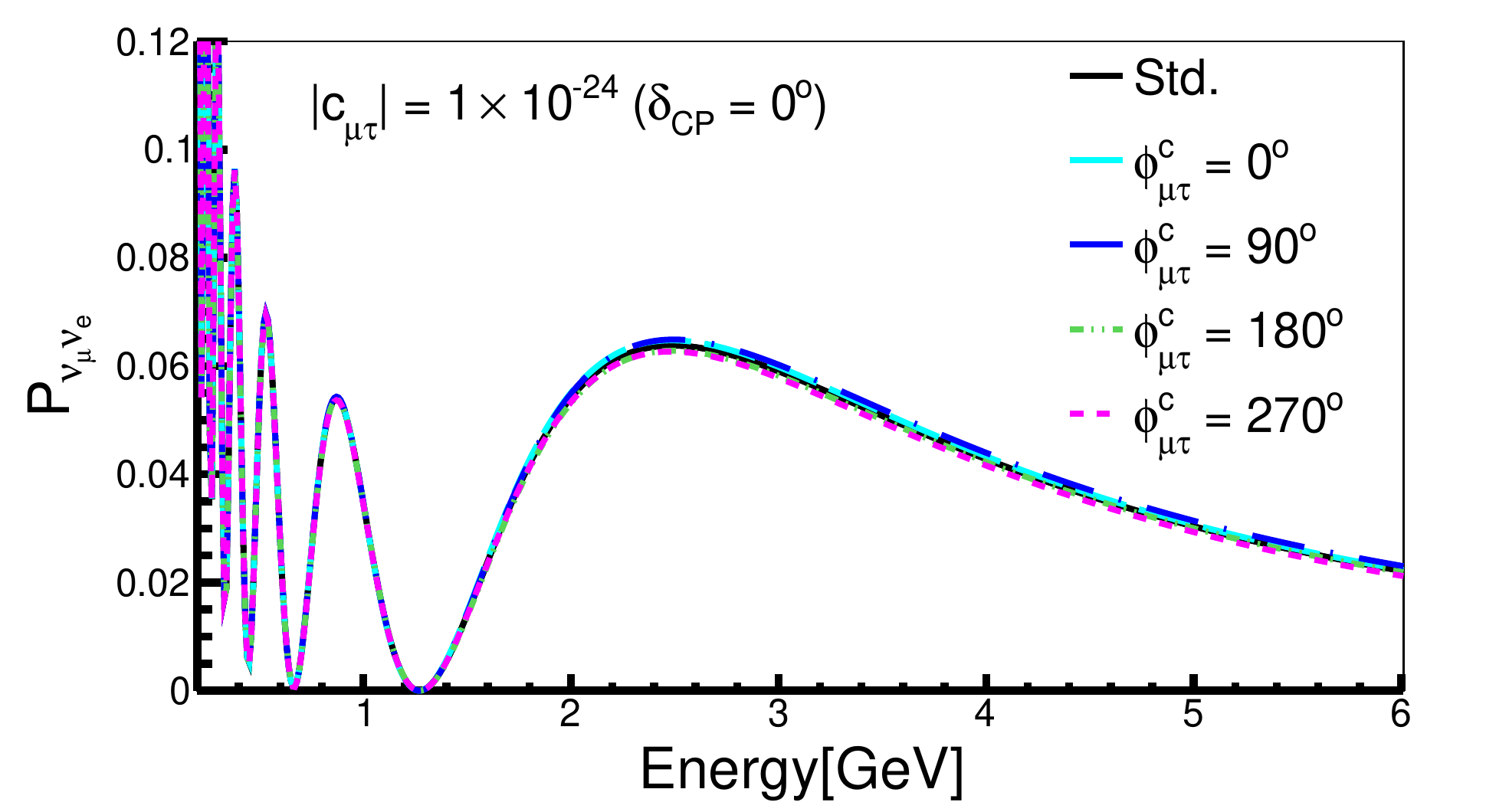}
\end{minipage}%

\begin{minipage}[t]{0.3\textwidth}
  \includegraphics[width=\linewidth]{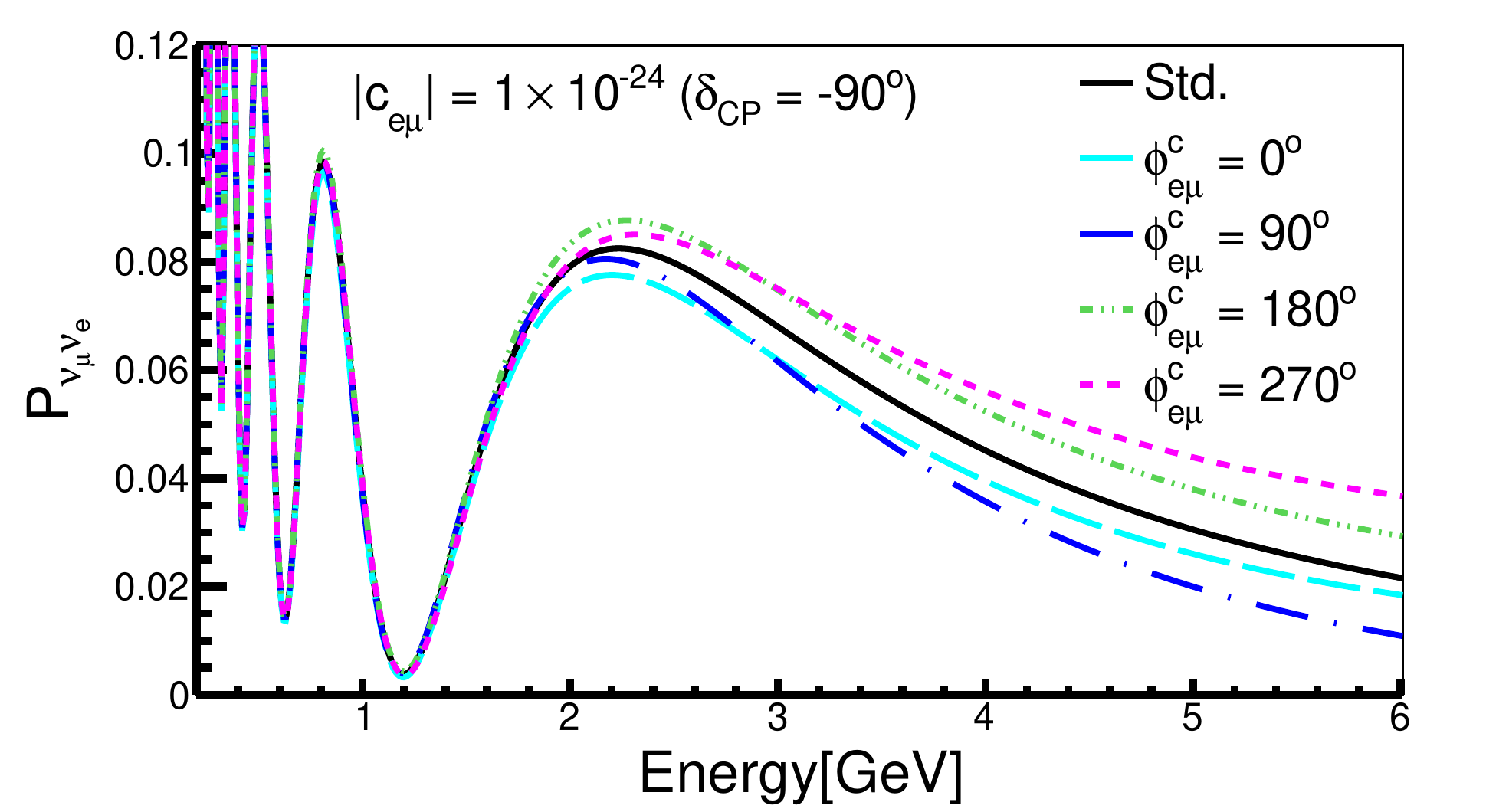}
 \end{minipage}
\begin{minipage}[t]{0.3\textwidth}
  \includegraphics[width=\linewidth]{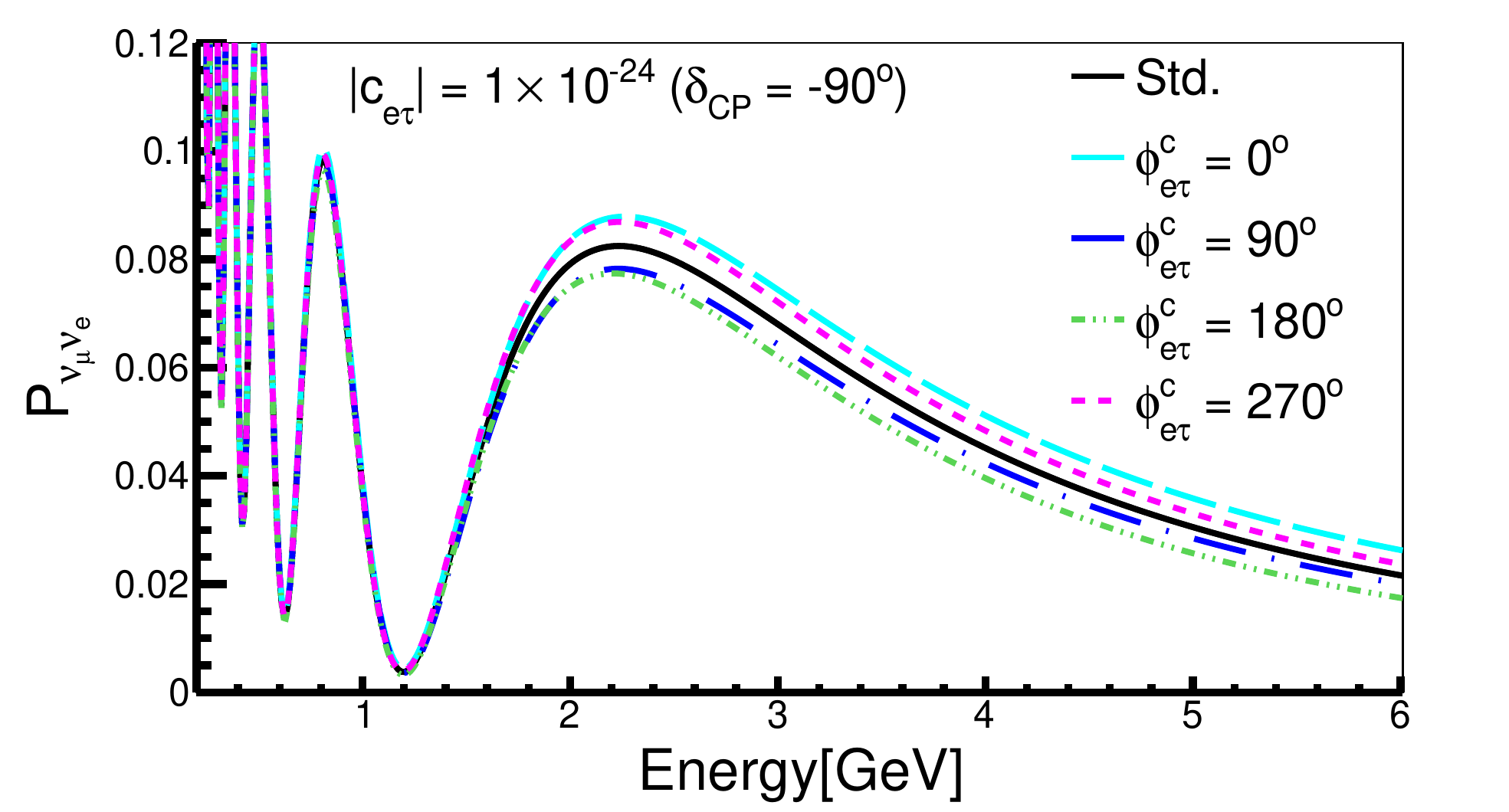}
 \end{minipage}
\begin{minipage}[t]{0.3\textwidth}
  \includegraphics[width=\linewidth]{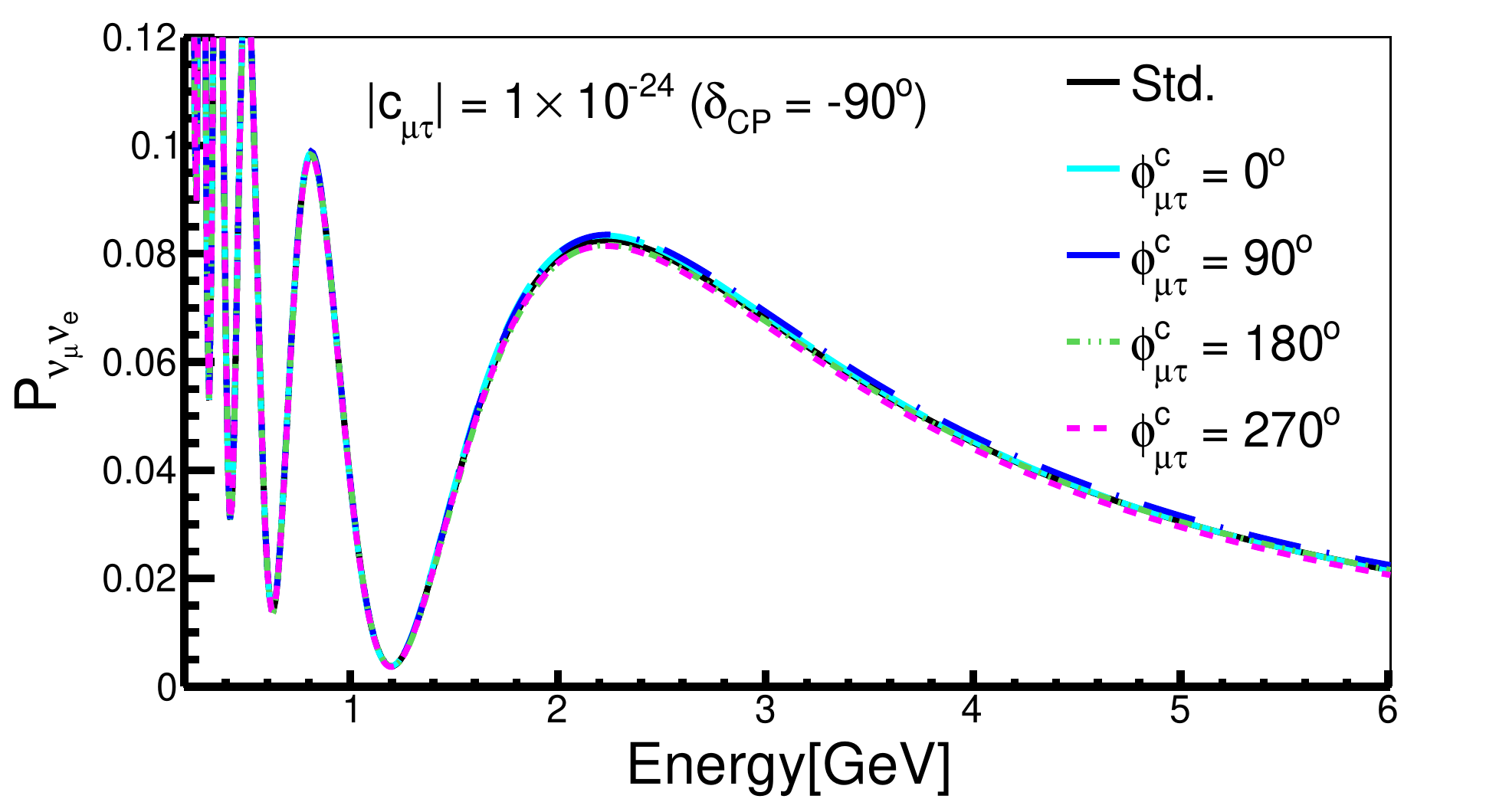}
\end{minipage}%
\caption{The $\nu_{e}$ appearance probability plots for $c_{ee}$, $c_{\mu\mu}$, $c_{\tau\tau}$ (first/second row is for $\delta_{\rm CP}=0^\circ/-90^\circ$) and $c_{e\mu}$, $c_{e\tau}$, $c_{\mu\tau}$ (third/fourth row is for $\delta_{\rm CP}=0^\circ/-90^\circ$) for DUNE setup. In all panels the probabilities for standard three generation oscillations are shown by the solid black curves.}
\label{fig:dune-prob-c}
\end{figure*}

In Fig.~\ref{fig:dune-prob-a} and Figure~\ref{fig:dune-prob-c}, we present the appearance channel ($\nu_\mu \rightarrow \nu_e$) probabilities as a function of energy ($E$) for DUNE baseline, considering $a_{\alpha \beta}$ and $c_{\alpha \beta}$, respectively. The first and second rows in both figures correspond to diagonal LIV parameters, while the third and fourth rows correspond to off-diagonal LIV parameters. In the first and third rows, we set the value of $\delta_{\rm CP}$ to $0^\circ$, whereas in the second and fourth rows, it is set to $-90^\circ$. In each panel, the black curve represents standard oscillations in the three-flavor scenario. For diagonal LIV parameters, we present curves for both positive and negative values, and for non-diagonal LIV parameters, we show curves for four values of $\phi^{a/c}_{\alpha \beta}$: $0^\circ$, $90^\circ$, $180^\circ$, and $270^\circ$. These figures are only for neutrinos. To understand the effect of LIV in the disappearance channel, we refer to Fig.\ref{fig:inooscillogrm_a} and Fig.\ref{fig:inooscillogrm_c}. The red line indicates the relevant regions in the oscillograms for DUNE. 

From these figures, we can deduce the following physics points. From the analytical calculation of the appearance channel probability it has been shown that the LIV parameters $a_{e\mu}$, $a_{e\tau}$ and $a_{ee}$ dominate at the leading order \cite{KumarAgarwalla:2019gdj}. Therefore at DUNE, we expect to obtain good sensitivity for these parameters. Amongst $a_{\alpha \alpha}$, we observe that DUNE is more sensitive to the parameters $a_{ee}$ and $a_{\tau\tau}$ as compared to $a_{\mu\mu}$. The effect of LIV for the parameter $a_{ee}$ is opposite to that of $a_{\tau\tau}$. For positive values of $a_{ee}$, the probability is higher than the probability in the standard three-flavour scenario, and for a negative value of $a_{ee}$, the probability is lower than the probability in the standard three-flavour scenario. This behaviour is opposite for $a_{\tau\tau}$. The effect of LIV on $a_{\mu \mu}$ is negligible. Regarding $a_{\alpha \beta}$ with $\alpha \neq \beta$, the effect of LIV in DUNE is greater in $a_{e\mu}$ and $a_{e\tau}$ compared to $a_{\mu\tau}$. For a given value of $a_{e\mu}$, the difference with the standard oscillation probability is greater for $\phi^a_{e\mu}=0^\circ$ and $180^\circ$ at $\delta_{\rm CP} = 0^\circ$, and it is greater for $\phi^a_{e\mu}=90^\circ$ and $270^\circ$ at $\delta_{\rm CP} = -90^\circ$. For $a_{e\tau}$, the behavior is opposite to that of $a_{e\mu}$, i.e., for $\delta_{\rm CP} = 0^\circ$, the separation of the probability from the standard oscillation probability in the presence of $a_{e\tau}$ is greater at $\phi^a_{e\tau}=90^\circ$ and $270^\circ$, and for $\delta_{\rm CP} = -90^\circ$, the separation is greater for $\phi^a_{e\tau}=0^\circ$ and $180^\circ$. The effect of $a_{\mu\tau}$ for the appearance channel is seen to be negligible for DUNE.

In general, the appearance channel in DUNE is less sensitive to CPT-conserving LIV parameters as compared to CP-violating LIV parameters. The behavior of $c_{\alpha\alpha}$ is seen to be similar to that of $a_{\alpha\alpha}$. There is more sensitivity to parameters $c_{ee}$ and $c_{\tau\tau}$ as compared to $c_{\mu\mu}$. The effect of the parameter $c_{ee}$ is opposite to that of $c_{\tau\tau}$. For a positive values of $c_{ee}$, the probability is lower compared to the probability in the standard three-flavor scenarios, while for negative values of $c_{ee}$, the probability is higher than the probability in the standard three-flavor scenarios. The trend is opposite for $c_{\tau\tau}$. The effect of $c_{\mu\mu}$ is negligible.

The behavior of $c_{\alpha\beta}$ with $\alpha\neq\beta$ is similar to that of $a_{\alpha\beta}$. The effect is more pronounced for $c_{e\mu}$ and $c_{e\tau}$ as compared to $c_{\mu\tau}$. For a given value of $c_{e\mu}$, the difference with the standard oscillation probability is higher for $\phi^c_{e\mu}=0^\circ$ and $180^\circ$ when $\delta_{\rm CP} = 0^\circ$, and it is higher for $\phi^c_{e\mu}=90^\circ$ and $270^\circ$ when $\delta_{\rm CP} = -90^\circ$. For $c_{e\tau}$ the situation is opposite to what we see for $c_{e\mu}$. The effect of $c_{\mu\tau}$ is negligible.


Upon examining the previous discussion it becomes clear that the appearance channel of DUNE is weakly sensitive to the parameters $a_{\mu \mu}$, $a_{\mu\tau}$, $c_{\mu \mu}$, and $c_{\mu\tau}$. However, the disappearance channel  exhibits better sensitivity to these parameters. Thus, by combining information from both channels, it will be possible to achieve some sensitivity for these four parameters.

\subsection{T2HK}

\begin{figure*}

\begin{minipage}[t]{0.3\textwidth}
  \includegraphics[width=\linewidth]{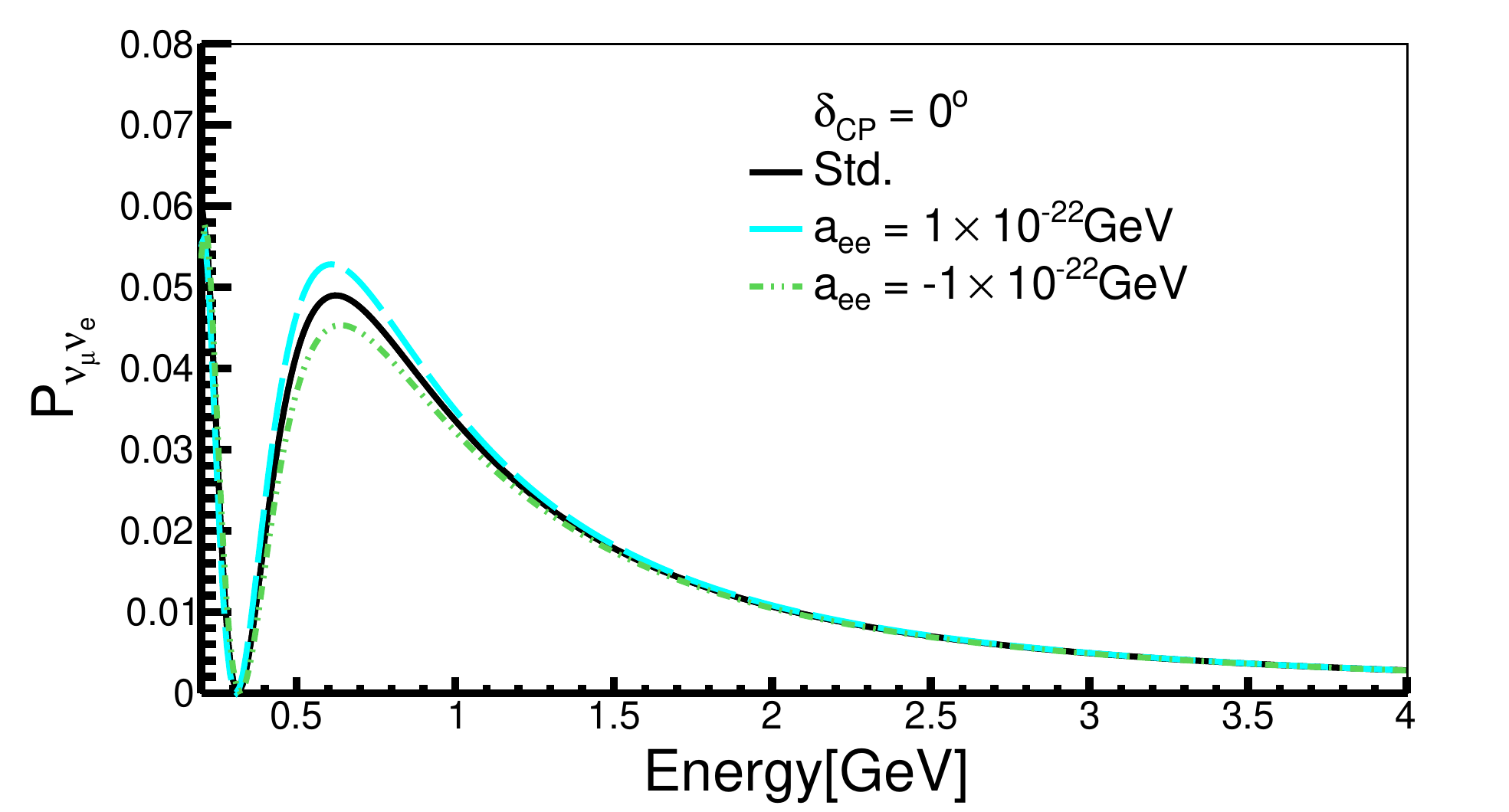}
 \end{minipage}
\begin{minipage}[t]{0.3\textwidth}
  \includegraphics[width=\linewidth]{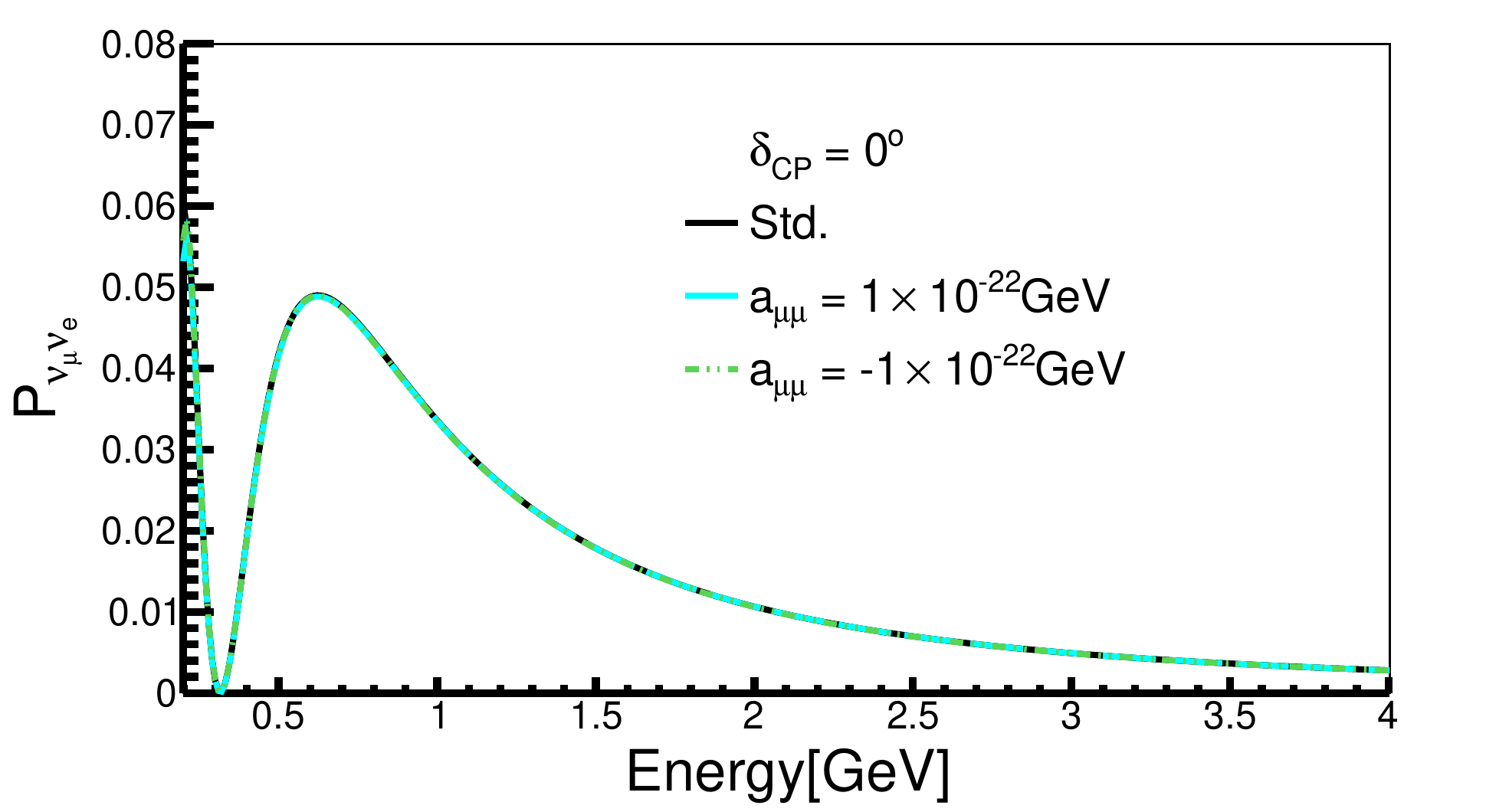}
 \end{minipage}
\begin{minipage}[t]{0.3\textwidth}
  \includegraphics[width=\linewidth]{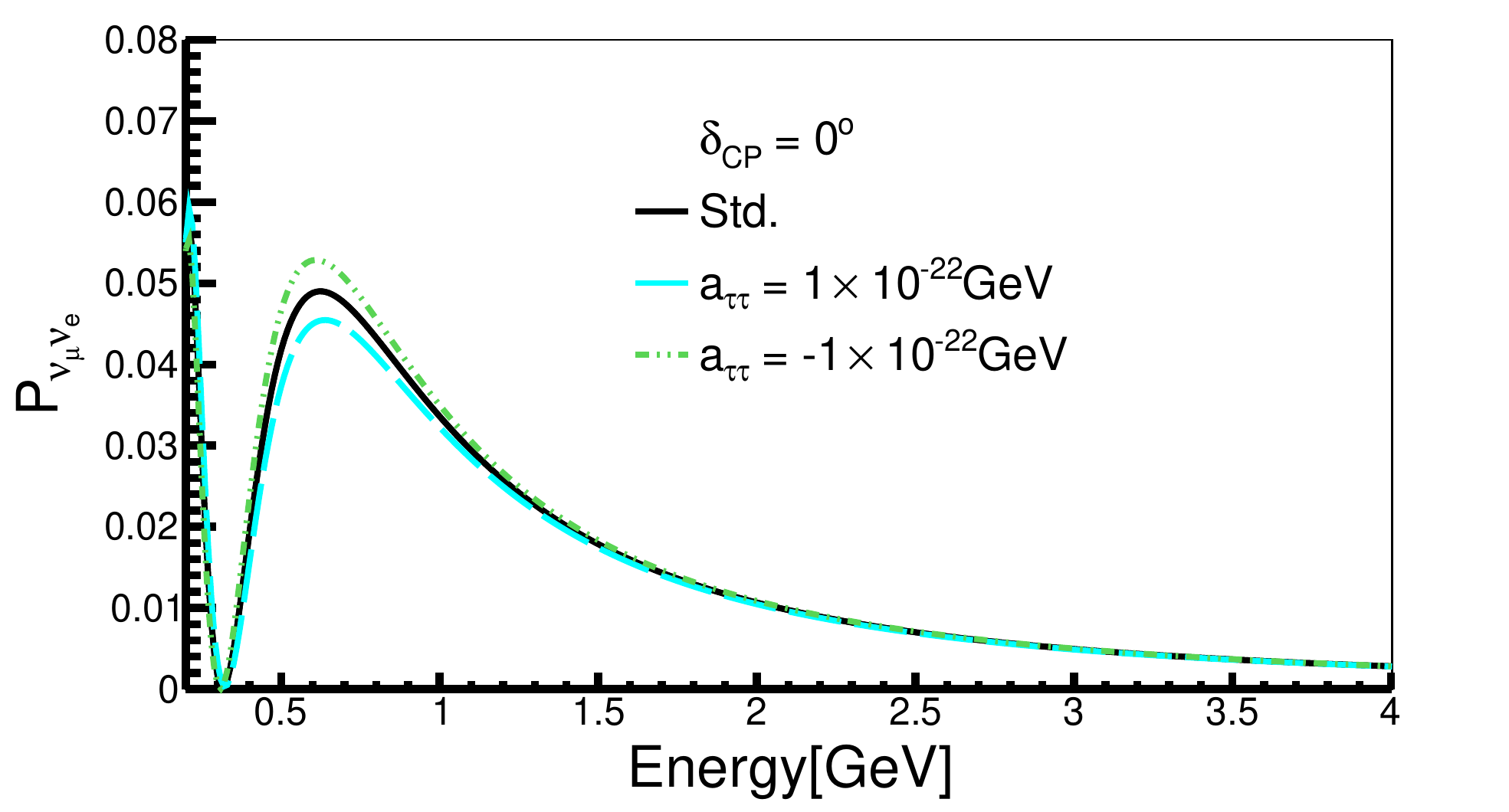}
\end{minipage}%

\begin{minipage}[t]{0.3\textwidth}
  \includegraphics[width=\linewidth]{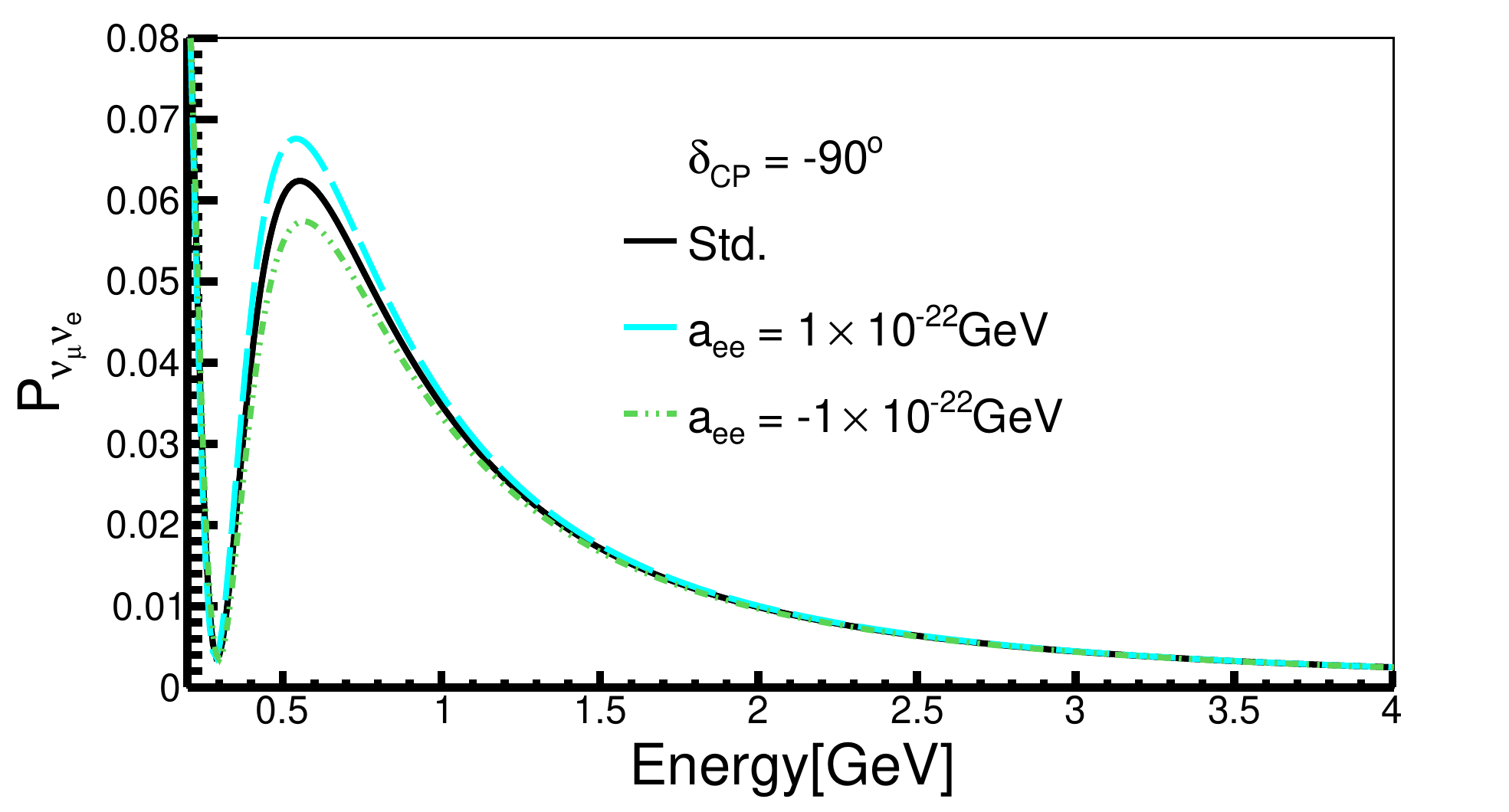}
 \end{minipage}
\begin{minipage}[t]{0.3\textwidth}
  \includegraphics[width=\linewidth]{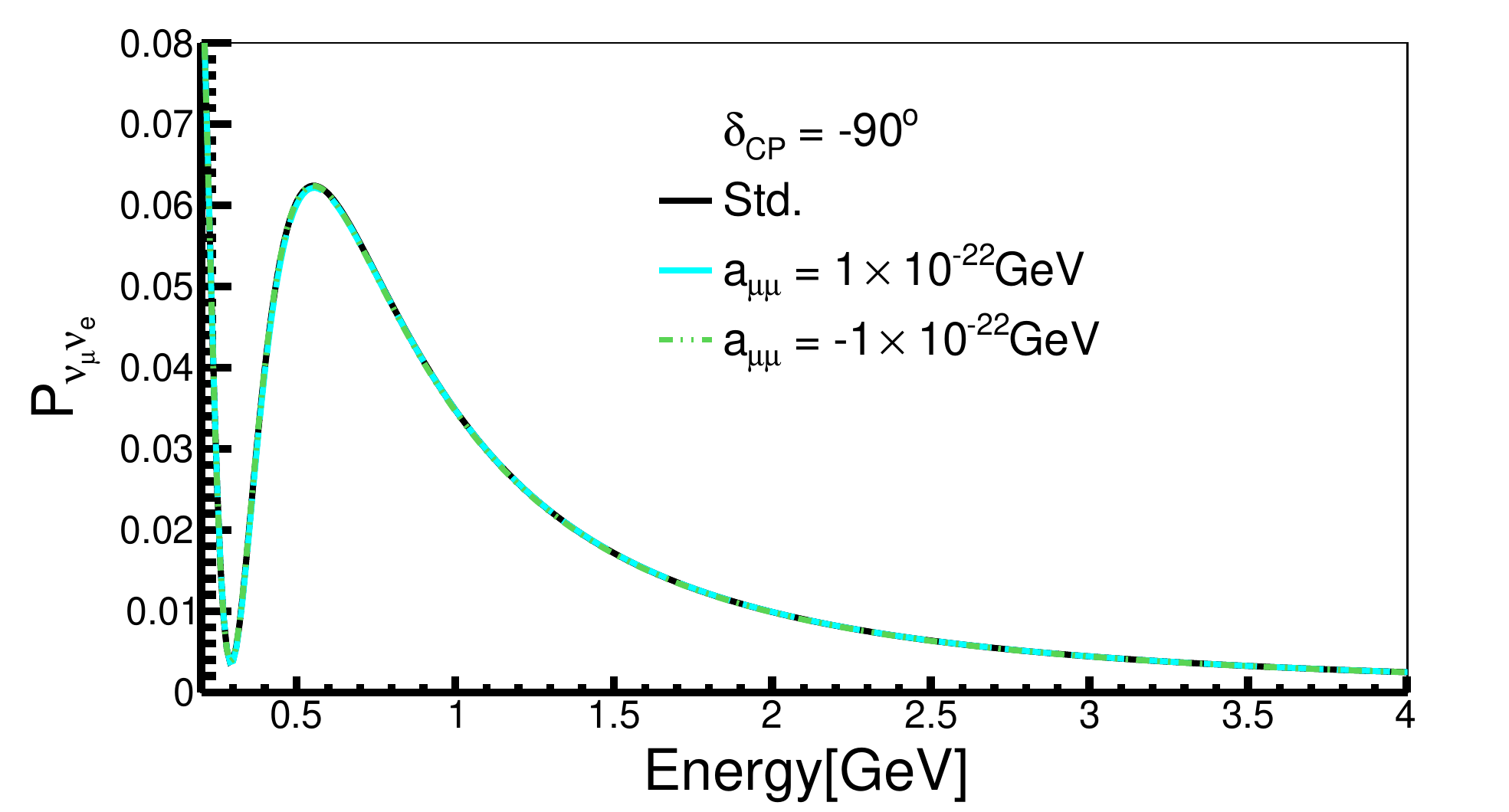}
 \end{minipage}
\begin{minipage}[t]{0.3\textwidth}
  \includegraphics[width=\linewidth]{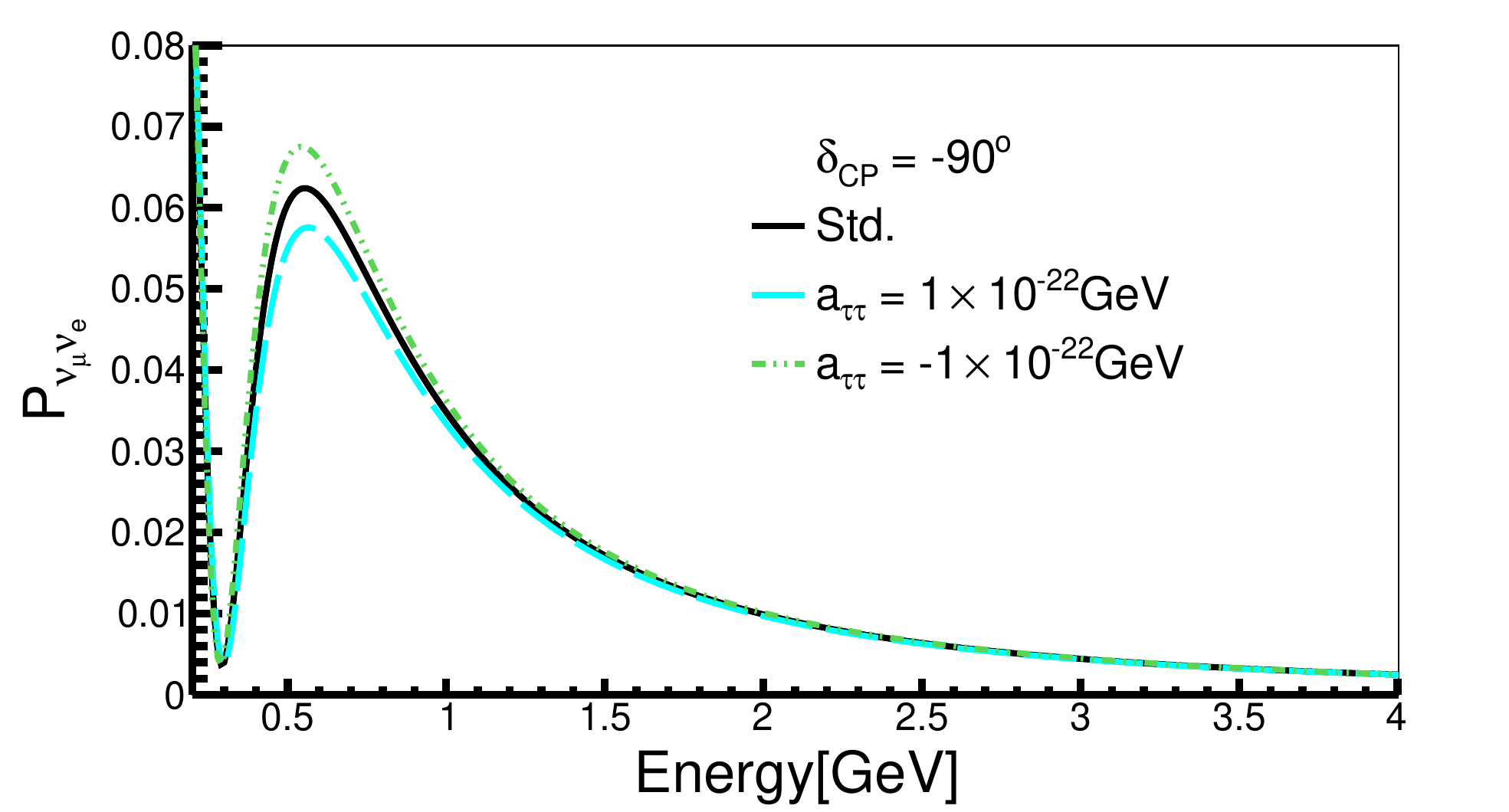}
\end{minipage}%

\begin{minipage}[t]{0.3\textwidth}
  \includegraphics[width=\linewidth]{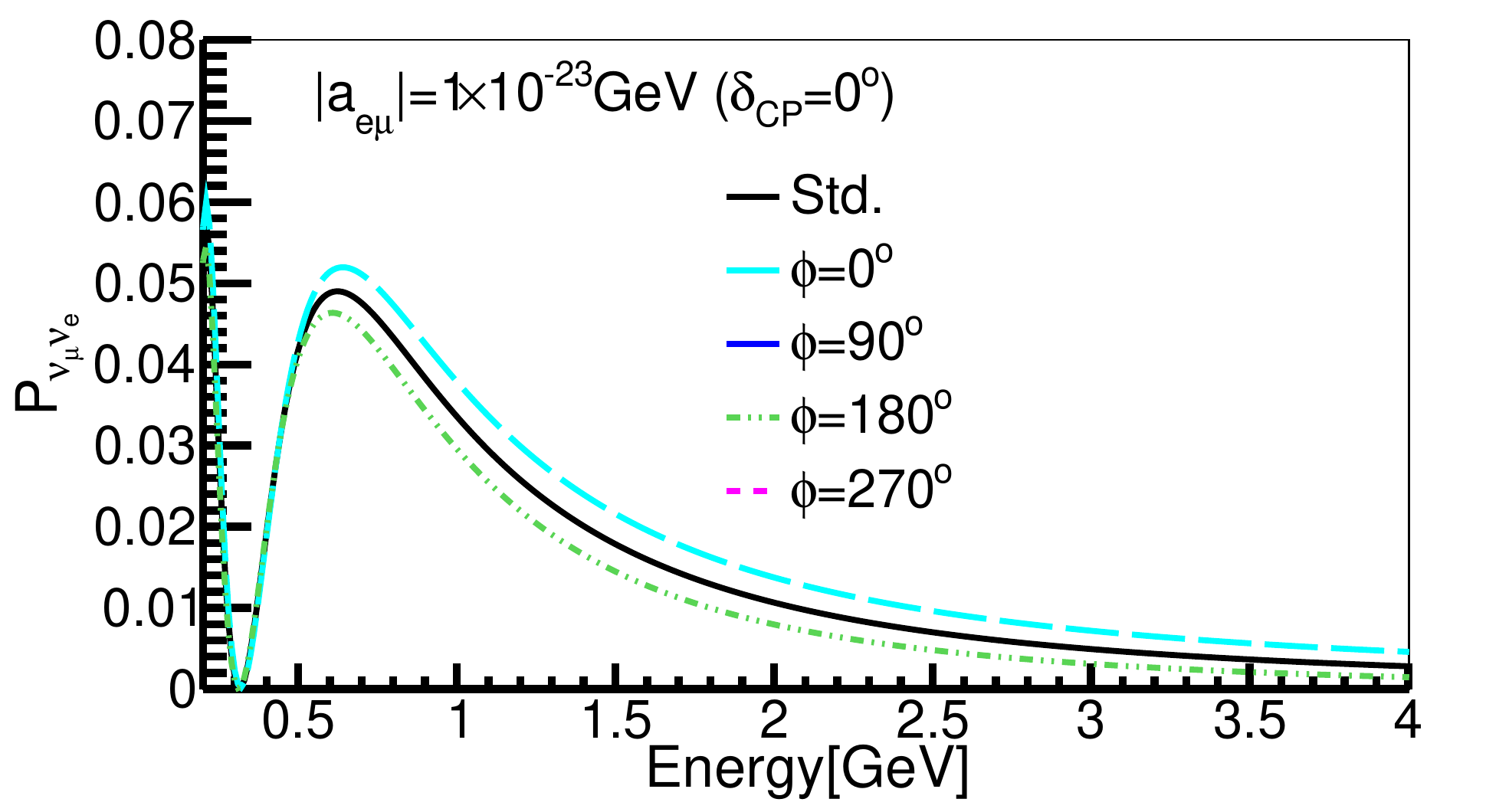}
 \end{minipage}
\begin{minipage}[t]{0.3\textwidth}
  \includegraphics[width=\linewidth]{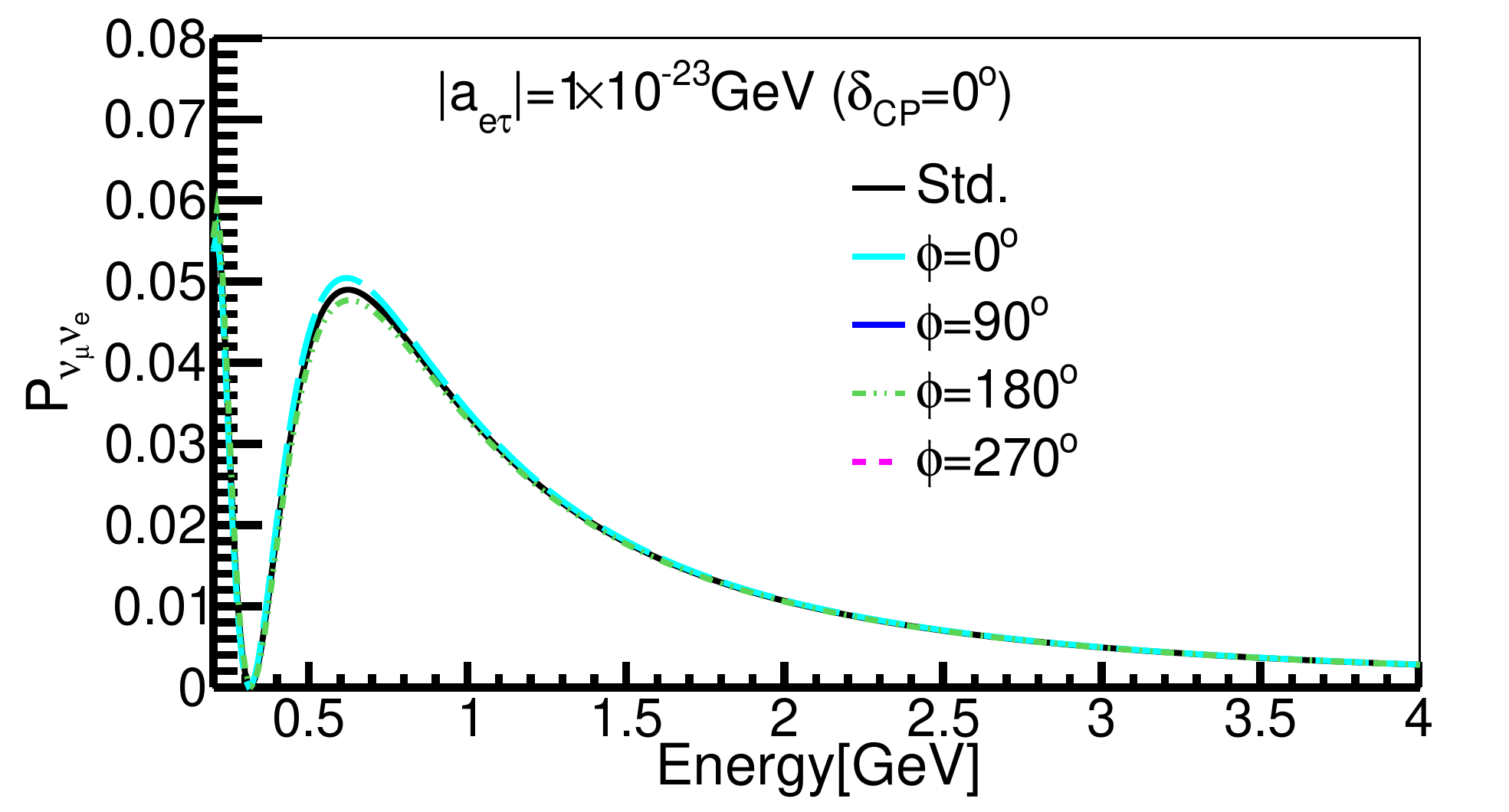}
 \end{minipage}
\begin{minipage}[t]{0.3\textwidth}
  \includegraphics[width=\linewidth]{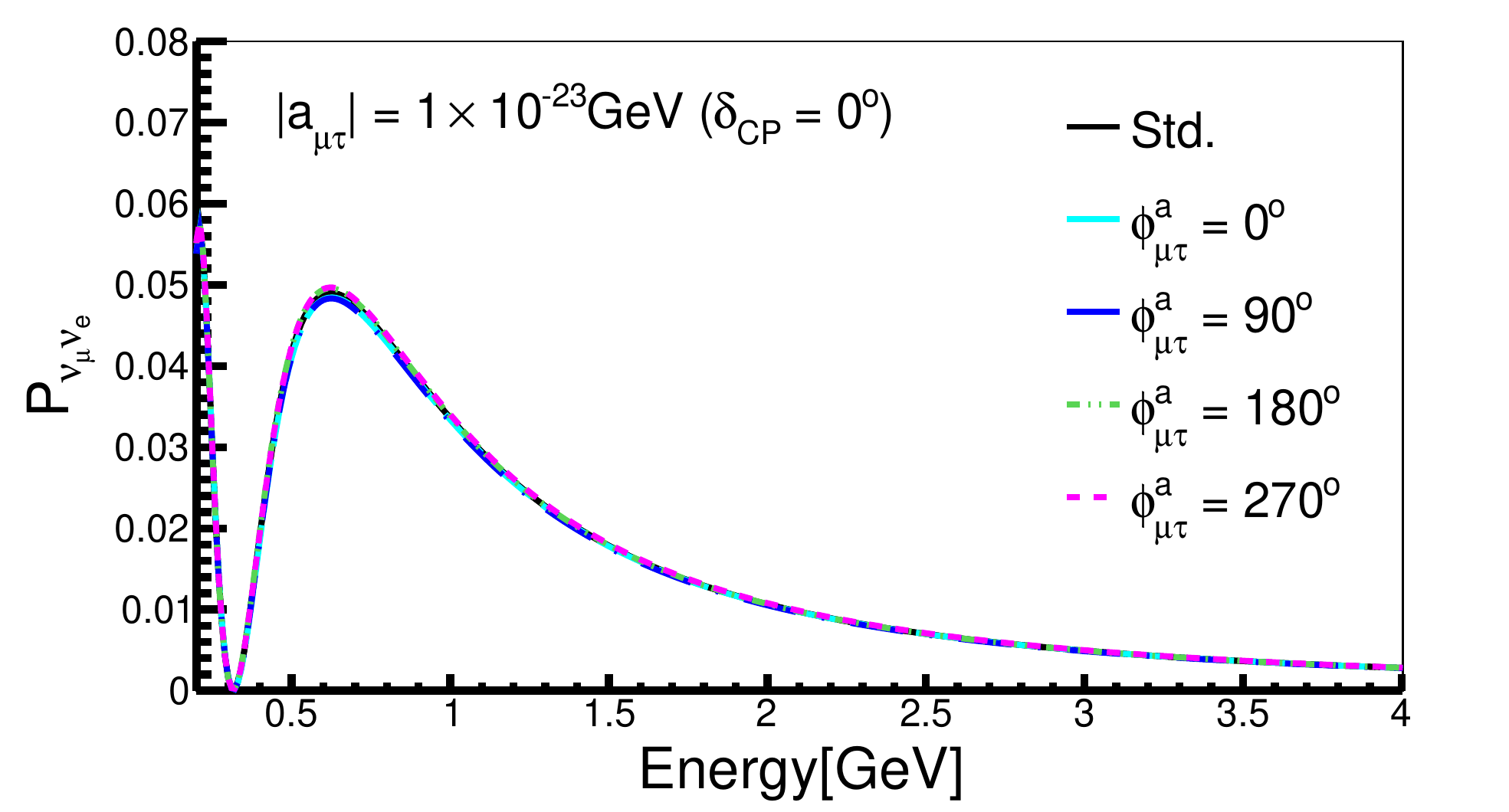}
\end{minipage}%

\begin{minipage}[t]{0.3\textwidth}
  \includegraphics[width=\linewidth]{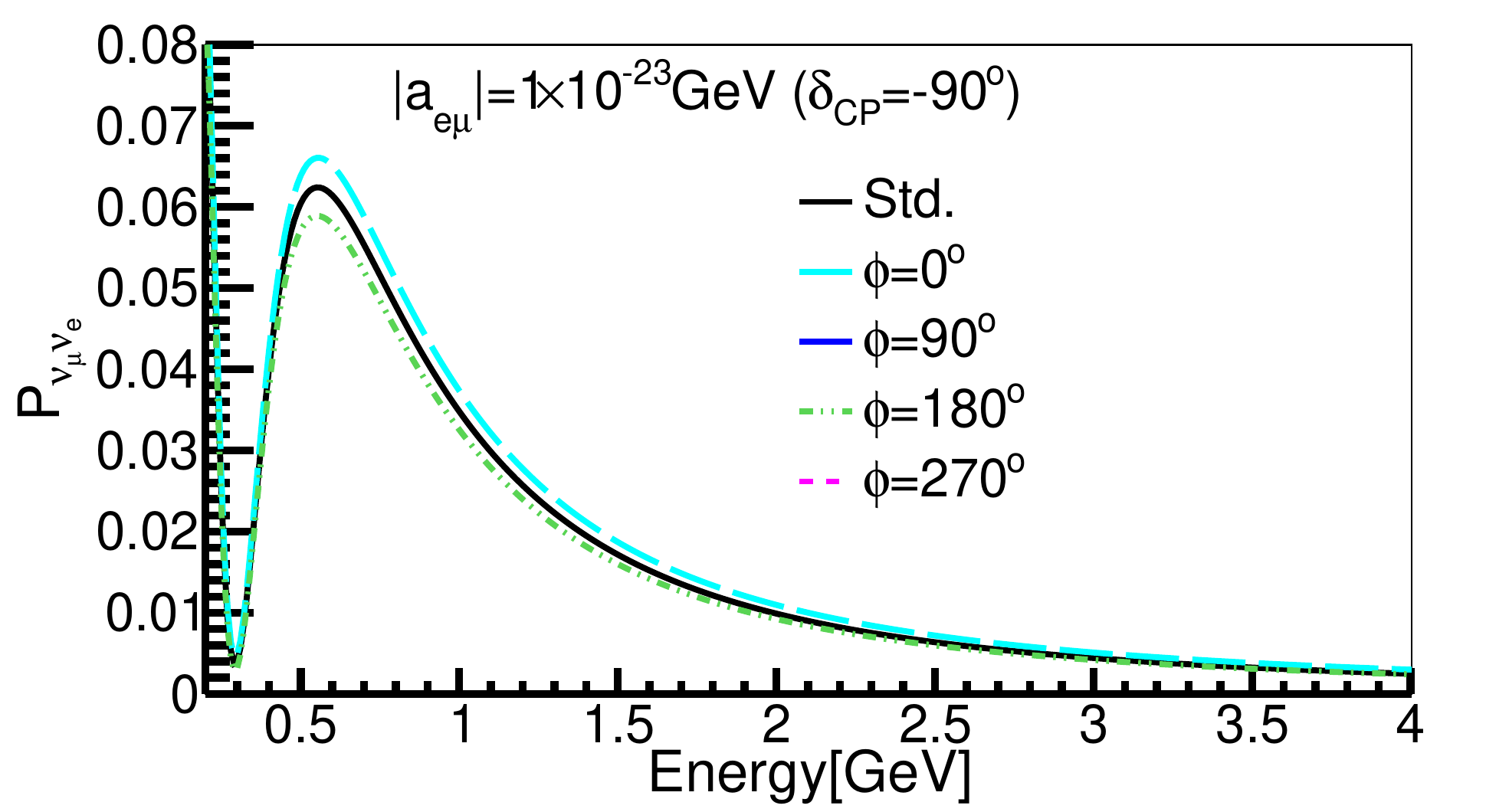}
 \end{minipage}
\begin{minipage}[t]{0.3\textwidth}
  \includegraphics[width=\linewidth]{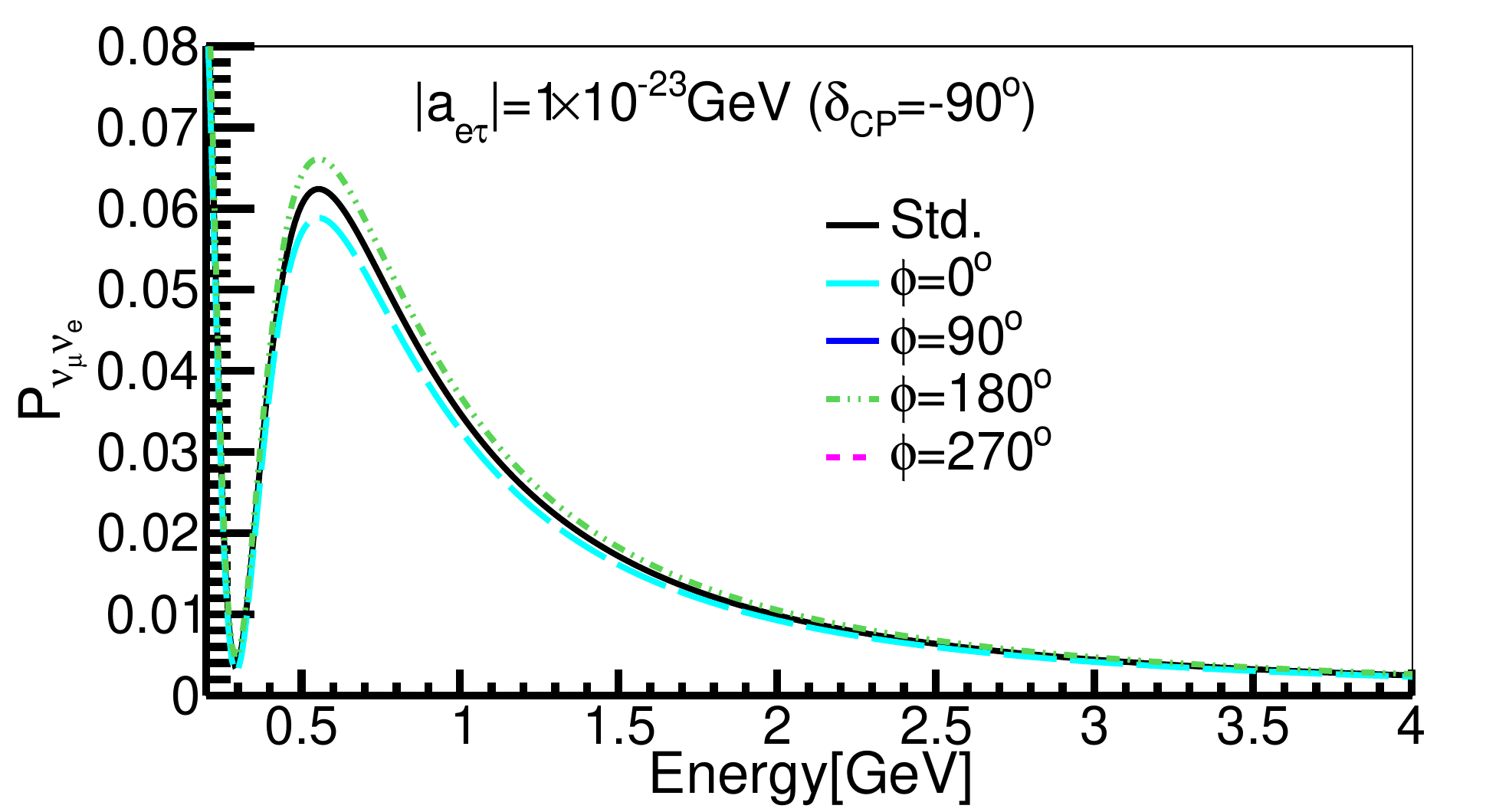}
 \end{minipage}
\begin{minipage}[t]{0.3\textwidth}
  \includegraphics[width=\linewidth]{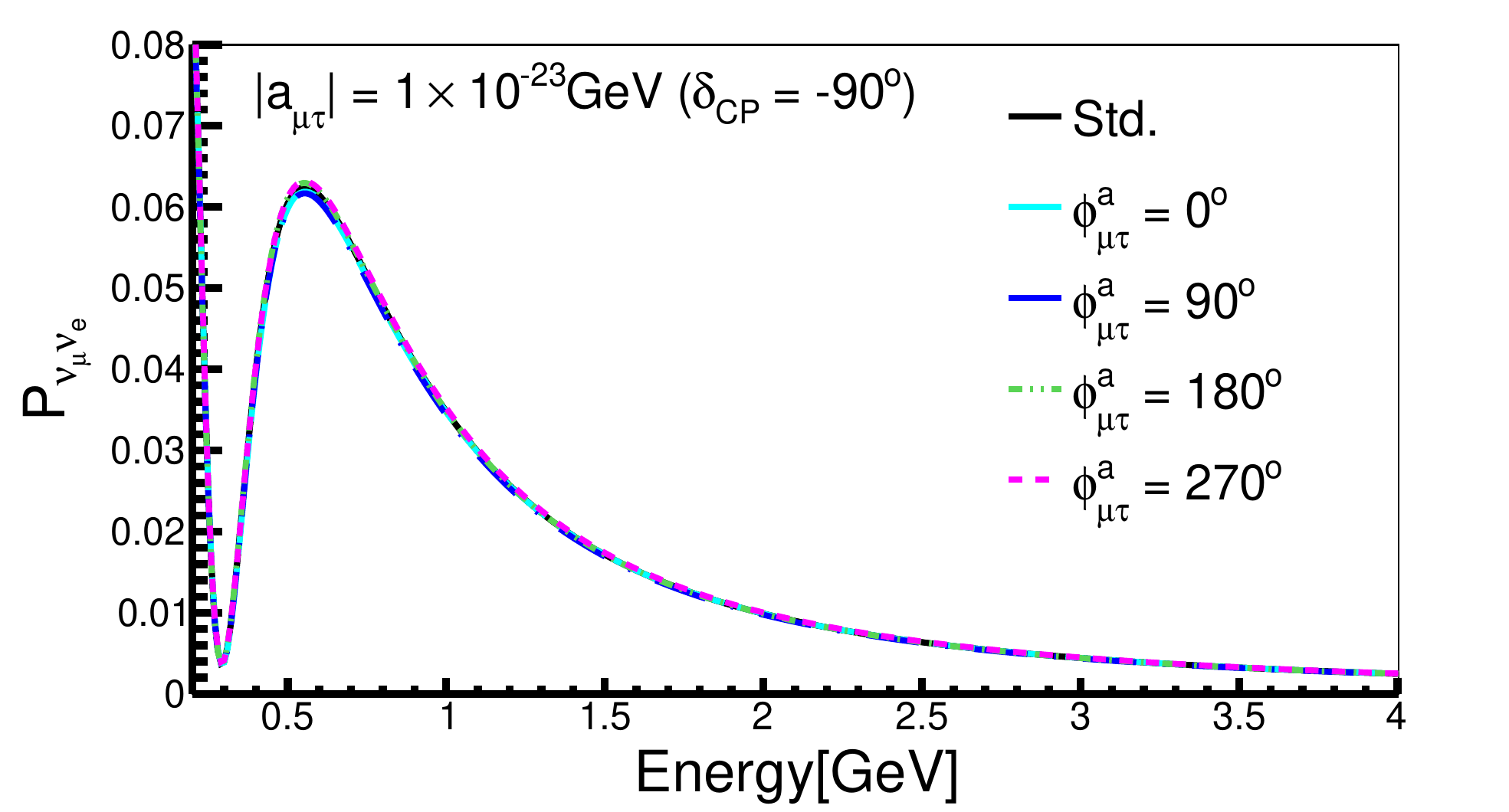}
\end{minipage}%
\caption{The $\nu_{e}$ appearance probability plots for $a_{ee}$, $a_{\mu\mu}$, $a_{\tau\tau}$ (first/second row is for $\delta_{\rm CP}=0^\circ/-90^\circ$) and $a_{e\mu}$, $a_{e\tau}$, $a_{\mu\tau}$ (third/fourth row is for $\delta_{\rm CP}=0^\circ/-90^\circ$) for T2HK setup. In all panels the probabilities for standard three generation oscillations are shown by the solid black curves.}
\label{fig:t2hk-prob-a}
\end{figure*}

Fig. \ref{fig:t2hk-prob-a} depicts the same information for the appearance channel as shown in Fig. \ref{fig:dune-prob-a}, but for the T2HK baseline. For the disappearance channel in T2HK we refer to the black line in Fig. \ref{fig:inooscillogrm_a}. From the figures we note that the effect of LIV parameters in T2HK has features similar  to what we saw in the case of DUNE. However, for T2HK the separation between the probabilities corresponding to LIV and standard three-flavor oscillation is considerably less than that in DUNE. Thus, we expect to have the same behavior as in DUNE, but with significantly less sensitivity.

We do not present any figures for CP conserving LIV parameters for T2HK because we have explicitly checked that for the T2HK baseline and energy, the probability curves corresponding to standard oscillation and the probability curves for different values of $c_{\alpha \beta}$ are almost inseparable.

\section{Results}
\label{sens}

In this section we will present the capability of T2HK, DUNE and ICAL to limit the LIV parameters. We will estimate the sensitivity by calculating a $\chi^2$ function defined as~\cite{moon:2014}:

\begin{equation}
\resizebox{.9\hsize}{!}{$
\chi^{2} = \sum_{i} 2\left[(T_{i}-D_{i}) - D_{i} ln(\frac{T_{i}}{D_{i}})\right] + \sum_{j}\xi^{2}_{j}$
}
\end{equation}

where the sum is over the energy bins for T2HK and DUNE whereas in ICAL the sum is over $E_{\mu}^{obs}$, $\theta_{\mu}^{obs}$ and $E_{\rm had}^{obs}$ bins. The number of events in each bin for theory (data) is given by $T_{i}$ $(D_{i})$. To implement the systematic errors, the theory events are varied as:
\begin{equation}
T_{i} = T^{0}_{i}\left(1+\sum_{j}\pi^{j}_{i}\xi_{j} \right)
\end{equation}
where $T^{0}_{i}$ is the corresponding number of events in theory without systematic errors. The pull parameters $\xi_{ j}$ correspond to different sources of systematic uncertainties.

The oscillation parameters that we use in our calculations are specified in Table \ref{table:para}.  In our analysis, we use $\Delta m^{2}_{\rm eff}$ given by~\cite{Nunokawa:2005nx,Raut:2012dm}
\begin{eqnarray}
\resizebox{.85\hsize}{!}{$
\Delta m^{2}_{\rm eff}= \Delta m^{2}_{31}-(\cos^{2}\theta_{12}-\cos\delta_{\rm CP}\sin\theta_{13}\sin2\theta_{12}\tan\theta_{23})\Delta m^{2}_{21}$
}
\end{eqnarray}
We perform minimisation of the $\chi^2$ over $\theta_{23}$ in the range $40^\circ$ to $51^\circ$, $|\Delta m^{2}_{\rm eff}|$ within its current 3$\sigma$ range, and the mass ordering. The phase $\delta_{\rm CP}$ is varied over their full range for DUNE and T2HK. For ICAL, we have set $\delta_{\rm CP}=0^\circ$ both in theory and data since the $\chi^{2}$ for ICAL depends weakly on $\delta_{\rm CP}$. For T2HK and DUNE, we consider two values of $\delta_{\rm CP}$(true), namely $0^\circ$ and $-90^\circ$. Our analysis keeps $\Delta m^{2}_{21}$, $\theta_{13}$, and $\theta_{12}$ fixed. We present our results for true normal ordering of the neutrino masses. For true inverted ordering the plots are qualitatively similar, so we do not present them here for brevity. 

\begin{table}[h!]
\begin{center}
\scalebox{0.9}{
\begin{tabular}{ |c|c|c|c|c|c| } 
 \hline
$\Delta m^{2}_{21}$(eV$^{2})$  & $\Delta m^{2}_{\rm eff}$(eV$^{2})$ & $\sin^{2}\theta_{12}$ & $\sin^{2}\theta_{23}$ & $\sin^{2}2\theta_{13}$ & $\delta_{\rm CP}$\\ 
\hline
 7.42$\times 10^{-5}$ & 2.49$\times 10^{-3}$  & 0.33 & 0.5    & 0.0875  & $0^\circ/-90^\circ$\\ 		\hline
 fix & 2.38-2.62$\times 10^{-3}$  & fix & 0.41-0.6   & fix  & $0-360^\circ$\\ 		\hline
\end{tabular}}
\caption{ Oscillation parameters used in analysis }
\label{table:para}
\end{center}
\end{table}


\subsection{Sensitivity for $a_{\alpha\alpha}$ }

\begin{figure*}

\begin{minipage}[t]{0.45\textwidth}
  \includegraphics[width=\linewidth]{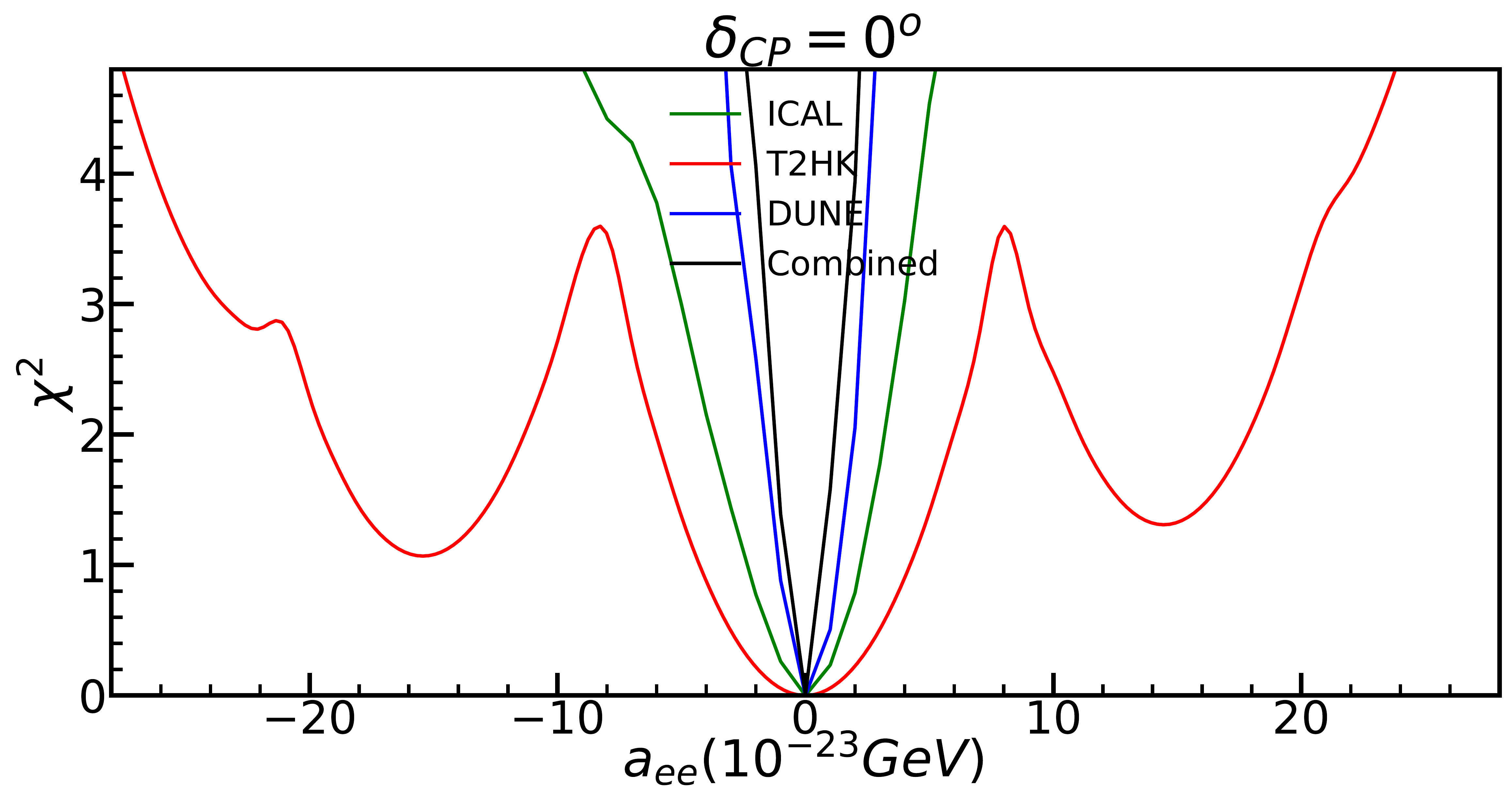}
\end{minipage}%
\hfill 
\begin{minipage}[t]{0.45\textwidth}
  \includegraphics[width=\linewidth]{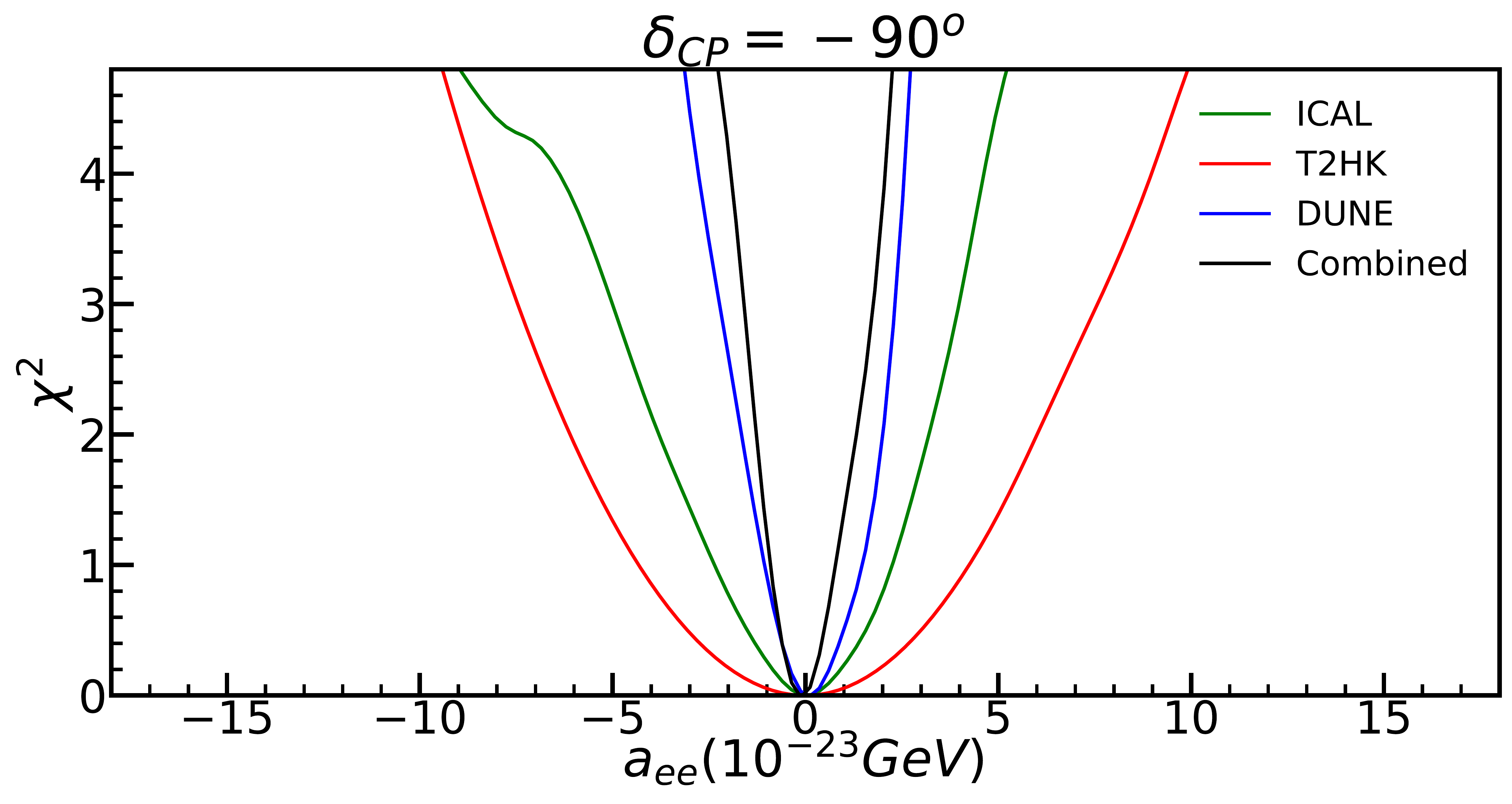}
\end{minipage}%

\begin{minipage}[t]{0.45\textwidth}
  \includegraphics[width=\linewidth]{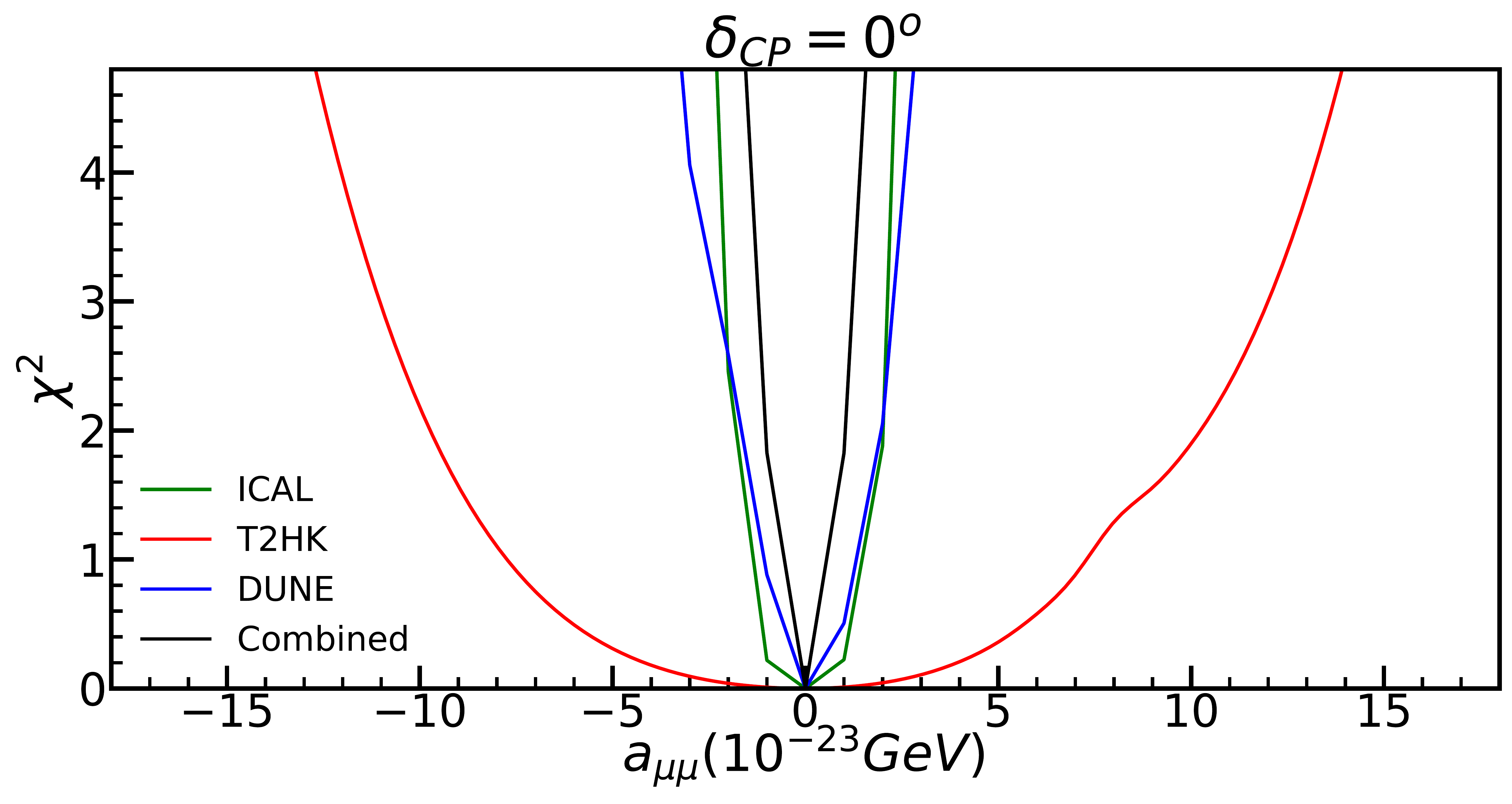}
\end{minipage}%
\hfill 
\begin{minipage}[t]{0.45\textwidth}
  \includegraphics[width=\linewidth]{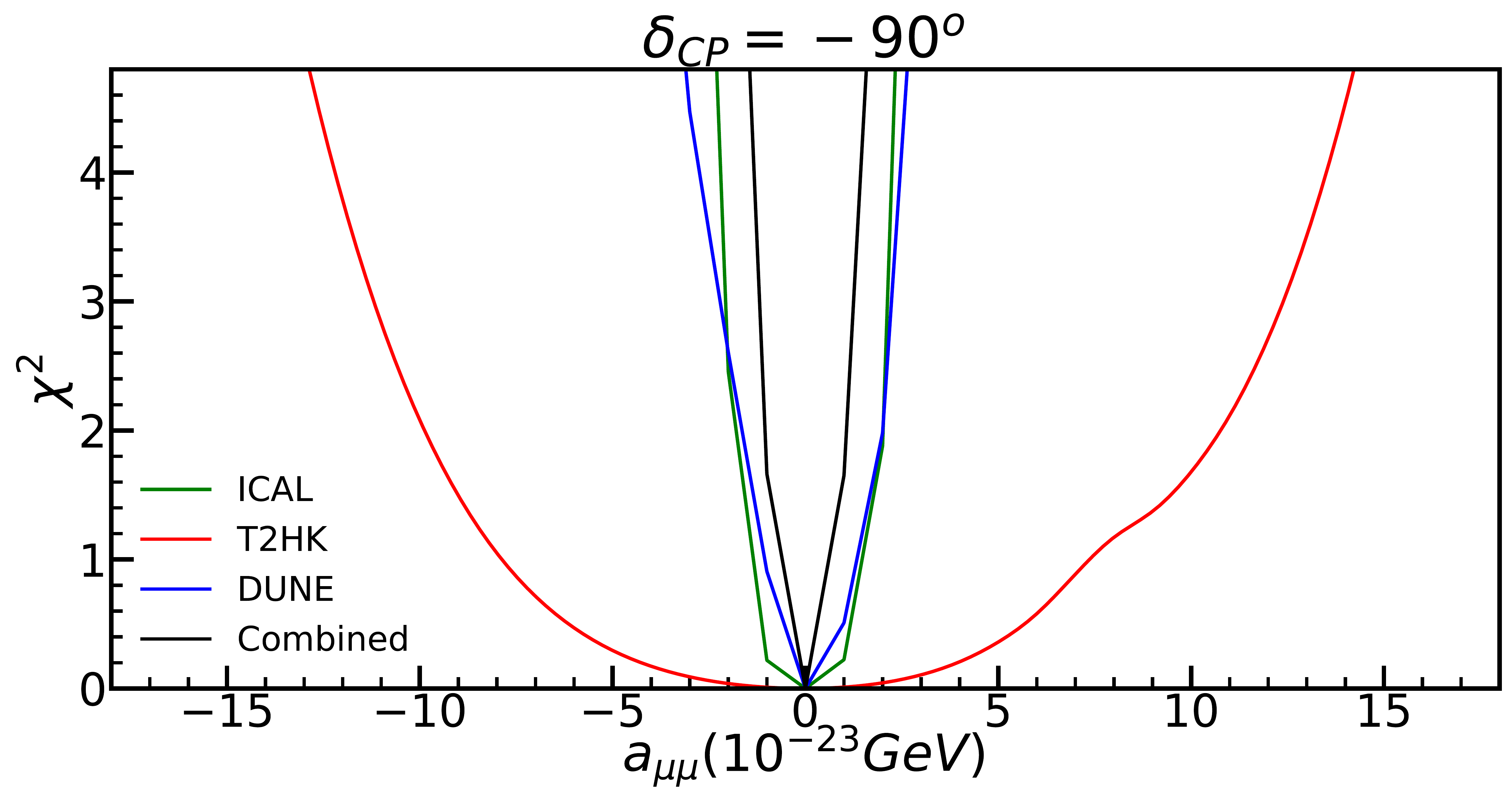}
\end{minipage}%

\begin{minipage}[t]{0.45\textwidth}
  \includegraphics[width=\linewidth]{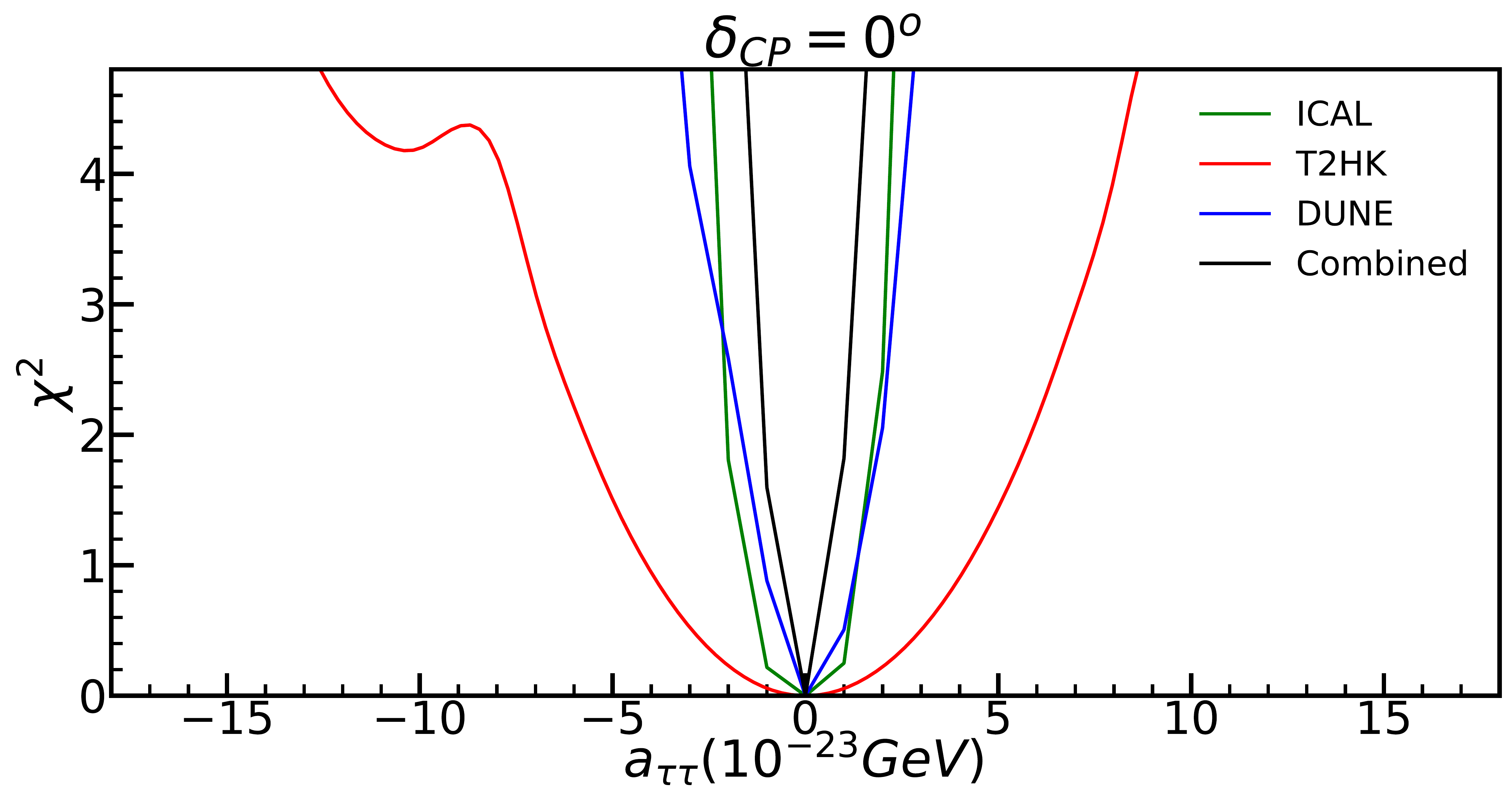}
\end{minipage}%
\hfill 
\begin{minipage}[t]{0.45\textwidth}
  \includegraphics[width=\linewidth]{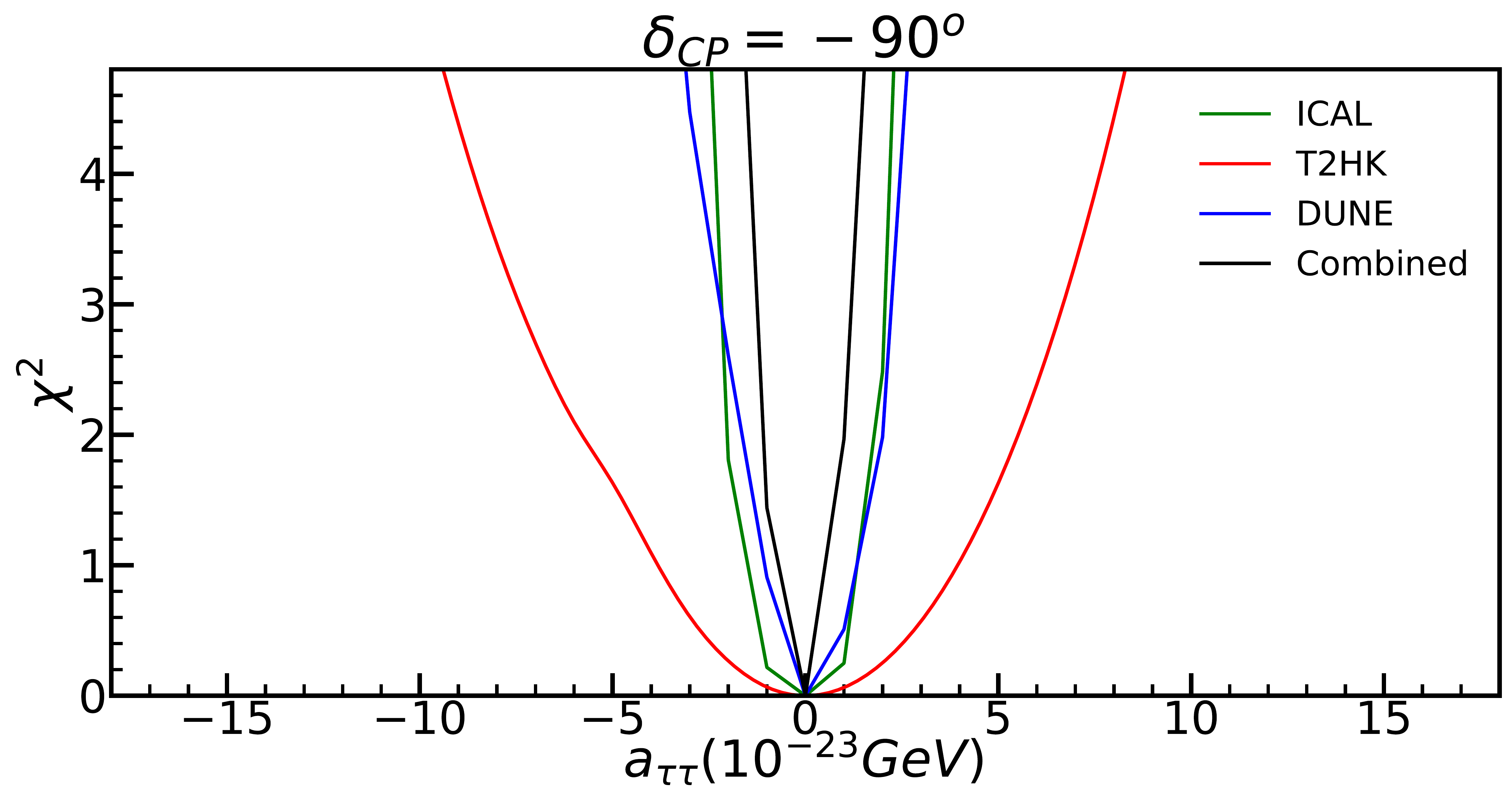}
\end{minipage}%

\caption{$\chi^{2}$ as a function of LIV parameters $a_{ee}$, $a_{\mu\mu}$ and $a_{\tau\tau}$ for true $\delta_{\rm CP}$ = $0^{\circ}$ (left column) and $-90^{\circ}$ (right column) in ICAL, DUNE, T2HK and combined.}
\label{fig:all-chi-no_aaa}
\end{figure*}

Let us begin our discussion with the diagonal CPT-violating LIV parameters. In Fig.~\ref{fig:all-chi-no_aaa}, we present the bounds on the parameters in the $\chi^2$ vs $a_{\alpha \alpha}$ plane. The left column is for $\delta_{\rm CP} = 0^\circ$, and the right column is for $\delta_{\rm CP} = -90^\circ$. In each column, different panels correspond to different $a_{\alpha \alpha}$ parameters. In each panel, we present the individual sensitivities of T2HK, DUNE, ICAL, and the combined sensitivity of all these three experiments.

From the discussion on the probabilities, we understood that ICAL has weaker sensitivity on $a_{ee}$ as compared to DUNE. On the other hand the sensitivity to $a_{\mu \mu}$ is poor in the appearance channel for DUNE, making it more difficult to measure this parameter in DUNE as compared to $a_{ee}$. For $a_{\tau \tau}$ both DUNE and ICAL are sensitive. T2HK was seen to have the weakest dependence on LIV. These features are clearly visible in Fig.~\ref{fig:all-chi-no_aaa}. From the panels we see that DUNE has the best sensitivity for $a_{ee}$, with better sensitivity expected for $\delta_{\rm CP} = -90^\circ$ as compared to $\delta_{\rm CP} = 0^\circ$. For $a_{\mu \mu}$ and $a_{\tau \tau}$, both ICAL and DUNE have comparable sensitivities. Also, for these parameters we expect similar sensitivity for both choices of $\delta_{\rm CP}$. For all the cases, T2HK has the weakest sensitivity. As expected, when we combine all the three experiments, we obtain the best possible sensitivity for all the parameters. In Table \ref{table:para-all-1} we have listed the 95$\%$ C.L. sensitivity limit of these parameters for both $\delta_{\rm CP}=0^\circ$ and $-90^\circ$. 
 
\subsection{Sensitivity for $a_{\alpha\beta}$}

\begin{figure*}

\begin{minipage}[t]{0.45\textwidth}
  \includegraphics[width=\linewidth]{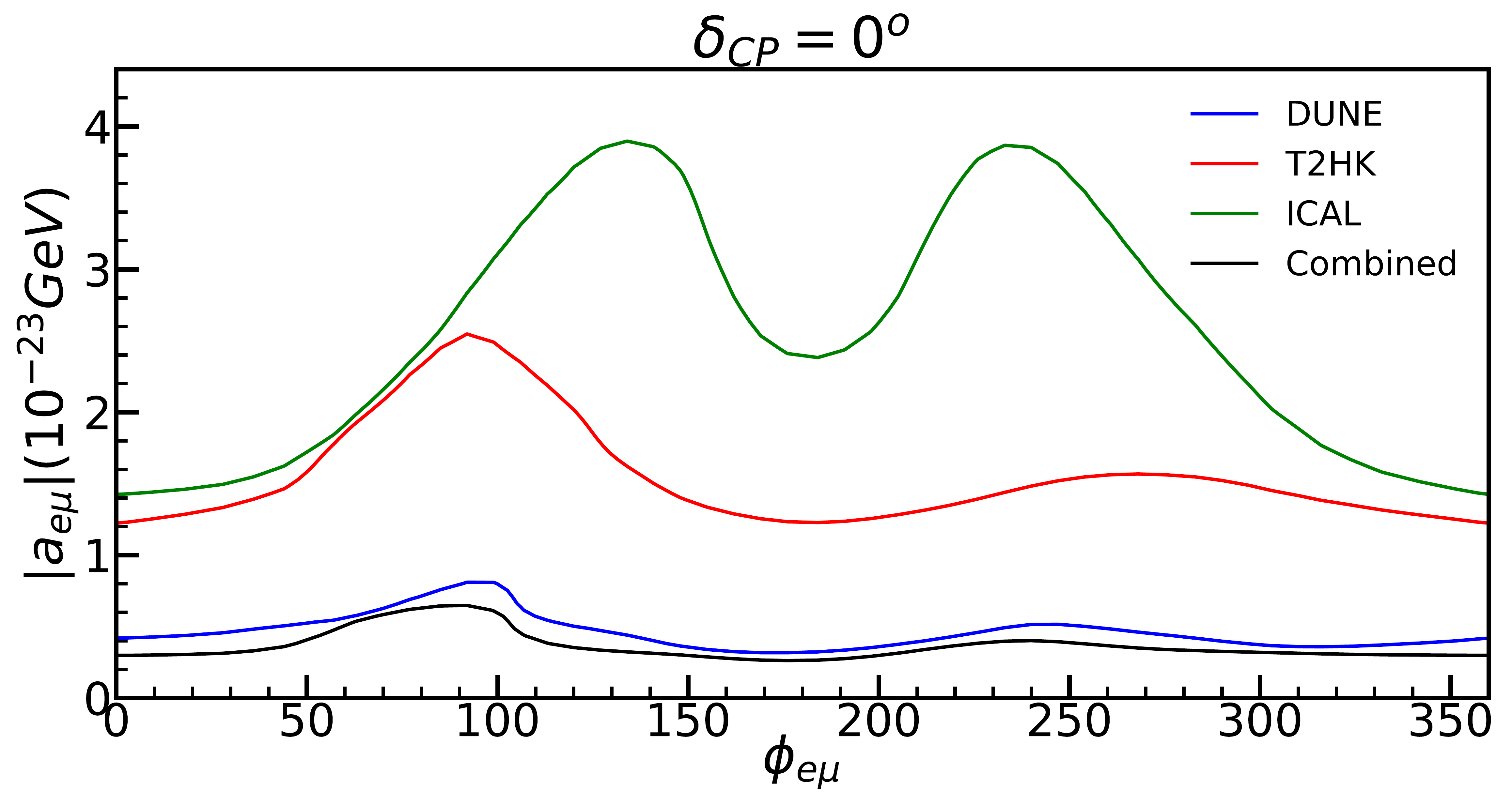}
\end{minipage}%
\hfill 
\begin{minipage}[t]{0.45\textwidth}
  \includegraphics[width=\linewidth]{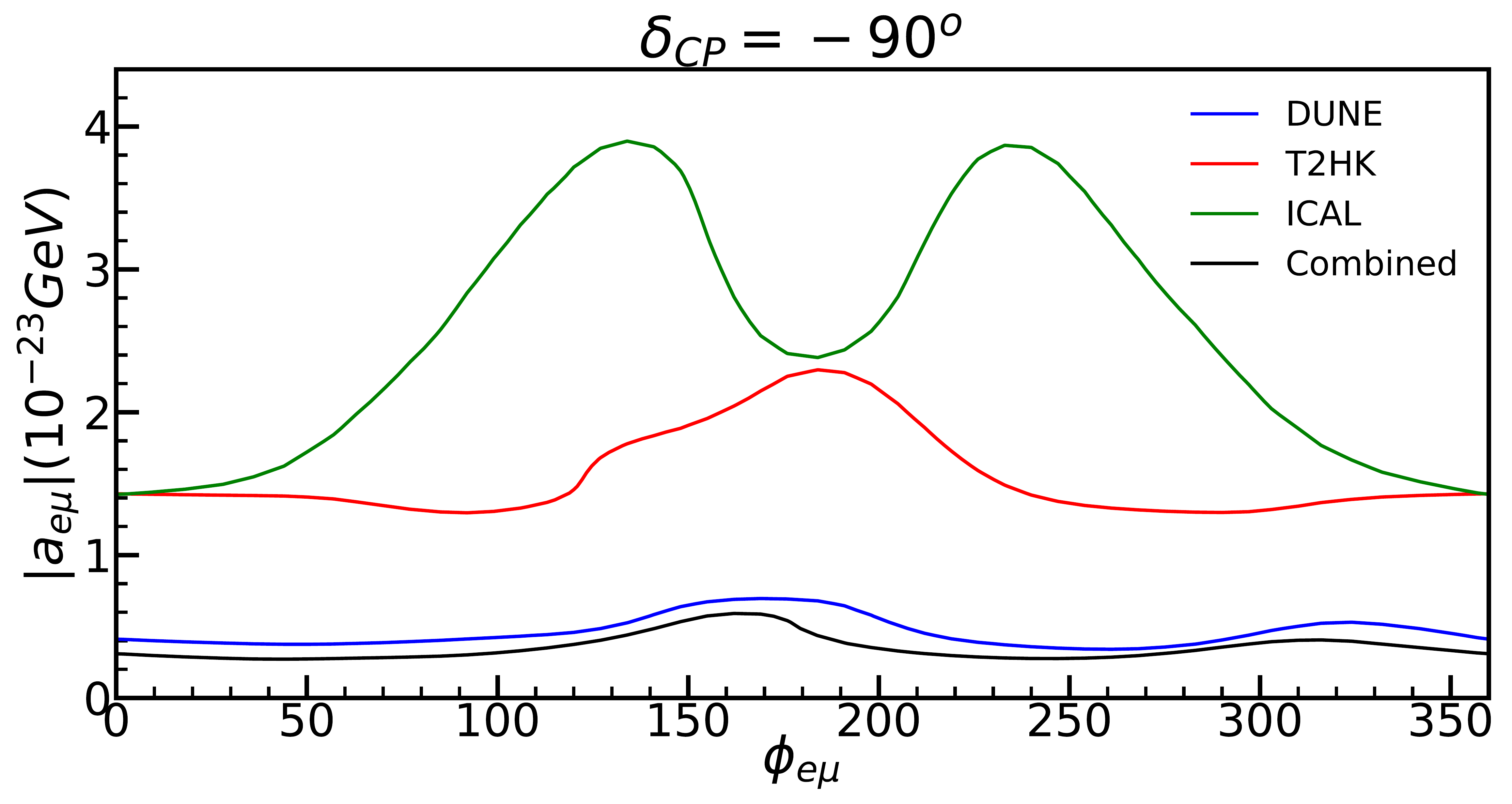}
\end{minipage}%

\begin{minipage}[t]{0.45\textwidth}
  \includegraphics[width=\linewidth]{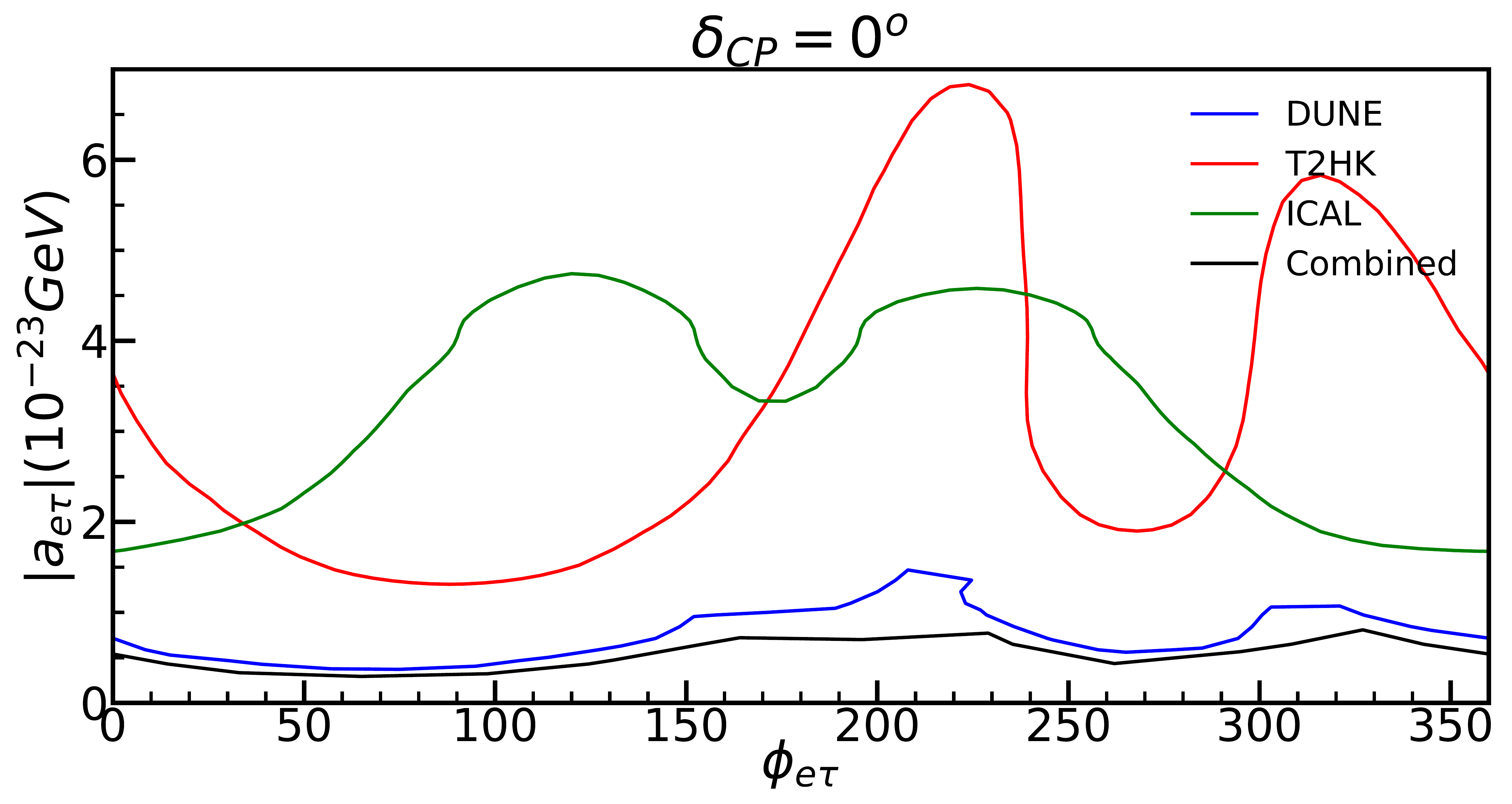}
\end{minipage}%
\hfill 
\begin{minipage}[t]{0.45\textwidth}
  \includegraphics[width=\linewidth]{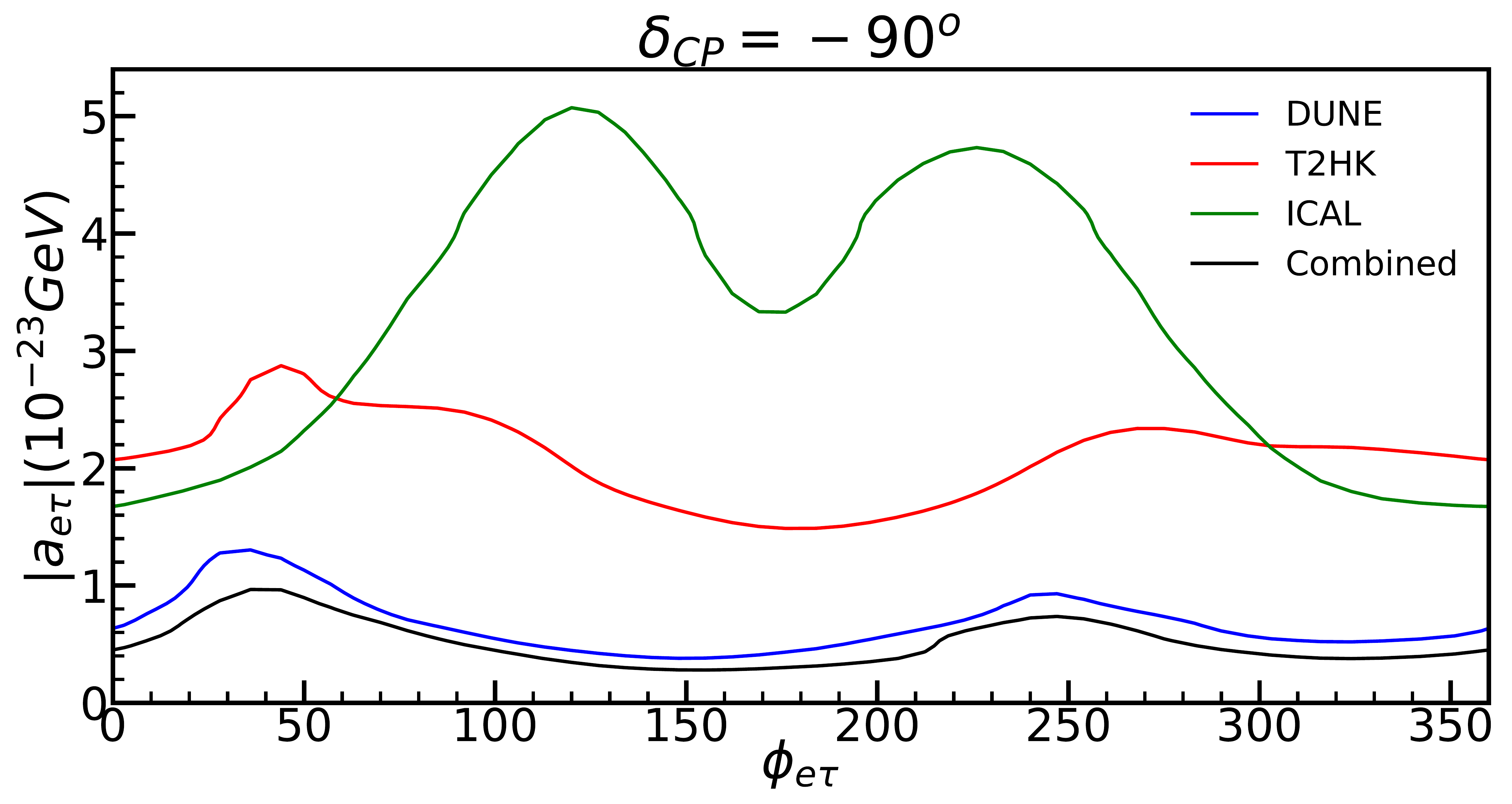}
\end{minipage}%

\begin{minipage}[t]{0.45\textwidth}
  \includegraphics[width=\linewidth]{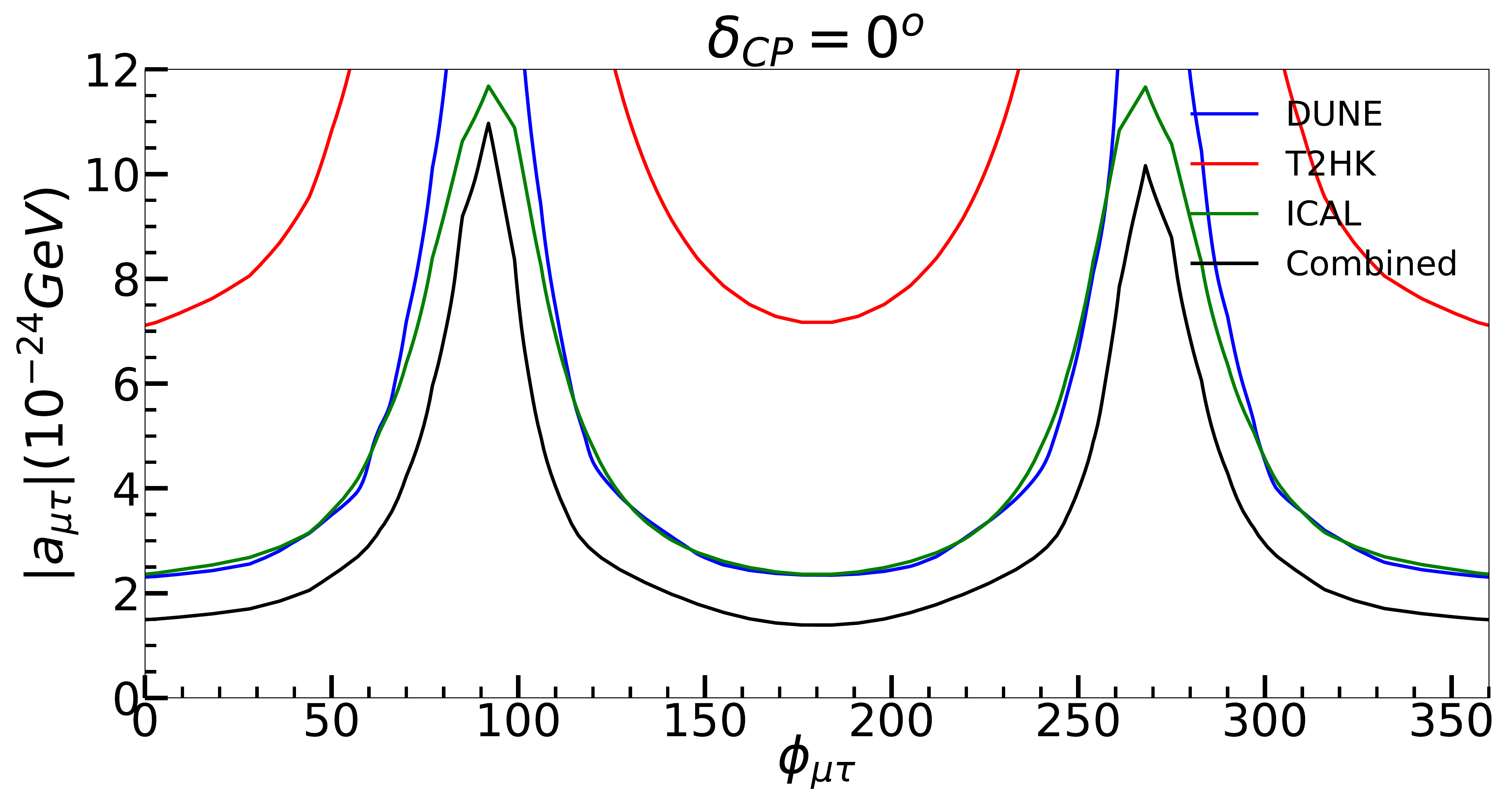}
\end{minipage}%
\hfill 
\begin{minipage}[t]{0.45\textwidth}
  \includegraphics[width=\linewidth]{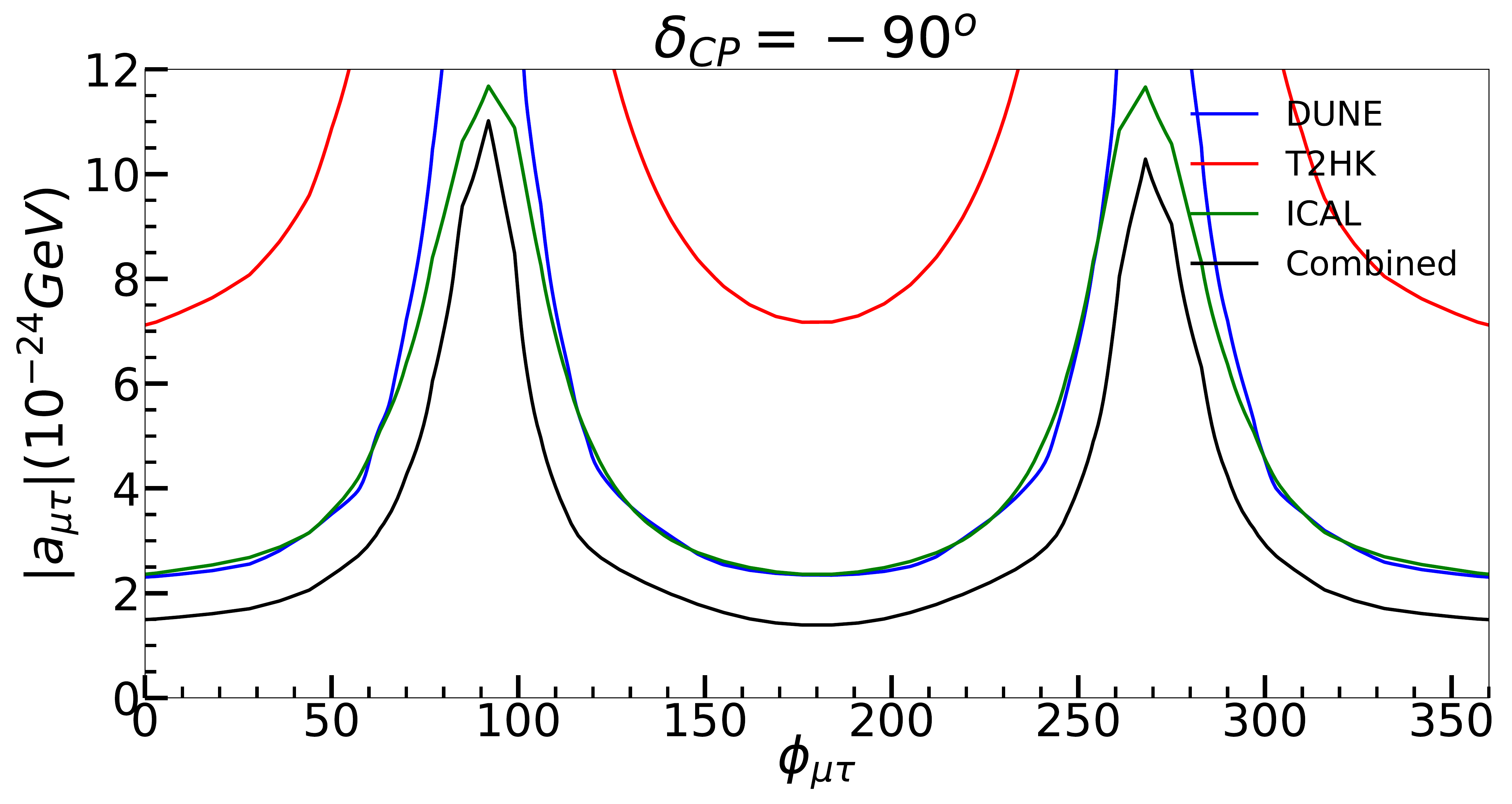}
\end{minipage}%

\caption[]{95$\%$ C.L. (2 dof) contour plots between $|a_{\alpha\beta}|$ and $\phi^a_{\alpha \beta}$. Left column is for $\delta_{\rm CP}=0^\circ$ and right column is for $\delta_{\rm CP}=-90^\circ$.}
\label{fig:all-chi-no-aab}
\end{figure*}

Let us now discuss the sensitivities for the non-diagonal CPT violating LIV parameters. In Fig.~\ref{fig:all-chi-no-aab} we present contour plots at 95$\%$ C.L. (2 dof) in the $|a_{\alpha \beta}|$ vs $\phi^a_{\alpha \beta}$ plane. The left column is for $\delta_{\rm CP} = 0^\circ$ and the right column is for $\delta_{\rm CP} = -90^\circ$. In each column, different panels corresponds to different $a_{\alpha \beta}$ parameters. In each panel, we have presented the individual sensitivities of T2HK, DUNE, ICAL and the combined sensitivity of all these three experiments.

From the discussion on the probability level, it is evident that ICAL has good sensitivity for $a_{\mu\tau}$ while long-baseline experiments are expected to have good sensitivity for parameters $a_{e\mu}$ and $a_{e\tau}$. Furthermore, it has been observed that between the two long-baseline experiments, T2HK has weaker sensitivity than DUNE. This is consistent with the features seen in Fig.\ref{fig:all-chi-no-aab}. For $a_{\mu\tau}$, ICAL has the best sensitivity, while DUNE is also has reasonably sensitive. The upper bound for this parameter is obtained at $\phi^a_{\mu\tau} = 90^\circ$ for all the three experiments and both values of $\delta_{\rm CP}$. For $a_{e\mu}$, T2HK has weaker sensitivity than DUNE but better than ICAL. In ICAL, the upper bounds for $a_{e\mu}$ and $a_{e\tau}$ correspond to $\phi^a_{e\mu}/\phi^a_{e \tau}=120^\circ$. As long-baseline experiments and ICAL depend differently on $\phi^a_{\alpha \beta}$ for $a_{e\mu}$ and $a_{e\tau}$, combining all three experiments is expected to provide the best possible sensitivity. Table \ref{table:para-all-1} lists the 95$\%$ C.L.(1 dof) sensitivity limit for these parameters for $\delta_{\rm CP}=0^{\circ}$ and $-90^{\circ}$. To obtain the limits quoted in this table, we have minimised the $\chi^2$ over $\phi^a_{\alpha \beta}$.

\begin{table}[h]
 \begin{center}
 \scalebox{0.7}{
\begin{tabular}{ |c|c|c|c|c| } 
\hline
LIV parameters  & ICAL & DUNE & T2HK & Combined \\
\hline
$a_{ee}$&   -6/4.5 & -2.84/2.56(-2.7/2.5)& -26/21.6(-8.5/8.8) & -1.95/2.0(-1.9/2.0)\\ 
$a_{\mu\mu}$  & -2.2/2.2 & -2.9/2.5(-2.7/2.4) & -12/13(-12/13.4) &-1.3/1.3(-1.3/1.4)\\
$a_{\tau\tau}$  & -2.3/2.2 &-2.9/2.5(-2.7/2.4) & -7.7/8.0(-8.4/7.5) &-1.5/1.5(-1.4/1.4)\\
$a_{e\mu}$ & 3.3 & 0.64(0.56) & 2.2(2.0)& 0.45(0.4) \\ 
$a_{e\tau}$ & 4.4 & 1.1(0.9)& 5.8(2.3)& 0.6(0.63)\\ 
$a_{\mu\tau}$ & 1.0 & 1.95(2.0) & 6.8(6.9) & 0.95(0.96) \\ 
\hline
\end{tabular}}
\caption{95$\%$ C.L. (1 dof) bounds for ICAL, DUNE, T2HK and their combination in the units of $10^{-23}$ GeV. For $a_{\alpha\alpha}$, we have given two values. One corresponds to $+$ve values of the parameters and the other for $-$ve values. For DUNE, T2HK and Combined, we have given two different limits, one is for true $\delta_{\rm CP}=0^{\circ}$ (outside parentheses) and other is for true $\delta_{\rm CP}=-90^{\circ}$ (inside parentheses). }
\label{table:para-all-1}
\end{center}
\end{table}

\subsection{Sensitivity for $c_{\alpha\alpha}$ }

\begin{figure*}

\begin{minipage}[t]{0.45\textwidth}
  \includegraphics[width=\linewidth]{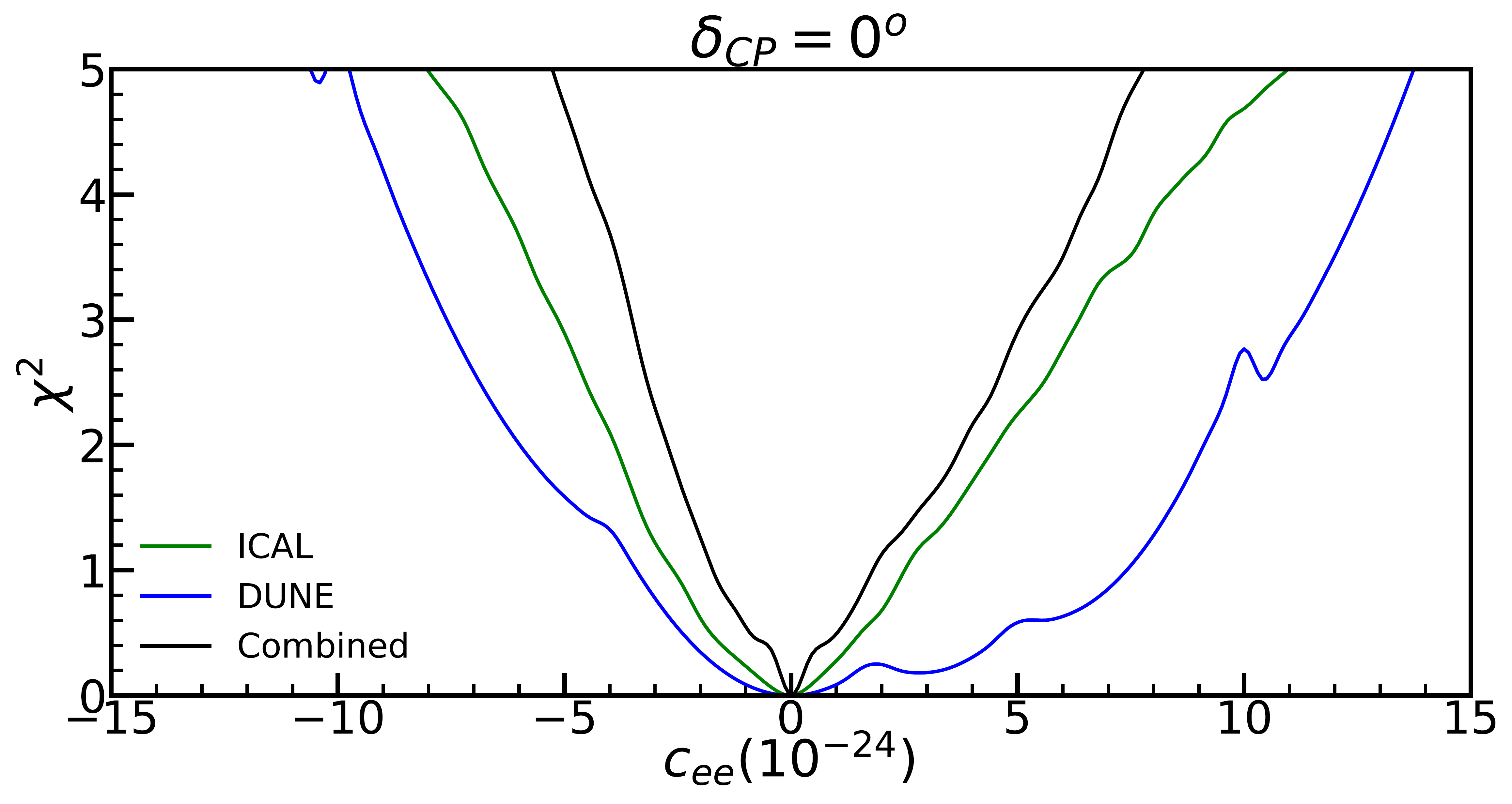}
\end{minipage}%
\hfill 
\begin{minipage}[t]{0.45\textwidth}
  \includegraphics[width=\linewidth]{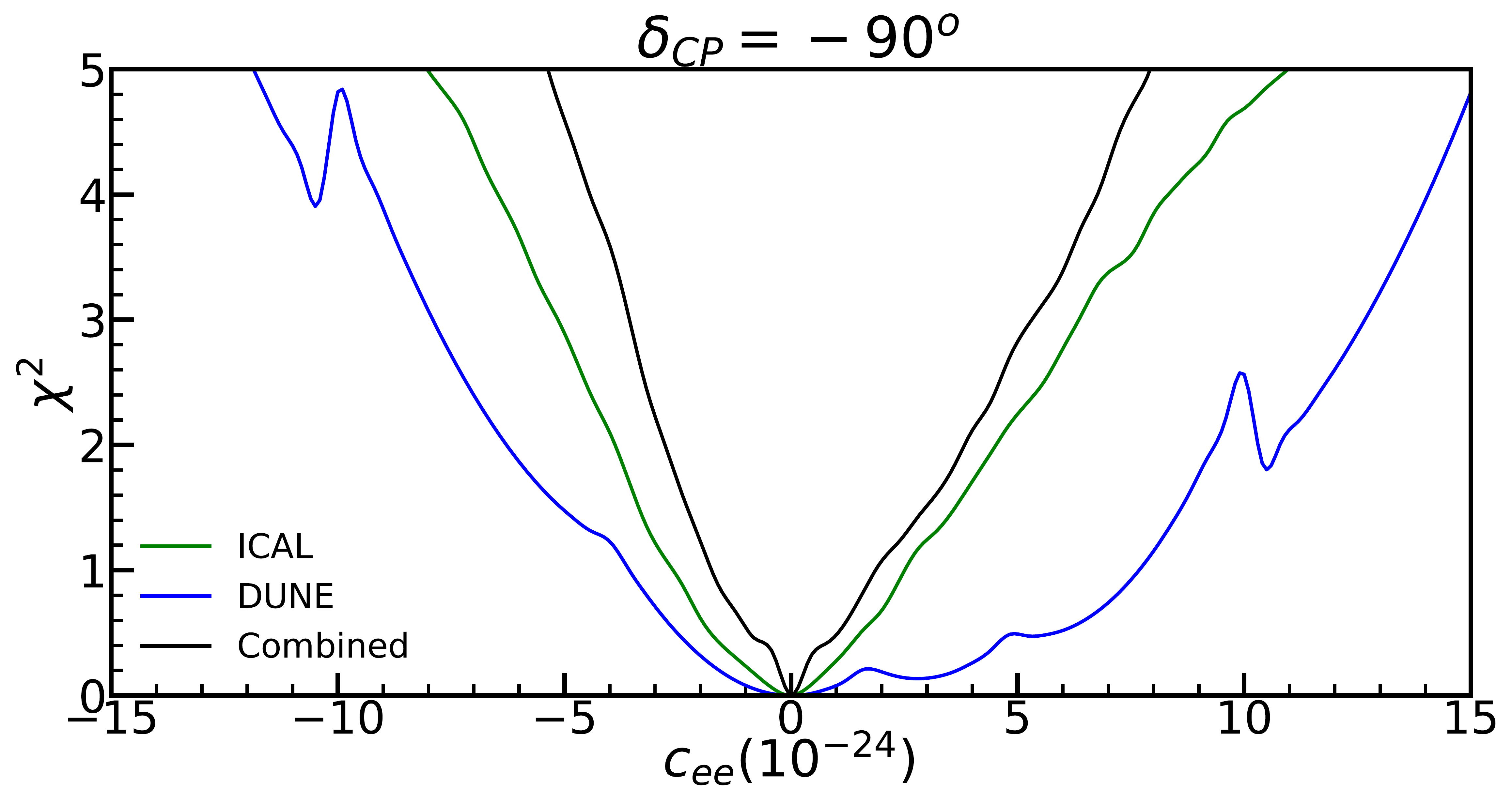}
\end{minipage}%

\begin{minipage}[t]{0.45\textwidth}
  \includegraphics[width=\linewidth]{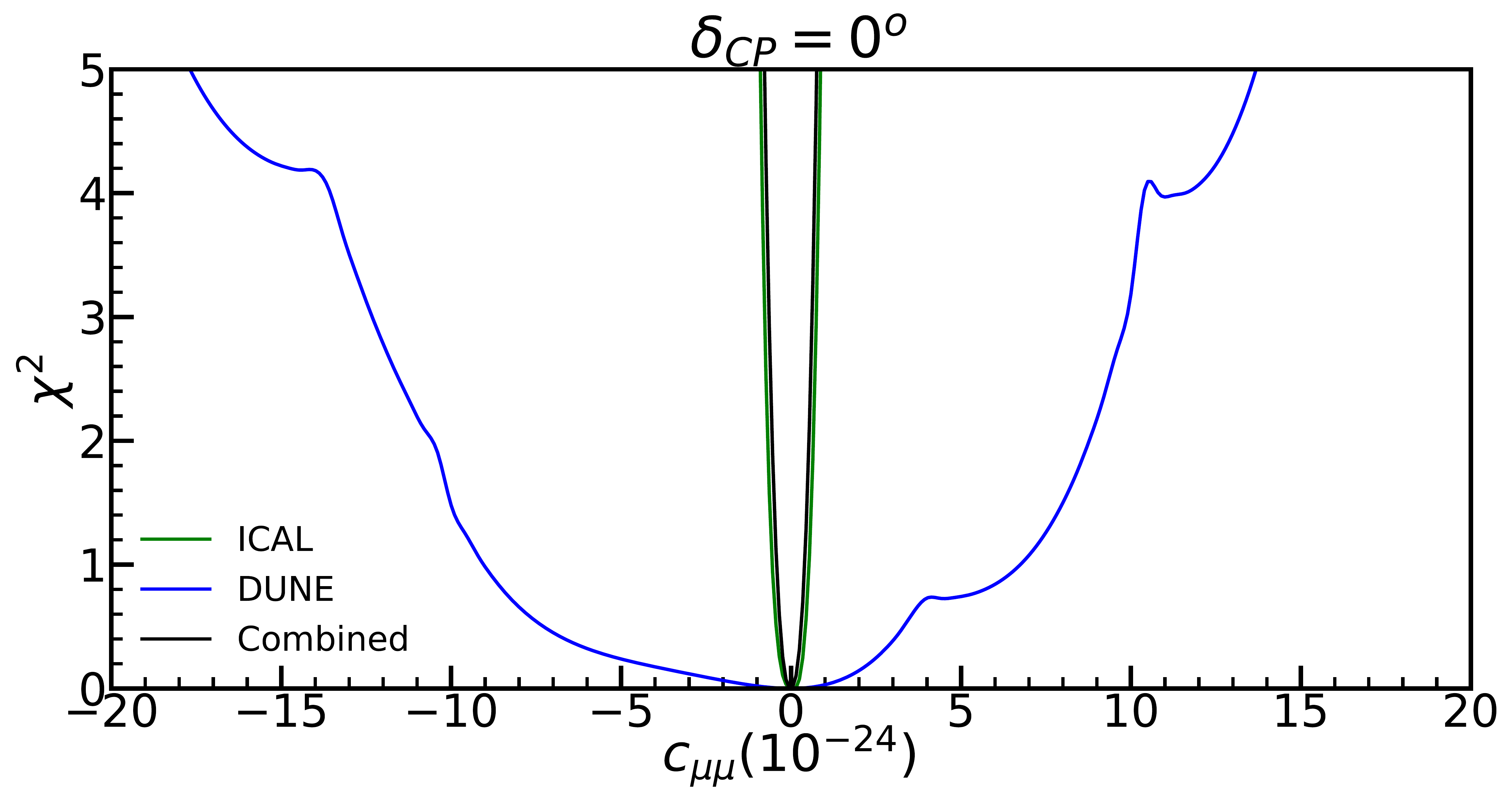}
\end{minipage}%
\hfill 
\begin{minipage}[t]{0.45\textwidth}
  \includegraphics[width=\linewidth]{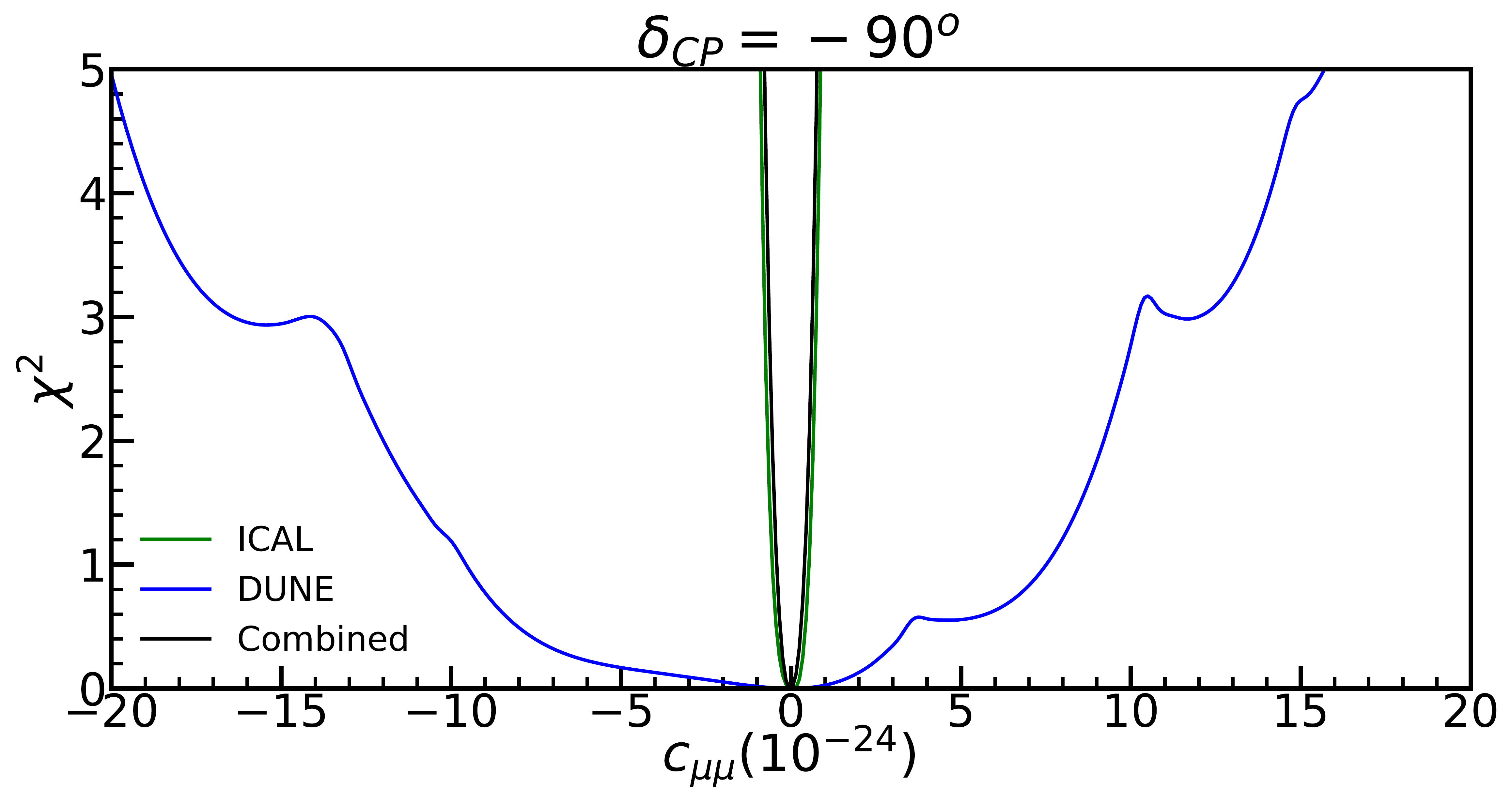}
\end{minipage}%

\begin{minipage}[t]{0.45\textwidth}
  \includegraphics[width=\linewidth]{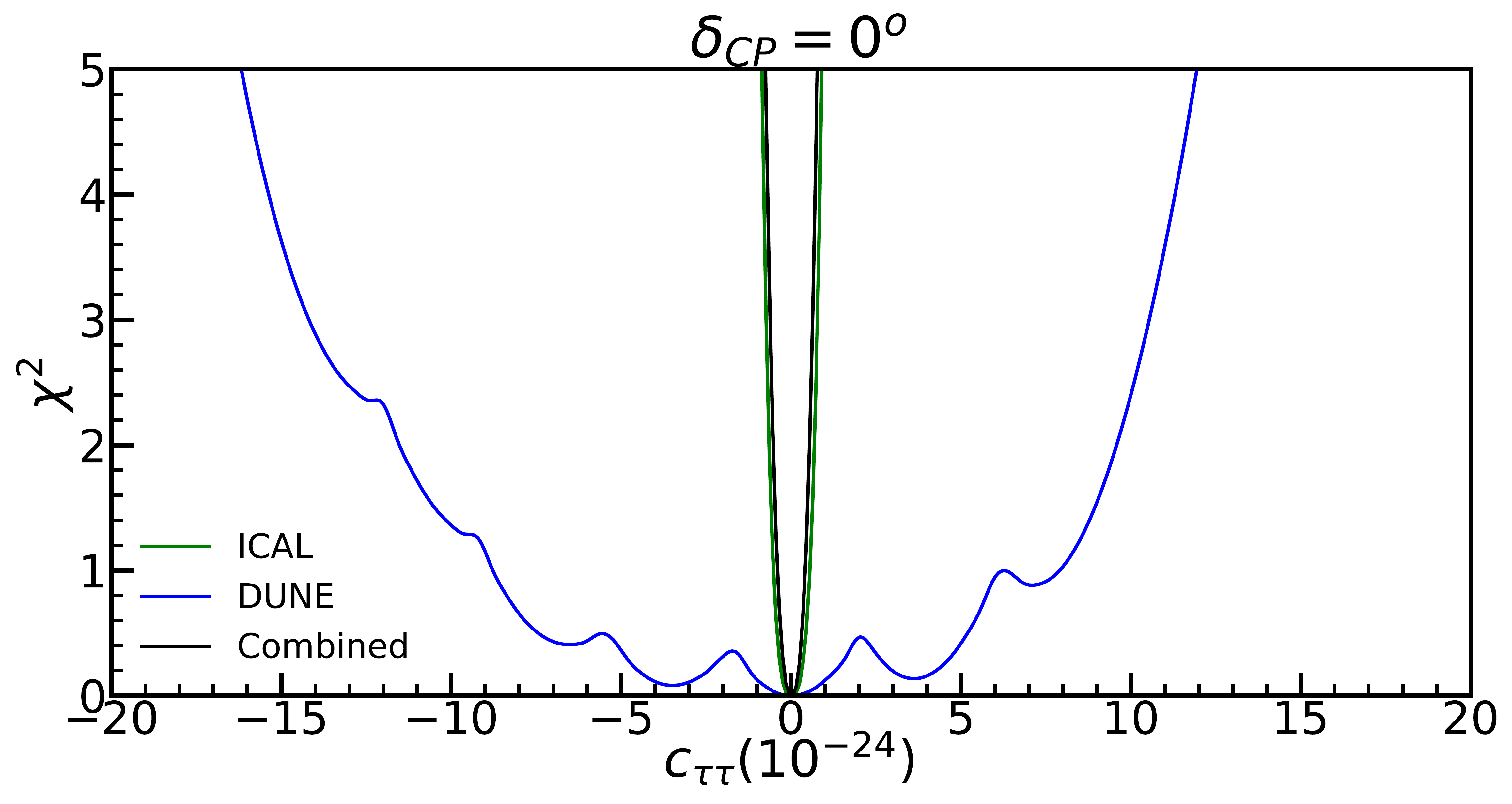}
\end{minipage}%
\hfill 
\begin{minipage}[t]{0.45\textwidth}
  \includegraphics[width=\linewidth]{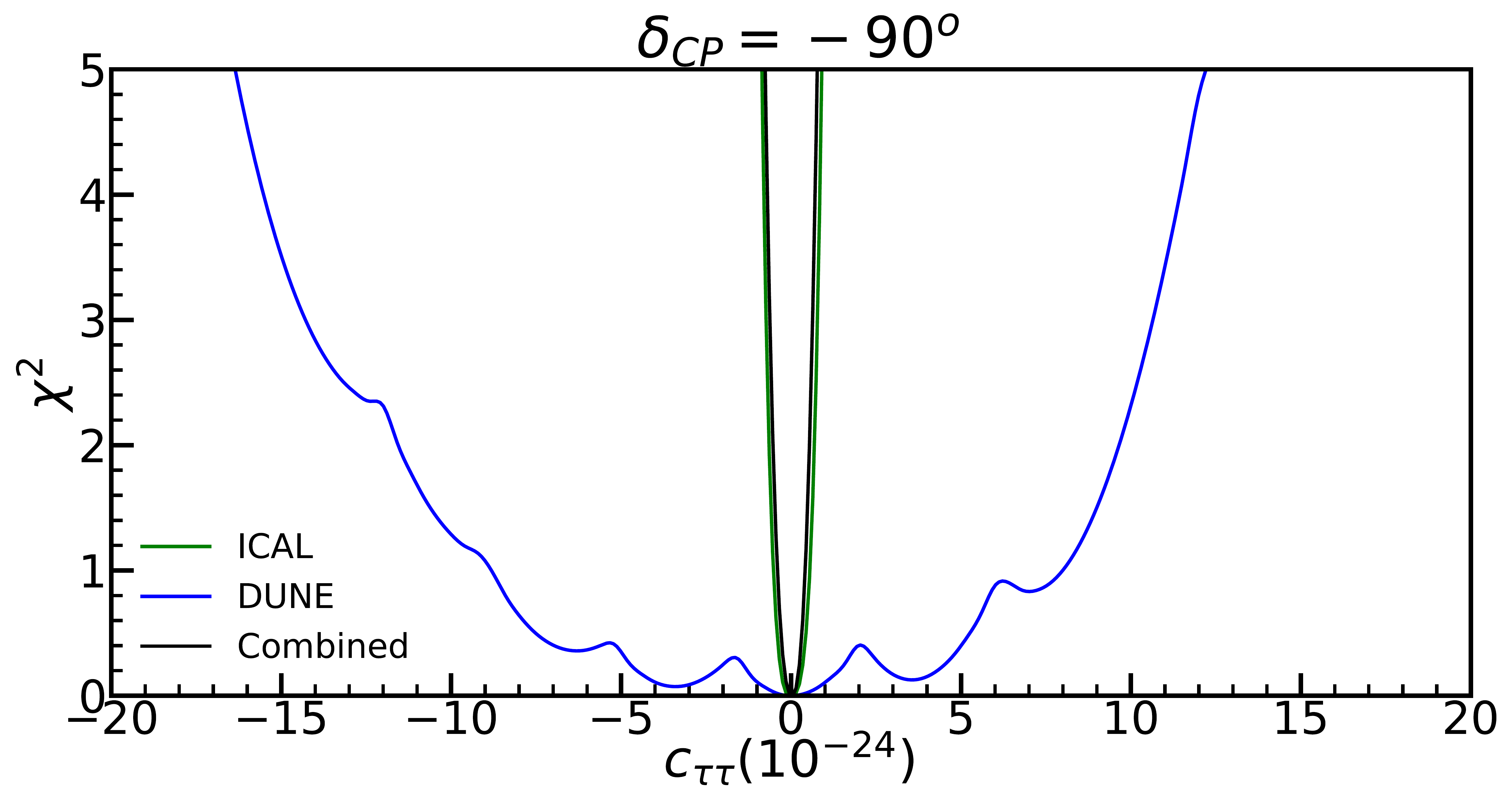}
\end{minipage}%

\caption{$\chi^{2}$ as a function of LIV parameters $c_{ee}$, $c_{\mu\mu}$ and $c_{\tau\tau}$ for true $\delta_{\rm CP}$ = $0^{\circ}$ (left column) and $-90^{\circ}$ (right column) in ICAL, DUNE and combined.}
\label{fig:t2hk-chi-no_c}
\end{figure*}

In this subsection, we will discuss the sensitivity of the different experiments to the diagonal CPT conserving LIV parameters. Fig. \ref{fig:t2hk-chi-no_c} presents the same information as Fig. \ref{fig:all-chi-no_aaa} but for $c_{\alpha \alpha}$. We have not provided any curves for T2HK in these panels, as its sensitivity to CPT conserving LIV parameters is very weak.

The probability discussion had shown that ICAL has better sensitivity to diagonal CPT-conserving LIV parameters than DUNE, except for $c_{ee}$. We had also seen that for long-baseline experiments the behavior of $c_{\alpha \alpha}$ with respect to the appearance channel probabilities is similar to that of $a_{\alpha \alpha}$, but the changes in the probability due to $c_{\alpha \alpha}$ are somewhat lower than those due to $a_{\alpha \alpha}$. From Fig. \ref{fig:t2hk-chi-no_c} DUNE gives the best sensitivity for $c_{ee}$, while ICAL is better for the two. Combining the  two experiments improves the sensitivity for $c_{ee}$, but for the other two parameters adding DUNE provides only a slight improvement. Sensitivities for all three parameters are similar for both values of $\delta_{\rm CP}$. Table \ref{table:para-all-2} lists the 95$\%$ C.L. (1 dof) sensitivity limit for these parameters for both choices of $\delta_{\rm CP}=0^\circ$ and $-90^\circ$.

\subsection{Sensitivity for $c_{\alpha\beta}$}

\begin{figure*}

\begin{minipage}[t]{0.45\textwidth}
  \includegraphics[width=\linewidth]{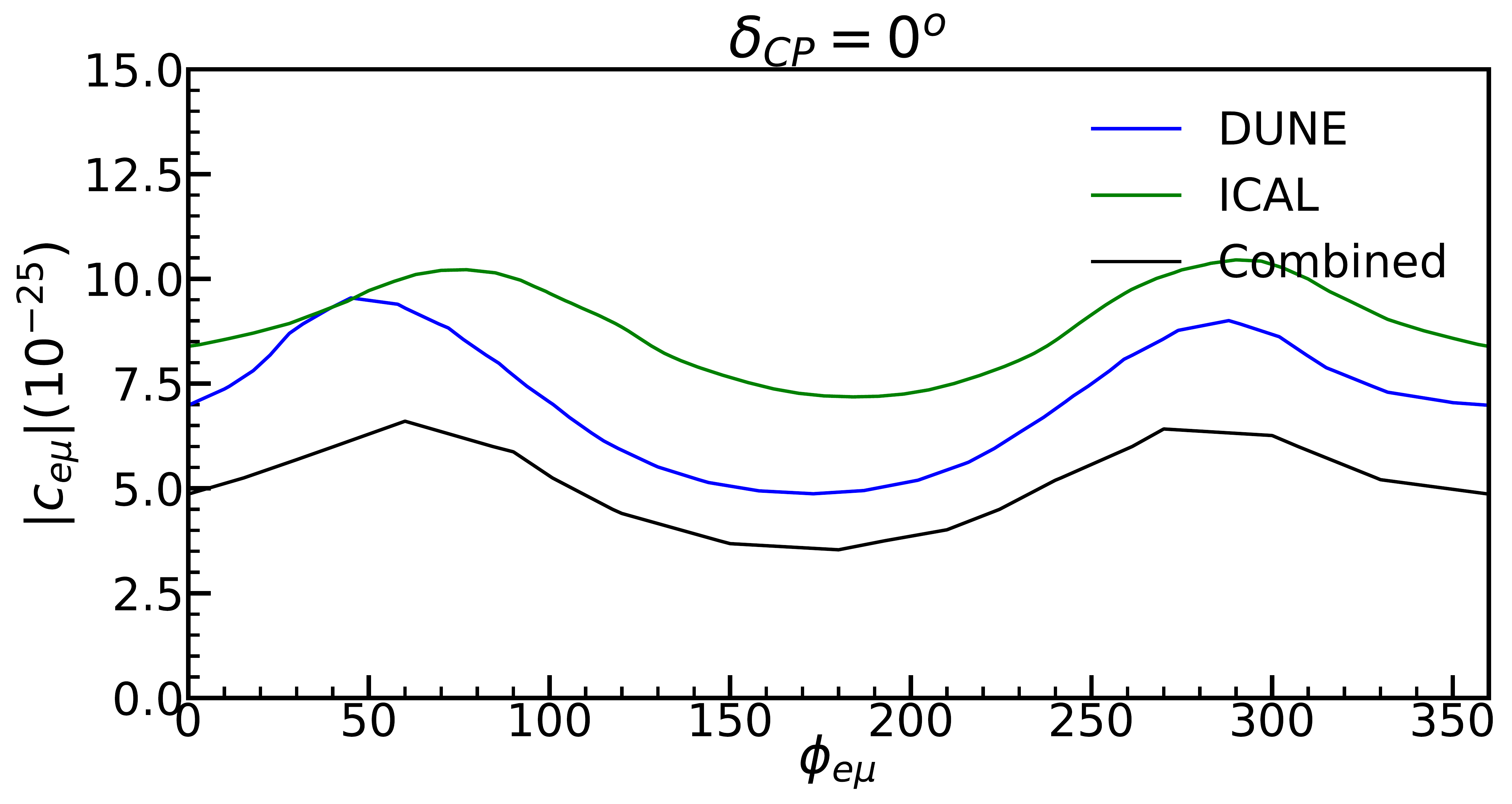}
\end{minipage}%
\hfill 
\begin{minipage}[t]{0.45\textwidth}
  \includegraphics[width=\linewidth]{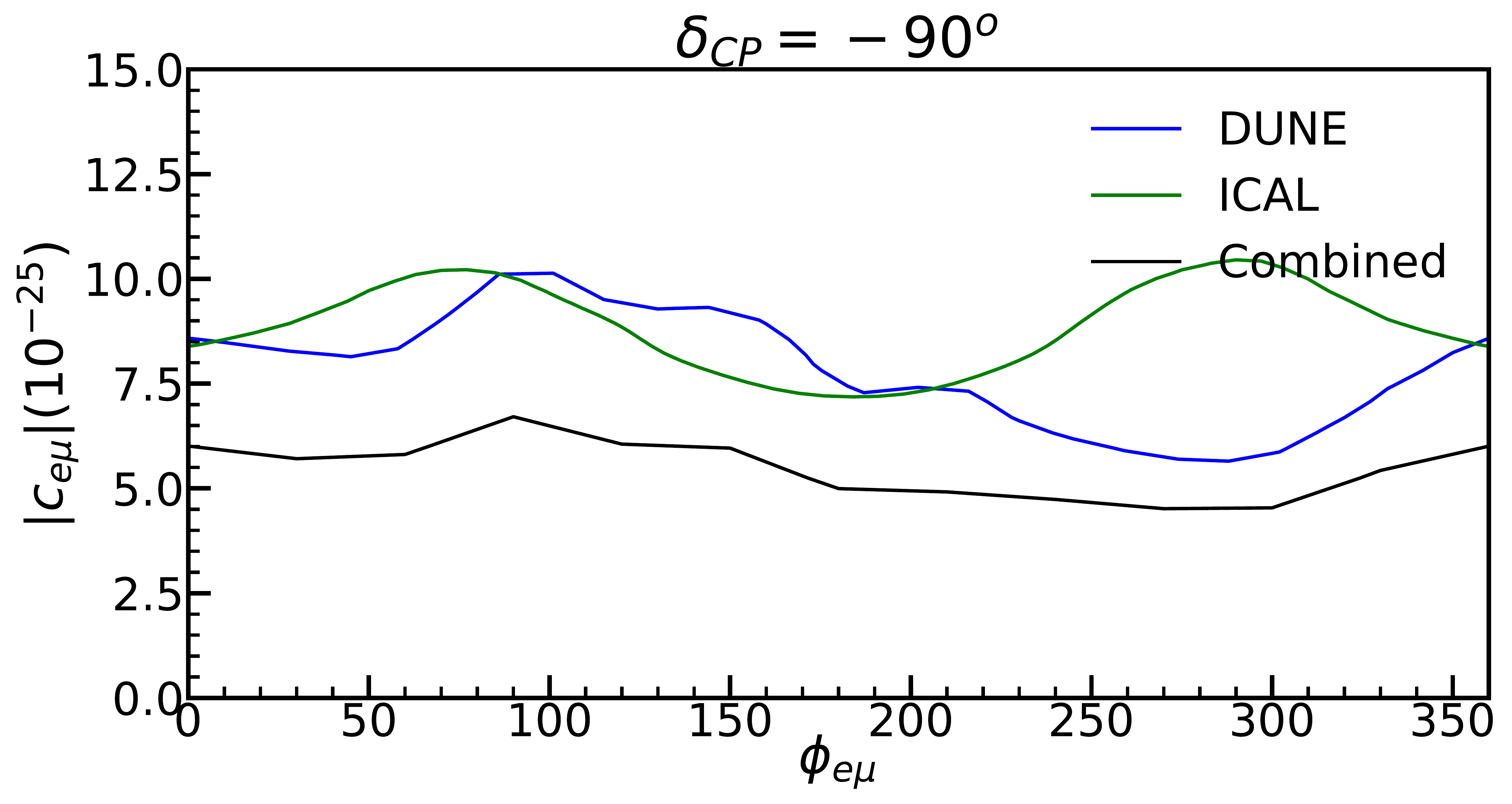}
\end{minipage}%

\begin{minipage}[t]{0.45\textwidth}
  \includegraphics[width=\linewidth]{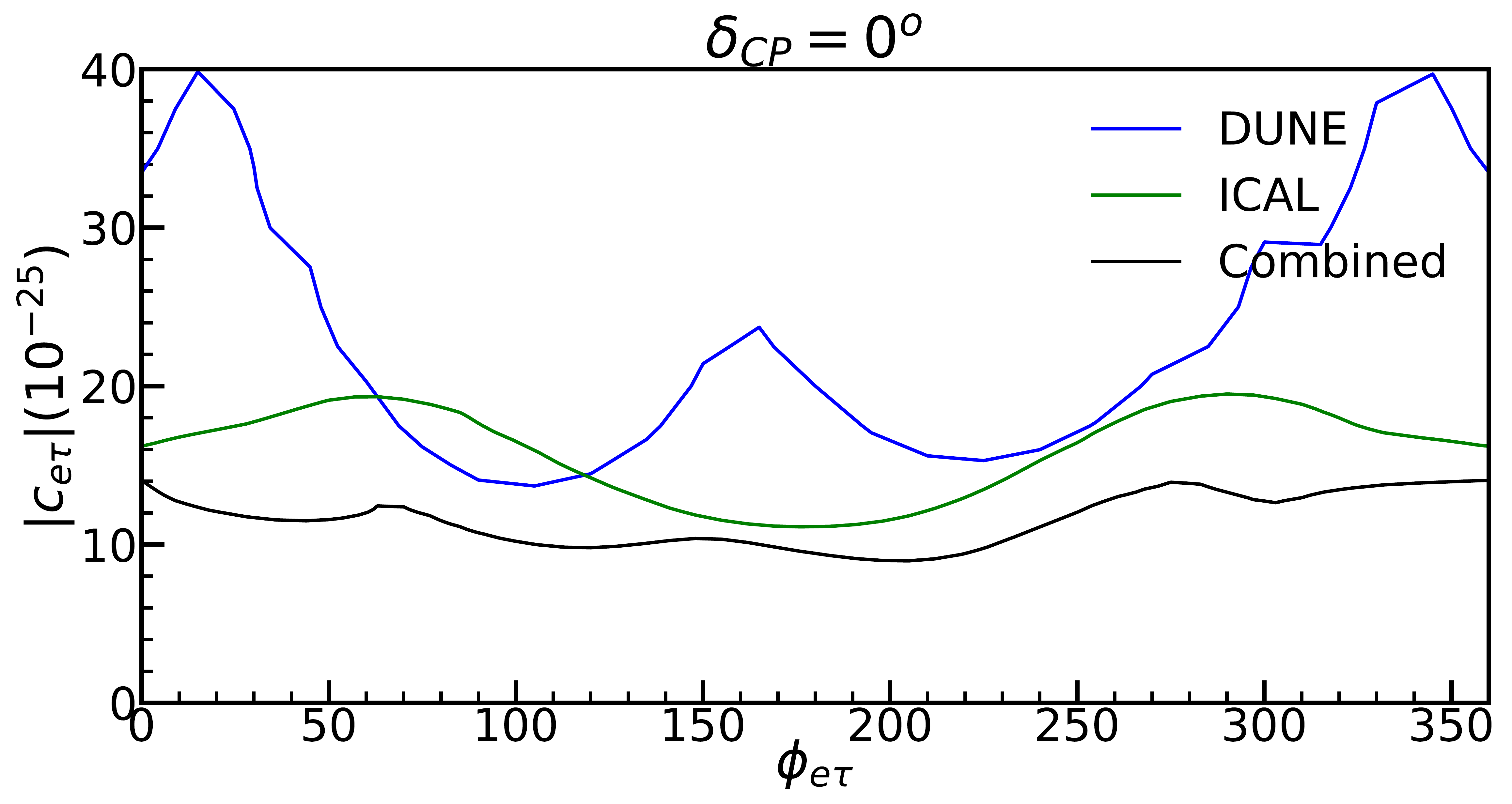}
\end{minipage}%
\hfill 
\begin{minipage}[t]{0.45\textwidth}
  \includegraphics[width=\linewidth]{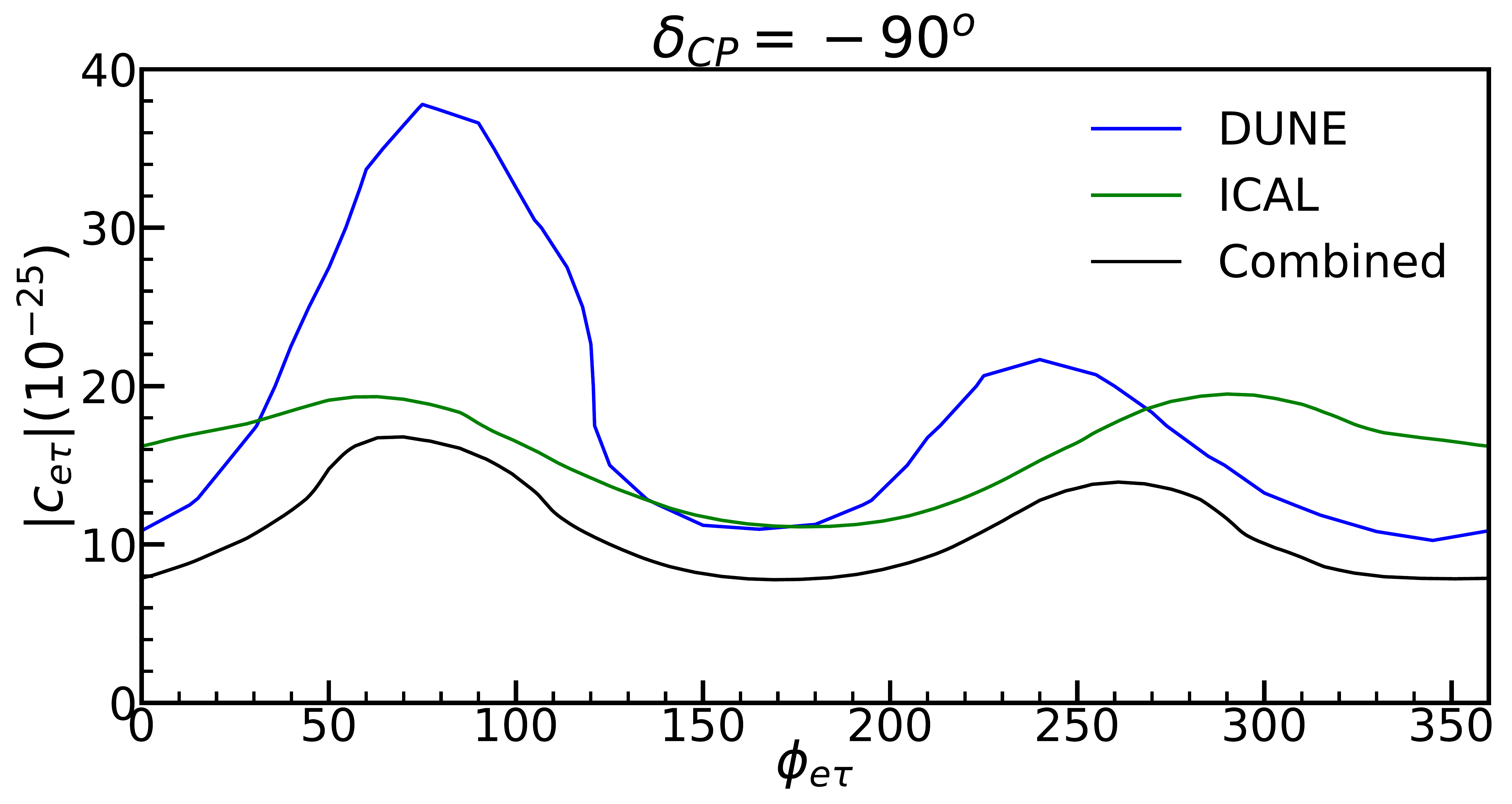}
\end{minipage}%

\begin{minipage}[t]{0.45\textwidth}
  \includegraphics[width=\linewidth]{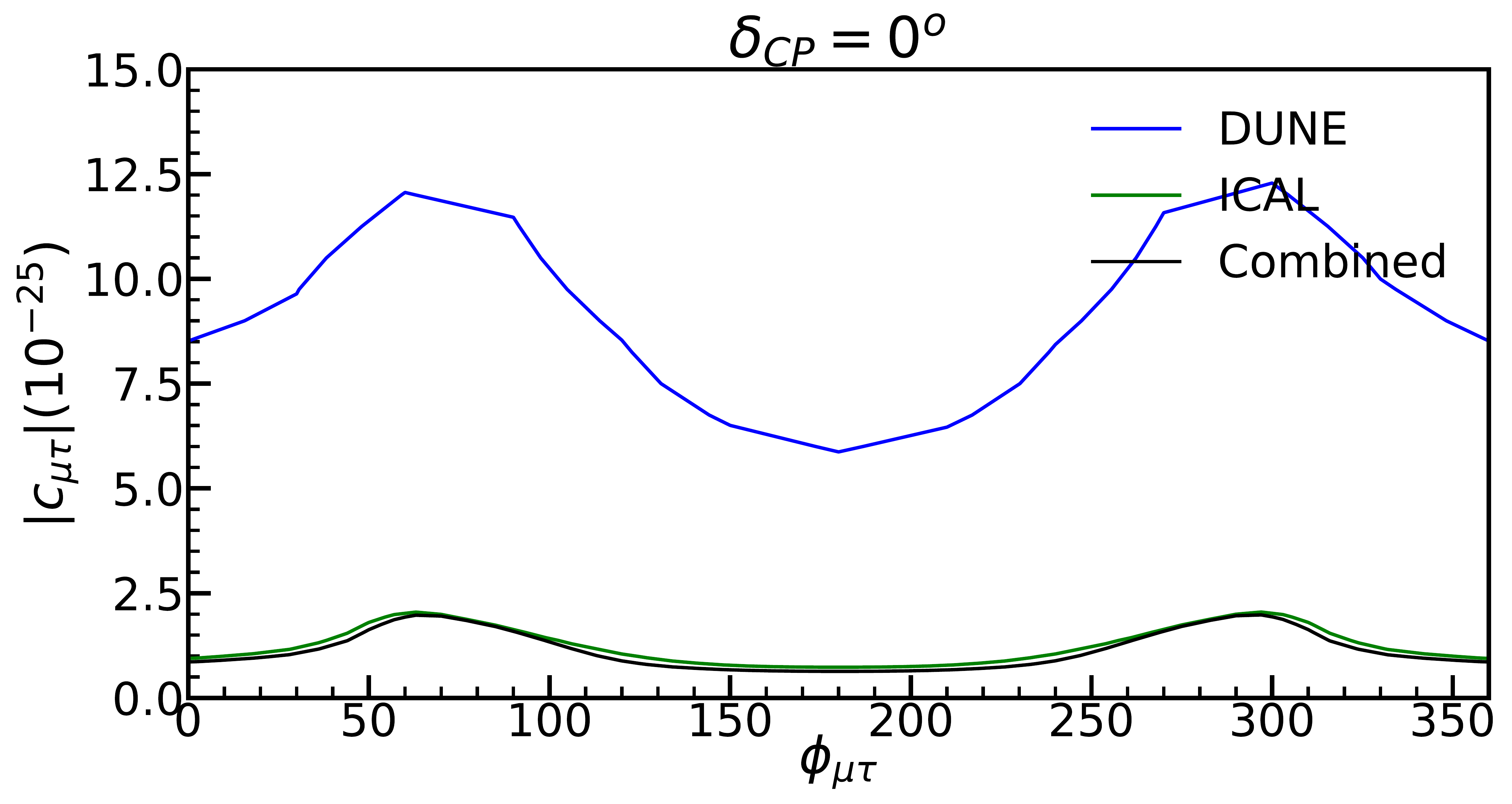}
\end{minipage}%
\hfill 
\begin{minipage}[t]{0.45\textwidth}
  \includegraphics[width=\linewidth]{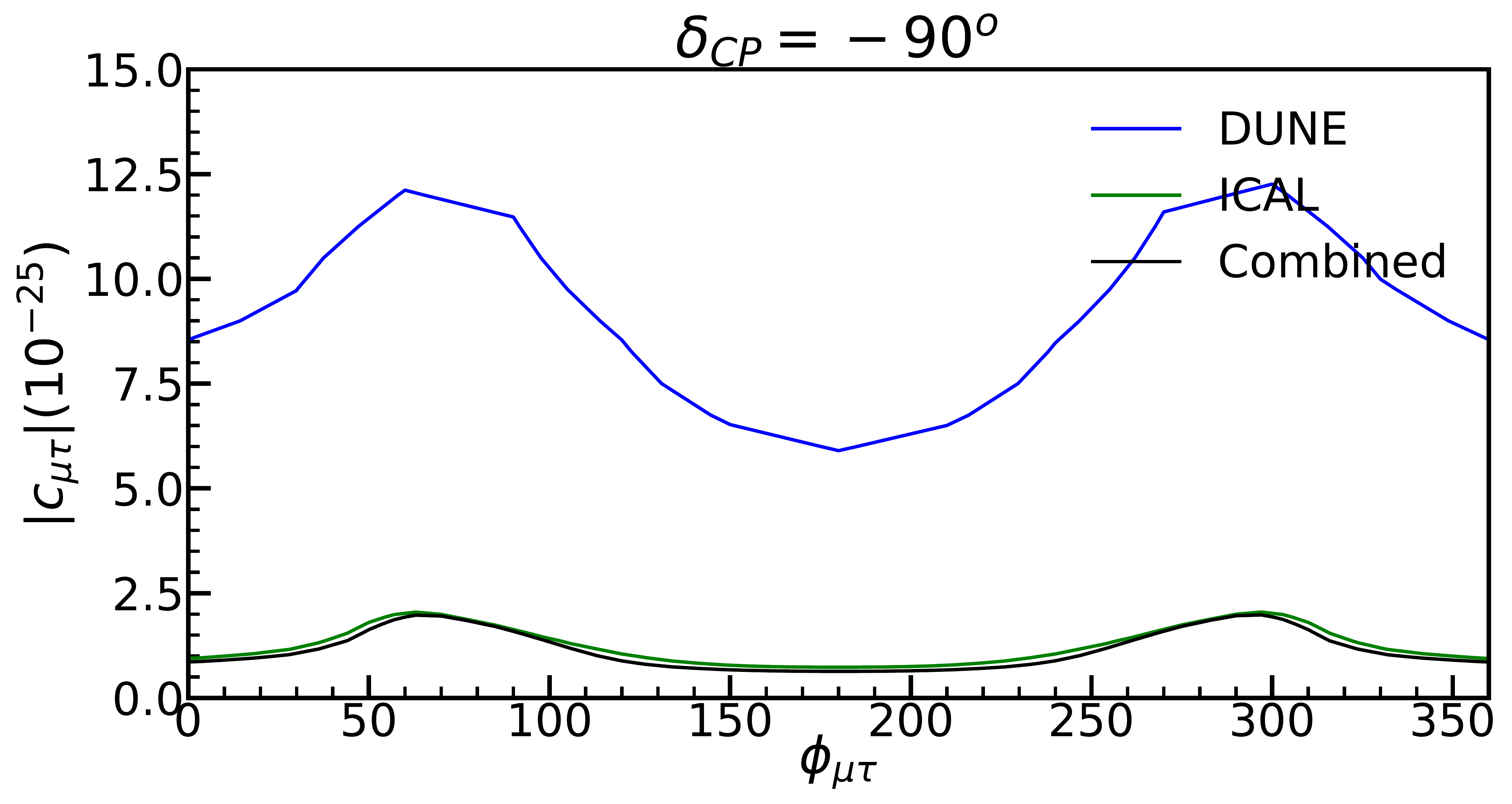}
\end{minipage}%

\caption{95$\%$ C.L. (2 dof) contour plots between $|c_{\alpha\beta}|$ and $\phi^c_{\alpha \beta}$. Left column is for $\delta_{\rm CP}=0^{\circ}$ and right column is for $\delta_{\rm CP}=-90^{\circ}$.}
\label{fig:ino-cab}
\end{figure*}

In this subsection, we will discuss the sensitivity for the off-diagonal CPT conserving LIV parameters. Fig.~\ref{fig:ino-cab} is the same as Fig.~\ref{fig:all-chi-no-aab} but for $c_{\alpha \beta}$ (when $\alpha \neq \beta$). Again we do not present the curves for T2HK. 

For these set of parameters, we should get sizable amount of sensitivity from ICAL. Regarding the long-baseline experiments, as understood from the probabilities, the behavior of $c_{\alpha \beta}$ is similar as that of $a_{\alpha \beta}$ with a lower sensitivity. For $c_{e\mu}$, DUNE gives better sensitivity than ICAL except $\delta_{\rm CP} = -90^\circ$ and $\phi^c_{e\mu} = 180^\circ$. For $c_{e\tau}$, the sensitivity of DUNE is better than ICAL in some of the regions depending on the values of $\delta_{\rm CP}$ and $\phi^c_{e\tau}$. For DUNE, the upper bound corresponds to $\phi^c_{e\mu} = 50^\circ$ for $\delta_{\rm CP} = 0^\circ$ and $\phi^c_{e\mu} = 90^\circ$ for $\delta_{\rm CP} = -90^\circ$ in $c_{e\mu}$. For $c_{e\tau}$, the upper bound comes at $\phi^c_{e\tau} = 10^\circ$ for $\delta_{\rm CP} = 0^\circ$ and $\phi^c_{e\tau} = 90^\circ$ for $\delta_{\rm CP} = -90^\circ$ in DUNE. For ICAL, the upper bound comes at $\phi^c = 90^\circ$ for both $c_{e\mu}$ and $c_{e\tau}$. For these two parameters, the sensitivity improves when DUNE and ICAL are combined and in this case the upper bound corresponds to around $\phi^c_{e\mu} = 90^\circ$. For $c_{\mu\tau}$, the sensitivity of DUNE is weak as compared to ICAL. Therefore when combined the improvement in the sensitivity is very marginal. In this case the upper bound corresponds to $\phi^c_{\mu\tau} = 90^\circ$. For this parameter, the sensitivity is similar for both the values of $\delta_{\rm CP}$. In Table \ref{table:para-all-2} we have listed the 95$\%$ C.L.(1 dof) sensitivity limit of these parameters for both values of $\delta_{\rm CP}$. To obtain the limits quoted in this table, we have minimised the $\chi^2$ over $\phi^c_{\alpha \beta}$.

\begin{table}[h]
 \begin{center}
 \scalebox{0.8}{
\begin{tabular}{ |c|c|c|c| } 
\hline
LIV parameters  & ICAL & DUNE & Combined \\
\hline
$c_{ee}$&   -62/80  &-86/124(-90/140) &-41/80(-43/65)\\ 
$c_{\mu\mu}$  & -9.4/7.3 &-136/103(187/140) & -7.8/6(-7.5/7.6)\\ 
$c_{\tau\tau}$  & -8/8.5  & -155/112(-153/113) & -6.1/6.5(-6.9/6.3)\\
$c_{e\mu}$ & 7.8 & 8.2(8.9)& 5.5(5.2)\\ 
$c_{e\tau}$ & 16.8 & 34(32.02)& 11(13.8)\\ 
$c_{\mu\tau}$ & 1.6 & 11.0(11.0)&1.6(1.6)\\ 
\hline
\end{tabular}}
\caption{95$\%$ C.L.(1 dof) bounds for ICAL, DUNE and their combination in the units of $10^{-25}$. For $c_{\alpha\alpha}$ parameters, we have given two values. One is corresponding to $+$ve value and another is for $-$ve value. For DUNE, T2HK and their combination, we have given two different limits, one is for $\delta_{\rm CP}=0^{\circ}$(outside parenthesis) and other is for $\delta_{\rm CP}=-90^{\circ}$(inside parenthesis).}
\label{table:para-all-2}
\end{center}
\end{table}

\section{Comparison with previous results}
\label{comp}

\begin{figure*}
 \begin{minipage}[t]{0.45\textwidth}
  \includegraphics[width=\linewidth]{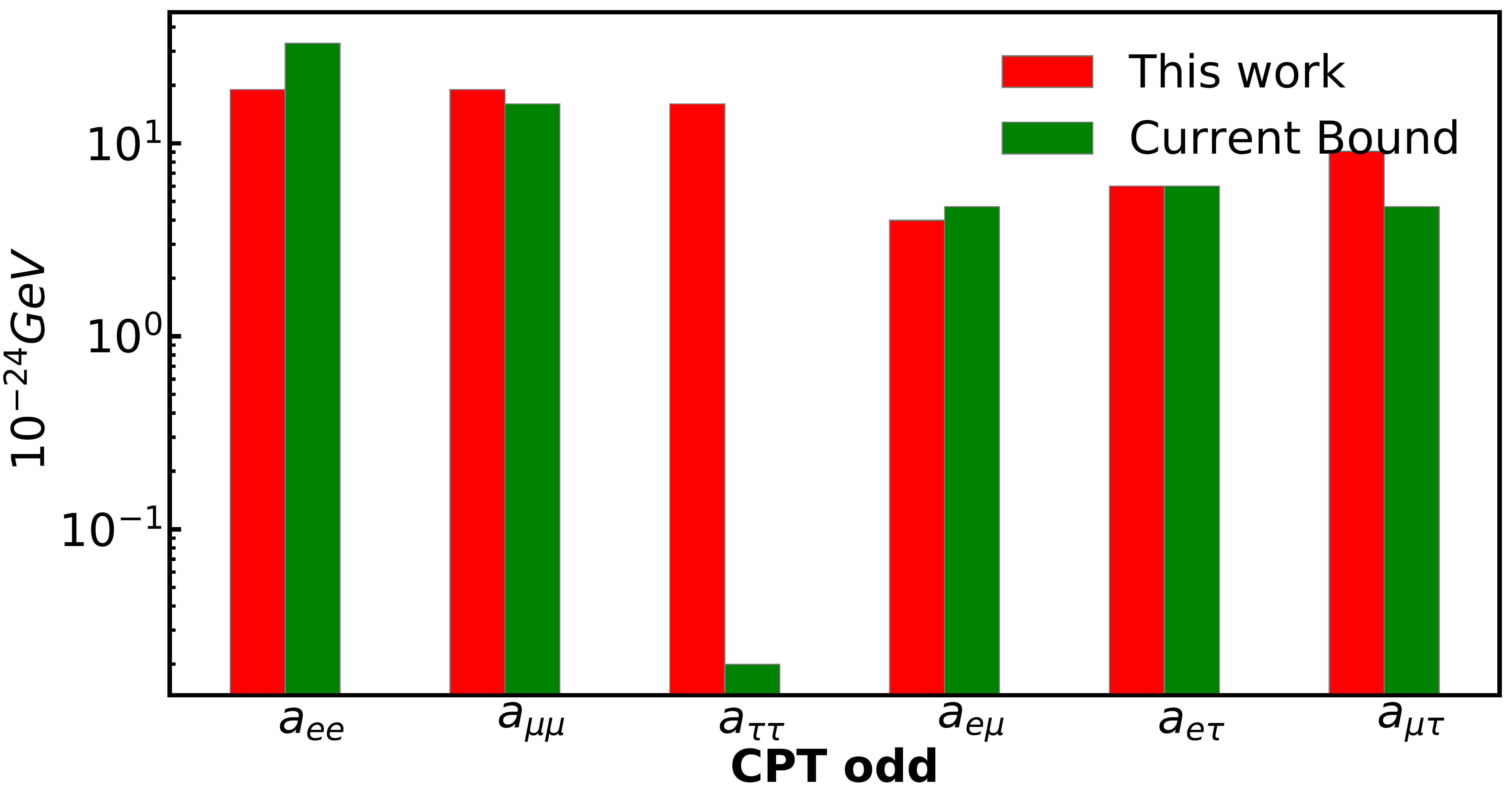}
\end{minipage}%
\begin{minipage}[t]{0.45\textwidth}
  \includegraphics[width=\linewidth]{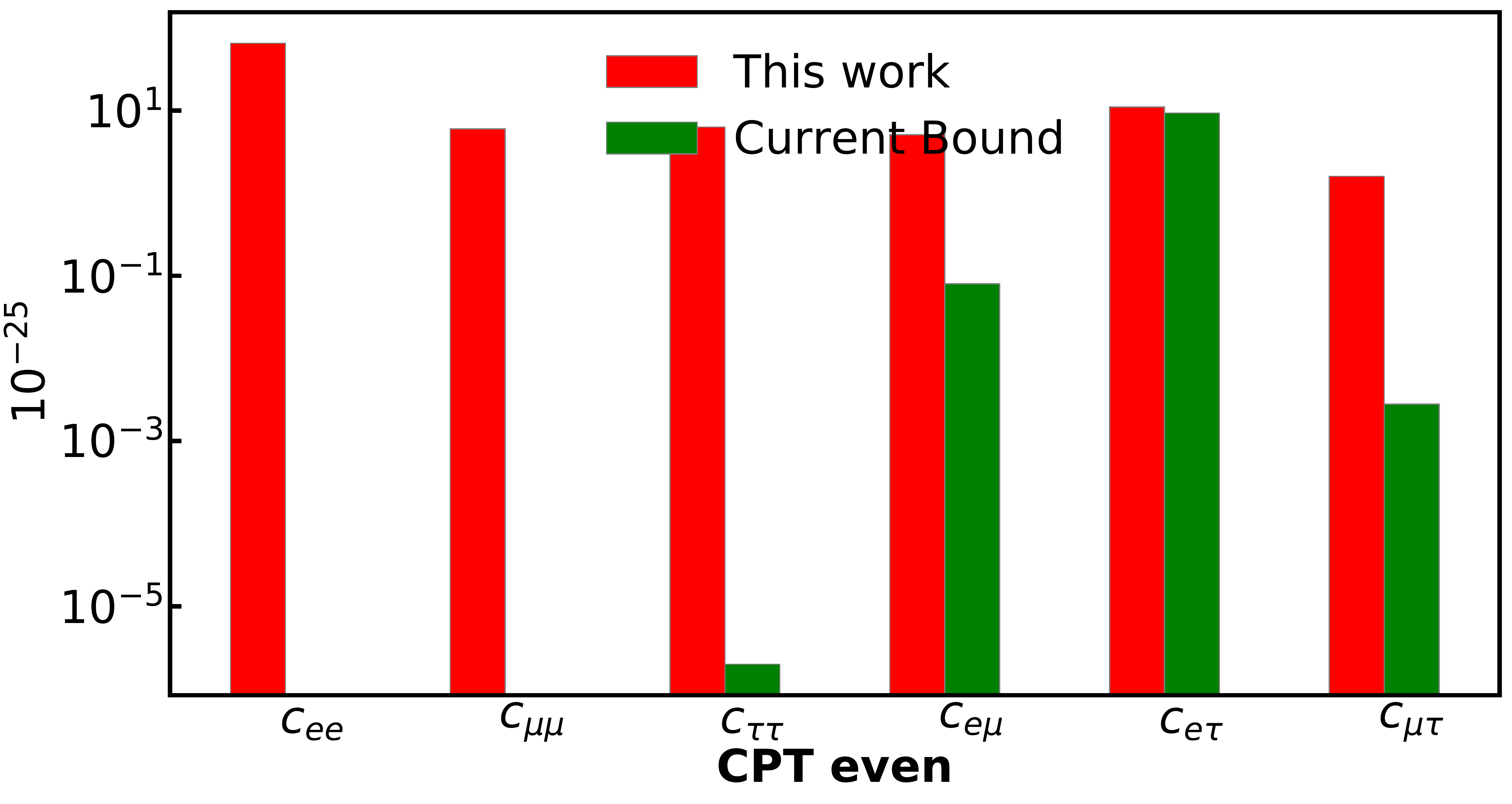}
\end{minipage}%
\caption{In the left panel, we have shown our limit for CPT odd parameters (red bars) and current best bounds (green bars) which are listed in Table \ref{table:all-exp}. In right panel, we have shown results for CPT even parameters. Note that the current bound for $c_{\tau\tau}$ is at 90\% C.L whereas all the other bounds are at 95\% C.L. }
\label{fig:bar-plot}
\end{figure*}

In this section, we will compare the results obtained in our study with previous results. It is important to note that in our work Lorentz invariance violation is isotropic in nature and this manifests as an alteration of the neutrino oscillation probabilities. However, Lorentz invariance violation can be also anisotropic and its effect can be studied by looking at the sidereal variations in the event rates in an experiment. All the bounds that we provide in this section are for isotropic LIV. For bounds on the anisotropic LIV, we refer to Ref.~\cite{Kostelecky:2008ts}. In Table~\ref{table:all-exp}, we have compiled all existing bounds on LIV parameters, along with our bounds obtained from the combined analysis of ICAL, T2HK, and DUNE. For $a_{\alpha \beta}$, we present the bounds obtained from ICAL+T2HK+DUNE, while for $c_{\alpha \beta}$, we present the bounds obtained from ICAL+DUNE. Individual bounds from these experiments can be found in Tables~\ref{table:para-all-1} and \ref{table:para-all-2}. Additionally, we have presented a bar chart in Fig.~\ref{fig:bar-plot} to show the comparison between our bounds and the current best available bounds on LIV parameters.

\begin{table*}[t]
 \begin{center}
 \scalebox{0.9}{
\begin{tabular}{ |p{3.6cm}|p{6.6cm}|p{7.2cm}|p{1.0cm}|} 
\hline
Experiments & Details & 95$\%$ C.L.(1 dof) $a_{\alpha\beta}$ in GeV $(c_{\alpha\beta})$ & Ref.  \\
\hline
SK (Atmospheric) & SK atmospheric data & $e\mu$ = 1.8$\times 10^{-23}$ (8$\times 10^{-27}$)    & \cite{Super-Kamiokande:2014exs}  \\
&&$e\tau$ = 2.8$\times 10^{-23}$ (9.3$\times 10^{-25}$)& \\
&&$\mu\tau$ = 5.1$\times 10^{-24}$ (4.4$\times 10^{-27}$)& \\
\hline
IceCube & IceCube data analyzed for $\mu\tau$ LIV & $\mu\tau$ = 2.0$\times 10^{-24}$ (2.7$\times 10^{-28}$) $90\%$ C.L. &  \cite{IceCube:2017qyp}\\ 
    & parameters for 3, 4, 5, 6 and 7 dim. operators   & $\tau\tau$ = 2.0$\times 10^{-26}$ (2.0$\times 10^{-31}$) $90\%$ C.L.  &   \\
   
\hline
ICAL & Atmospheric neutrino simulated  & $e\mu$ = 1.34$\times 10^{-23}$ (N.A)   &   \cite{Sahoo:2021dit}\\
&between 1-25 GeV range for $a_{\alpha\beta}$&$e\tau$ = 1.58$\times 10^{-23}$ (N.A)& \\
&&$\mu\tau$ = 2.2$\times 10^{-24}$ (N.A)& \\
\hline
DUNE & Long-baseline neutrino simulated & $e\mu$ = 7$\times 10^{-24}$ (N.A) &  \cite{Barenboim:2018ctx} \\
&between 0.2-10 GeV range for $a_{\alpha\alpha}$&$e\tau$ = 1.0$\times 10^{-23}$ (N.A)& \\
&and $a_{\alpha\beta}$&$\mu\tau$ = 1.7$\times 10^{-23}$ (N.A)& \\
&&$ee$ = (-2.5, 3.2)$\times 10^{-23}$ (N.A)& \\
&&$\mu\mu$ = (-3.7, 2.8)$\times 10^{-23}$ (N.A)& \\
\hline
T2K+NOVA & NOVA and T2K long-baseline & $e\mu$ = 3.6$\times 10^{-23}$ (N.A) & \cite{Majhi:2019tfi}\\
&experiments  simulation for LIV &$e\tau$ = 1.08$\times 10^{-22}$ (N.A)& \\
&sensitivity for $a_{\alpha\alpha}$ and  $a_{\alpha\beta}$&$\mu\tau$ = 8$\times 10^{-23}$ (N.A)& \\
&&$ee$ = (-5.5, 3.4)$\times 10^{-22}$ (N.A)& \\
&&$\mu\mu$ = (-1.07, 1.18)$\times 10^{-22}$ (N.A)& \\
&&$\tau\tau$ = (-1.12, 0.9)$\times 10^{-22}$ (N.A)& \\
\hline
DUNE+P2O & Long-baseline DUNE and P2O data & $e\mu$ = 4.7$\times 10^{-24}$ (N.A) &  \cite{Fiza:2022xfw}\\
&simulated for LIV sensitivity for $a_{\alpha\alpha}$ &$e\tau$ = 6$\times 10^{-24}$ (N.A)& \\
&and $a_{\alpha\beta}$&$\mu\tau$ = 1.3$\times 10^{-23}$ (N.A) & \\
&&$ee$ = (-2.6, 3.3)$\times 10^{-23}$ (N.A)& \\
&&$\mu\mu$ = (-1.5, 1.6)$\times 10^{-23}$ (N.A)& \\
\hline
ICAL+T2HK+DUNE   & Combining atmospheric and & $e\mu$ = 4$\times 10^{-24}$ (5.1$\times 10^{-25}$) & This \\
($a_{\alpha\beta}$) &long-baseline experiments with $a_{\alpha\alpha}$,  &$e\tau$ = 6$\times 10^{-24}$ (1.1$\times 10^{-24}$)& work \\
&$a_{\alpha\beta}$, $c_{\alpha\alpha}~{\rm and}~c_{\alpha\beta}$. ICAL with extended &$\mu\tau$ = 9.5$\times 10^{-24}$ (1.6$\times 10^{-25}$) & \\
ICAL+DUNE  &range 1-100 GeV. DUNE with recent &$ee$ = (-2.1, 2.1)$\times 10^{-23}$ ((-4.2, 6.5)$\times 10^{-24}$)& \\
($c_{\alpha\beta}$)&TDR included. Exploring $\phi_{\alpha\beta}$ impact &$\mu\mu$ = (-1.8, 1.9)$\times 10^{-23}$ ((-7.4, 6)$\times 10^{-25}$)& \\
&on experiments and combined analysis.&$\tau\tau$ = (-1.5, 1.5)$\times 10^{-23}$ ((-6.7, 6.4)$\times 10^{-25}$)& \\
\hline
\end{tabular}}
\caption{95$\%$ C.L. (1 dof) limit of the LIV parameters from available literature, except for IceCube. For IceCube, the bounds are available only at 90\% C.L.}
\label{table:all-exp}
\end{center}
\end{table*}

It is worth noting that the study of LIV parameters in the context of DUNE and ICAL has been previously conducted. However, our analysis presents some new features, which we would like to highlight.
\begin{itemize}
    \item Ref.\cite{Sahoo:2021dit} previously studied CPT odd LIV parameters for off-diagonal elements ($a_{\alpha\beta}$, where $\alpha \neq \beta$) for ICAL, assuming that complex phases have a negligible effect on the sensitivity. They used the Nuance MC generator for event generation and limited the muon energy range from 1 to 25 GeV. However, in our analysis, we considered the impact of nontrivial $\phi_{\alpha\beta}$ on the LIV parameters and used the GENIE MC generator, which employs updated cross-section values. Moreover, we implemented an extended binning scheme, which expanded the muon energy range from 25 GeV to 100 GeV. As a result, our sensitivity bounds are more stringent than those obtained in Ref.\cite{Sahoo:2021dit}.
    
    \item In Ref.\cite{Barenboim:2018ctx}, the authors studied CPT odd LIV parameters in DUNE, while Ref.\cite{Fiza:2022xfw} combined P2O with DUNE. Both studies used the conceptual design report of DUNE (CDR) \cite{DUNE:2016rla} for detector configuration, efficiency, resolution, and systematic uncertainties. In our analysis, we used the recently published technical design report of DUNE (TDR) \cite{DUNE:2020txw} to build our detector setup and consider new efficiency, resolution, and systematic uncertainties. As a result, our sensitivity is improved compared to Ref.~\cite{Barenboim:2018ctx}.
\end{itemize}

Now let us compare the bounds for the individual parameters as obtained from different experiments.
\begin{itemize}
    \item The CPT odd diagonal parameters $a_{\alpha\alpha}$ have been studied in the context of DUNE~\cite{Barenboim:2018ctx}, T2K+NO$\nu$A~\cite{Majhi:2019tfi}, and DUNE+P2O~\cite{Fiza:2022xfw}. In Refs.\cite{Fiza:2022xfw} and \cite{Barenboim:2018ctx}, only two independent diagonal parameters, $a_{ee} - a_{\tau\tau}$ and $a_{\mu\mu}-a_{\tau\tau}$, have been considered. In Ref.\cite{Majhi:2019tfi}, all three diagonal parameters, $a_{ee}$, $a_{\mu\mu}$, and $a_{\tau\tau}$, have been taken as independent parameters. In our work, we have also considered all three diagonal parameters as independent parameters. From Table~\ref{table:all-exp}, we can see that our results provide the best sensitivity for $a_{ee}$. 
    
    \item The CPT-odd off-diagonal parameters $a_{\alpha\beta}$ (for $\alpha \neq \beta$) have been studied in all the experiments listed in Table~\ref{table:all-exp}. Among them, the combination of ICAL, T2HK, and DUNE provides the best sensitivity for the parameter $a_{e\mu}$.

    \item We have studied the CPT-odd diagonal parameter $c_{\alpha\alpha}$ in atmospheric neutrinos using the ICAL detector setup and in the long baseline experiment DUNE. For the parameter $c_{\tau\tau}$, IceCube~\cite{IceCube:2017qyp} provides a stronger bound than our results\footnote{Note that the bound for IceCube, listed in Table~\ref{table:all-exp}, is given at 90$\%$ C.L., whereas our results are given at 95$\%$ C.L.}. The other two CPT-even diagonal parameters $c_{ee}$ and $c_{\mu\mu}$ have not been studied in any literature. Our work provides the first-ever bounds on these parameters.

    \item The CPT-even off-diagonal parameters $c_{\alpha \beta}$ (for $\alpha \neq \beta$) have been studied in the context of SK~\cite{Super-Kamiokande:2014exs} and IceCube~\cite{IceCube:2017qyp}. In our work, we have simulated results for DUNE and ICAL detector setups. Our results are comparable to SK for $c_{e\tau}$ only. For $c_{e\mu}$, SK has given the best limit, and for $c_{\mu\tau}$, IceCube has provided the strongest limit.
\end{itemize}

\section{Summary and conclusion}
\label{conc}

In this paper, we presented a comprehensive study of Lorentz invariance violation (LIV) in the context of atmospheric neutrino experiment ICAL and long-baseline experiments T2HK and DUNE. We considered the full parameter space of LIV parameters, which includes six CPT-violating LIV parameters ($a_{\alpha\beta}$) and six CPT-conserving LIV parameters ($c_{\alpha\beta}$). Our objective was to calculate the upper bound on all LIV parameters for individual experiments and their combination. Note that while some LIV parameters have been studied in the context of ICAL and DUNE before, the specifications used in our study differ from previous ones. For DUNE, we used the configuration from the latest technical design report, while for ICAL, we considered an extended energy region of 1 GeV to 100 GeV. Furthermore, we studied the effect of phases associated with LIV parameters in ICAL, which was not considered in earlier studies. As a result, our results are better as compared to previous ones.

The sensitivity of ICAL to LIV parameters mainly comes from the disappearance channel, whereas for DUNE and T2HK, the sensitivity can come from both the appearance and the disappearance channels. At the probability level, we showed that ICAL is mainly sensitive to $a_{\mu\mu}$, $a_{\tau\tau}$, and $a_{\mu\tau}$, amongst the CPT-violating LIV parameters. For CPT-conserving LIV parameters, ICAL is sensitive to all parameters except $c_{ee}$. For DUNE, we showed that the appearance channel is not very sensitive to the parameters $a_{\mu \mu}$, $a_{\mu\tau}$, $c_{\mu \mu}$, and $c_{\mu\tau}$, while being reasonably sensitive to the others. Sensitivity to some of these could still come from DUNE's disappearance channel. For T2HK, the sensitivity to the CPT odd parameters is weak, while the sensitivity to the CPT even parameters is negligible. 

At the $\chi^2$ level, we found that the results are consistent with the conclusions that we derived from the oscillation probabilities. Among the three experiments, DUNE gives the best sensitivity on $a_{ee}$, $a_{e\mu}$, $a_{e\tau}$, and $a_{\mu\tau}$, whereas ICAL gives the best sensitivity on $a_{\mu\mu}$, $c_{ee}$, $c_{\mu \mu}$, $c_{\tau\tau}$, $c_{e \mu}$, $c_{e \tau}$, and $c_{\mu\tau}$. For $a_{\tau\tau}$, the sensitivity of DUNE and ICAL is similar. When comparing the existing bounds on the LIV parameters from different experiments with the combination of T2HK, DUNE, and ICAL (i.e., bounds from ICAL+T2HK+DUNE on $a_{\alpha \beta}$ and bounds from ICAL+DUNE on $c_{\alpha \beta}$), we find that for the parameters $a_{ee}$ ($a_{\mu\mu}$), our results are better (comparable) with DUNE+P2O. For the parameter $a_{e\mu}$, the strongest bounds are from our results. For $c_{e\mu}$, SK has given the best limit, and for $c_{\mu\tau}$, IceCube has given the strongest limit. The bounds from SK on $c_{e\tau}$ are similar to what we obtained from the combination of DUNE and ICAL. For the diagonal CPT-conserving isotropic LIV parameters, our work provides the first ever bounds on $c_{ee}$ and $c_{\mu\mu}$. \\

\begin{acknowledgments}
This work is performed by the members of the INO-ICAL collaboration. We thank the members of the INO-ICAL collaboration for their valuable comments and constructive inputs. The HRI cluster computing facility (http://cluster.hri.res.in) is gratefully acknowledged. This work has been in part funded by Ministry of Science and Education of Republic of Croatia grant No. KK.01.1.1.01.0001. This project has received funding/support from the European Union’s Horizon 2020 research and innovation programme under the Marie Skłodowska -Curie grant agreement No 860881-HIDDeN.
\end{acknowledgments}

\bibliography{LIV-INO-DUNE-T2HK}

\end{document}